\documentclass[twocolumn, times]{aastex63} 
\usepackage{gensymb} 
\usepackage{amsmath}
\usepackage{soul}
\usepackage{natbib}
\usepackage{mathtools}
\graphicspath{{./}{./Pictures/}{./Pictures/ChannelMaps/}}
\allowdisplaybreaks 

\shorttitle{An SMA Survey of Chemistry in Disks around Herbig AeBe Stars}
\shortauthors{J. Pegues et al.}

\begin{document}

\title{An SMA Survey of Chemistry in Disks around Herbig AeBe Stars}

\author{Jamila Pegues}
\altaffiliation{Space Telescope Science Institute (STScI) Postdoctoral Fellow}
\affiliation{
Space Telescope Science Institute, Baltimore, MD 21218, USA}
\affiliation{
Center for Astrophysics $\mid$ Harvard \& Smithsonian,  Cambridge, MA 02138, USA}

\author{Karin I. \"Oberg}
\affiliation{
Center for Astrophysics $\mid$ Harvard \& Smithsonian,  Cambridge, MA 02138, USA}

\author{Chunhua Qi}
\affiliation{
Center for Astrophysics $\mid$ Harvard \& Smithsonian,  Cambridge, MA 02138, USA}

\author{Sean M. Andrews}
\affiliation{
Center for Astrophysics $\mid$ Harvard \& Smithsonian,  Cambridge, MA 02138, USA}

\author{Jane Huang}
\altaffiliation{NASA Hubble Fellowship Program Sagan Fellow}
\affiliation{
Department of Astronomy, University of Michigan, 323 West Hall, 1085 S. University Avenue, Ann Arbor, MI 48109, USA}

\author{Charles J. Law}
\affiliation{
Center for Astrophysics $\mid$ Harvard \& Smithsonian,  Cambridge, MA 02138, USA}

\author{Romane Le Gal}
\affiliation{Univ. Grenoble Alpes, CNRS, Institut de Planétologie et d’Astrophysique de Grenoble (IPAG), 38000 Grenoble, France}
\affiliation{Institut de Radioastronomie Millimétrique (IRAM), 38406 Saint-Martin d’Hères, France}
\affiliation{Center for Astrophysics \textbar\, Harvard \& Smithsonian, 60 Garden St., Cambridge, MA 02138, USA}

\author{Luca Matr\`{a}}
\affiliation{Centre for Astronomy, School of Physics, National University of Ireland, Galway, University Road, Galway, Ireland}

\author{David J. Wilner}
\affiliation{
Center for Astrophysics $\mid$ Harvard \& Smithsonian,  Cambridge, MA 02138, USA}


\begin{abstract}

Protoplanetary disks around Herbig AeBe stars are exciting targets for studying the chemical environments where giant planets form.  Save for a few disks, however, much of Herbig AeBe disk chemistry is an open frontier. We present a Submillimeter Array (SMA) $\sim$213-268 GHz pilot survey of mm continuum, CO isotopologues, and other small molecules in disks around five Herbig AeBe stars (HD 34282, HD 36112, HD 38120, HD 142666, and HD 144432).  We detect or tentatively detect $^{12}$CO 2--1 and $^{13}$CO 2--1 from four disks; C$^{18}$O 2--1 and HCO$^+$ 3--2 from three disks; HCN 3--2, CS 5--4, and DCO$^+$ 3--2 from two disks; and C$_2$H 3--2 and DCN 3--2 from one disk each.  H$_2$CO 3--2 is undetected at the sensitivity of our observations.  {The mm continuum images of HD 34282 suggest a faint, unresolved source $\sim$5\farcs0 away, which could arise from a distant orbital companion or an extended spiral arm.}  We fold our {sample} into a compilation of T Tauri and Herbig AeBe/F disks from the literature.  {Altogether}, most line fluxes generally increase with mm continuum flux. {Line flux ratios between CO 2--1 isotopologues are nearest to unity for the Herbig AeBe/F disks.  This may indicate emitting layers with relatively similar, warmer temperatures and more abundant CO relative to disk dust mass.  Lower HCO$^+$ 3--2 flux ratios may reflect} less ionization in Herbig AeBe/F disks.  {Smaller} detection rates and flux ratios for DCO$^+$ 3--2, DCN 3--2, and H$_2$CO 3--2 suggest smaller regimes of cold chemistry around the luminous Herbig AeBe/F stars.
\end{abstract}

%

\section{Introduction}
\label{sec_introduction}

Protoplanetary disks are the birthplaces of planetary systems.  Disks around pre-main-sequence A-type and B-type stars (``Herbig AeBe'' stars) are typically larger and brighter observational targets in {millimeter-wavelength (mm)} continuum emission than disks around less massive stars~\citep[see, e.g., empirical relationships between disk size derived from the millimeter-wavelength continuum, flux, and stellar mass in Figure 5 of][and references therein]{cite_andrewsetal2020}.  {Herbig} Ae stars are also expected to generally host more massive planets compared to less massive stars~\citep[e.g.,][]{cite_bowleretal2010, cite_johnsonetal2010}.  Protoplanetary disks around Herbig AeBe stars (``Herbig AeBe disks'') are therefore ideal targets for studying the formation environments of giant planets~\citep[e.g.,][]{cite_quanzetal2015a}.

Indeed, direct imaging has already detected gas giant planets around some A-type stars~\citep[e.g.,][]{cite_johnsonetal2007, cite_maroisetal2008, cite_carsonetal2013, cite_rameauetal2013, cite_quanzetal2015b}.  Depletions of heavy elements in the photospheres of Herbig AeBe stars with disks suggest that at least $\sim$30\% of their disks may host yet-undetected planets of $\sim$0.1-10 Jupiter masses~\citep{cite_kamaetal2015}.  Spatially-resolved observations from infrared to millimeter wavelengths have also revealed complex structures in the gas and dust of many Herbig AeBe disks, including spiral arms, asymmetries, cavities, and rings~\citep[e.g.,][]{cite_linetal2006, cite_isellaetal2007, cite_fukagawaetal2010, cite_hondaetal2012, cite_dongetal2018}.  Several studies have attributed such complex structures to nascent planets embedded within the disks~\citep[e.g.,][]{cite_gradyetal2013, cite_dongetal2015, cite_baeetal2016, cite_matteretal2016, cite_huangetal2018, cite_isellaetal2018, cite_pinteetal2018, cite_zhangetal2018}.

The compositions of these giant planets are seeded by the gaseous and icy molecular distributions of their ancestral protoplanetary disks.  It is currently unclear how similar or different Herbig AeBe disk compositions are to the better studied T Tauri disks.  Disks around Herbig AeBe stars are larger, hotter, and more highly irradiated in the ultraviolet than disks around lower mass stars.  As density, temperature, and radiation are key drivers of protoplanetary disk chemistry~\citep[e.g.,][]{cite_aikawaetal1999}, these environmental differences may foster different molecular {compositions and} distributions in Herbig AeBe disks compared to their lower mass counterparts.

While Herbig AeBe disk studies are quite common, there are few such studies of {their} disk chemistry.  {The majority of Herbig disk} studies have focused on CO gas and dust in order to characterize the disks' surface density profiles and structures~\citep[e.g.,][]{cite_dentetal2005, cite_boissieretal2011, cite_vanderplasetal2015, cite_gravitycollabetal2019, cite_stapperetal2022}.
Most existing chemistry surveys that do contain Herbig AeBe disks have targeted infrared wavelengths, which probe $<$10 au of these disks~\citep[e.g.,][]{cite_pontoppidanetal2010, cite_fedeleetal2011, cite_meeusetal2012}.

{For species other than CO, few Herbig AeBe disks have been surveyed at millimeter wavelengths, which probe $>$10 au.}
Notably \cite{cite_obergetal2010, cite_obergetal2011} conducted a millimeter-wavelength survey of a 12-disk sample containing one Herbig Ae disk {(MWC 480)} and three F-star disks.  Their survey included lines of HCO$^+$, CN, HCN, DCN, DCO$^+$, and H$_2$CO.  They found that small molecules (other than CO) were generally detected at lower rates for the Herbig AeBe/F disks than for the disks around lower mass stars.  They attributed the lower detection rates to a combination of (1) the brighter far-ultraviolet fields inherent to Herbig Ae stars and (2) the warmer disks themselves, which would cultivate less CO freeze-out and generally diminish abundances of molecules that form significantly in cold environments (e.g., deuterated molecules and H$_2$CO).

Beyond their survey, only a few Herbig AeBe disks have been well-characterized in {chemistry at millimeter (mm) wavelengths}: AB Aur, HD 163296, and MWC 480{~\citep[e.g.,][]{cite_thietal2004, cite_schreyeretal2008, cite_henningetal2010, cite_fuenteetal2010, cite_chapillonetal2012, cite_chapillonetal2012b, cite_guilloteauetal2013, cite_qietal2013, cite_qietal2013b, cite_graningeretal2015, cite_guzmanetal2015, cite_huangetal2017, cite_bergneretal2018, cite_bergneretal2019, cite_legaletal2019, cite_loomisetal2020, cite_peguesetal2020, cite_mapsI_2021, cite_mapsIII_2021, cite_mapsVI_2021, cite_mapsIX_2021, cite_mapsXII_2021}}.
{Otherwise, Oph IRS 48 has been detected in a mix of commonly and uncommonly targeted molecules, including H$_2$CO, CH$_3$OH, NO, SO, SO$_2$, and CH$_3$OCH$_3$~\citep{cite_vandermareletal2014, cite_boothetal2021, cite_vandermareletal2021, cite_brunkenetal2022}.}
Other molecule-specific exceptions include DCO$^+$ from HD 169142~\citep{cite_maciasetal2017}, {H$_2$CO, CH$_3$OH, and SO from HD 100546~\citep{cite_boothetal2018, cite_boothetal2023}}, HC$^{15}$N and H$^{13}$CO$^+$ from HD 97048~\citep{cite_boothetal2019}, and upper limits of CN, H$_2$CO, and SO from HD 36112~\citep[also known as MWC 758;][]{cite_guilloteauetal2013}.

{A subset of these studies~\citep[e.g.,][]{cite_guzmanetal2015, cite_legaletal2019, cite_loomisetal2020, cite_peguesetal2020} compared mm-wavelength observations between the Herbig Ae disks and T Tauri disks in their samples.  Altogether, they found tentative differences in the inventories of carbon, nitrogen, oxygen, and sulfur-bearing molecules between the two disk types.  The differences were attributed to differences in disk temperature, photochemical environment, and/or regions of CO freeze-out - all of which are properties derivative of the different luminosities and effective temperatures of the central stars.  These chemistry studies, however, have each considered only one or two Herbig Ae disks in total, and have often called for more observations of Herbig Ae disks to test these tentative hypotheses over larger samples.}

In this study, we present a molecular line survey of four Herbig Ae disks and one Herbig Be disk observed within $\sim$213-268 GHz with the Submillimeter Array (SMA).  We leverage the wide spectral range and high spectral resolution capabilities of the SMA to probe a suite of molecular lines in an efficient way, and to help enable follow-up of these lines at higher spatial resolution in future studies.
In Section~\ref{sec_sample}, we describe the target disks, SMA observations and data reduction, target molecular lines, and a comparative sample of disks from the literature.  In Section~\ref{sec_analysis}, we overview techniques used to analyze the molecular line emission.  In Section~\ref{sec_results}, we present our results, including molecular line detections, velocity-integrated emission maps, spectra, and extracted fluxes.  In Section~\ref{sec_discussion}, we interpret these results in the context of existing disk chemistry {observations} across the pre-main-sequence stellar mass distribution.  We summarize our findings in Section~\ref{sec_summary}.
%

\section{SAMPLE}
\label{sec_sample}

\subsection{Target Disks}
\label{sec_sample_systems}

\begin{deluxetable*}{lcccccccccccc}
\setlength{\tabcolsep}{3.4pt}
\tablecaption{Stellar and Disk Characteristics. \label{table_char}}
\tablehead{
Star+Disk           & Spectral  & R.A.$^{[1]}$ & Decl.$^{[1]}$ & Dist.$^{[1]}$     & $t_*^{[2]}$             & $L_*^{[2]}$               & $M_*^{[2]}$            & $T_\mathrm{eff}^{[2]}$ & $v_\mathrm{LSR}$\tablenotemark{\footnotesize{a}} & P.A.\tablenotemark{\footnotesize{b}} & Incl.\tablenotemark{\footnotesize{b}} & $R_\mathrm{mm,90}$ \\
System & Type    & {(}J2000{)}   & {(}J2000{)}     &  {(}pc{)} &   {(}Myr{)}    & ($L_\Sun$)    & ($M_\Sun$)    &  (K) & (km s$^{-1}$) & ($\degree$) & ($\degree$) & (au)}
 \colnumbers \startdata
\hline
HD 34282\tablenotemark{\footnotesize{c,d}}  & A3$^{[3]}$        & 05:16:00.48  & $-$09:48:35.41  & 309$\pm$2           & 6.5$^{+2.4}_{-0.6}$  & 9.55$^{+1.17}_{-0.84}$    & 1.45$^{+0.07}_{-0.07}$ & 9500$^{+250}_{-250}$   & $-$2.7                       & 117$^{[4]}$              & 59$^{[4]}$       &   260$^{[5]}$      \\
HD 36112\tablenotemark{\footnotesize{c,d}}  & A5$^{[6]}$          & 05:30:27.53  & $+$25:19:56.65  & 156$\pm$1           & 8.3$^{+0.4}_{-1.4}$  & 11.0$^{+3.49}_{-1.84}$   & 1.56$^{+0.11}_{-0.08}$ & 7605$^{+225}_{-297}$   & 5.9                        & 245$^{[7]}$             & 21$^{[7]}$        &   123$^{[5]}$        \\
HD 38120  & B9$^{[8]}$         & 05:43:11.89  & $-$04:59:49.91  & 385$\pm$5           & 3.5$^{+13.6}_{-1.4}$ & 52.5$^{+54.7}_{-19.4}$ & 2.37$^{+0.43}_{-0.24}$ & 10700$^{+800}_{-900}$  & 16.4                       & 165$^{[9]}$              & 49$^{[9]}$      &   $\sim$100        \\
HD 142666\tablenotemark{\footnotesize{c}} & A8$^{[10]}$           & 15:56:40.01  & $-$22:01:40.36  & 146$\pm$1           & 9.3$^{+0.8}_{-0.5}$  & 8.71$^{+0.84}_{-0.95}$    & 1.49$^{+0.08}_{-0.08}$ & 7500$^{+250}_{-250}$   & 4.0                          & 162$^{[11]}$              & 62$^{[11]}$   &   52$^{[5]}$           \\
HD 144432\tablenotemark{\footnotesize{d}} & A9$^{[10]}$        & 16:06:57.94  & $-$27:43:10.15  & 155$\pm$1           & 5.0$^{+0.3}_{-0.6}$  & 9.33$^{+0.90}_{-0.21}$    & 1.39$^{+0.07}_{-0.07}$ & 7500$^{+250}_{-250}$   & 5.8                       & 76$^{[9]}$             & 30$^{[9]}$      &   $\sim$100
\enddata
\tablecomments{Column 1: Star+disk system.  Column 2: Spectral type.  Columns 3, 4, and 5: Right ascension, declination, and distance, respectively, from \textit{Gaia}~\citep{cite_gaia2016, cite_gaia2018b}.  Columns 6, 7, 8, and 9: Stellar age, luminosity, mass, and effective temperature, respectively, from the compilation of~\cite{cite_vioqueetal2018}, which updated values from the literature as needed using parallaxes from the \textit{Gaia} Data Release 2.  Column 10: Systemic velocities in the local standard of rest frame, which were (1) converted from heliocentric values presented in~\cite{cite_dentetal2005} and then (2) adjusted as needed to better match the center of the $^{12}$CO 2--1 emission spectra.  We were unable to adjust the systemic velocity for HD 38120 (see Section~\ref{sec_results_detections}).  Columns 11 and 12: Position and inclination angles, respectively.  The position angles are measured East of North.  Column 13: disk radii from the literature, derived from 90\% of the $\sim$1.3 mm continuum emission.  No mm continuum radii are known for HD 38120 and HD 144432, so here we assume $\sim$100 au.}
\tablenotetext{a}{Adapted from~\cite{cite_dentetal2005}, as described in the main body of this caption.}
\tablenotetext{b}{Estimated from near-infrared, submillimeter, or millimeter continuum.}
\tablenotetext{c}{HD 34282, HD 36112, and HD 142666 are also known as V1366 Ori, MWC 758, and V1026 Sco, respectively, in the literature.}
\tablenotetext{d}{Orbital companions have been {reported} for HD 34282~\citep{cite_wheelwrightetal2010}, HD 36112~\citep{cite_thomasetal2007}, and HD 144432~\citep{cite_maheswaretal2002}.}
\tablerefs{[1]~\cite{cite_gaia2016, cite_gaia2018b}; [2]~\cite{cite_vioqueetal2018}; [3]~\cite{cite_merinetal2004}; [4]~\cite{cite_vanderplasetal2017}; [5]~\cite{cite_stapperetal2022}; [6]~\cite{cite_zuckermanetal1995}; [7]~\cite{cite_isellaetal2010}; [8]~\cite{cite_juhaszetal2010}; [9]~\cite{cite_gravitycollabetal2019}; [10]~\cite{cite_meeusetal1998}; [11]~\cite{cite_huangetal2018}.}
\end{deluxetable*}

We target four protoplanetary disks around Herbig Ae stars and one disk around a Herbig Be star.  {Table~\ref{table_char} summarizes their stellar and disk characteristics.}  These star+disk systems were chosen because they had previously been detected in at least one CO emission line~\citep[e.g., CO J=3--2 observations by][]{cite_dentetal2005, cite_vanderplasetal2017} but were otherwise not characterized in millimeter-wavelength chemistry.  The systems are associated with the Taurus-Auriga (HD 36112), Orion OB Ic (HD 38120), and Upper Scorpius (HD 142666 and HD 144432) star-forming regions, or are isolated (HD 34282)~\citep[see the compilations by][]{cite_sandelletal2011, cite_alecianetal2013_arxiv}.
{The stars were selected to span a range of spectral types, with four stars of types A3-A9 and the fifth star (HD 38120) of type B9}~\citep{cite_zuckermanetal1995, cite_meeusetal1998, cite_merinetal2004, cite_juhaszetal2010}.
Three of the star+disk systems are reported to have orbital companions: HD 34282~\citep{cite_wheelwrightetal2010}, HD 36112~\citep{cite_thomasetal2007}, and HD 144432~\citep{cite_maheswaretal2002}.

{Altogether, these five Herbig AeBe disks have lower CO luminosities than the Herbig Ae disks that have been relatively well characterized in chemistry so far~\citep[i.e., AB Aur, HD 163296, and MWC 480, found to be brightest in CO 3--2 emission in the survey by][]{cite_dentetal2005}.  Our observations of these disks serve as 1) an investigative probe into a larger and more representative sample of Herbig AeBe disk chemistry and 2) a foundation for potential follow-up observations at higher spatial resolution in the future.}

Studies have found complex morphologies in the dust disks of a subset of these systems, including a cavity within $\sim$88 au, a bright point-like source, and a single spiral arm in HD 34282~\citep[millimeter-wavelength continuum and scattered light observations of][respectively]{cite_vanderplasetal2017, cite_Deboeretal2020}; bright dust clumps and at least three spiral arms in HD 36112{~\citep[millimeter-wavelength continuum and scattered light observations of][respectively]{cite_boehleretal2018, cite_dongetal2018, cite_reggianietal2018}}; and an inner cavity of radius $\sim$10 au, rings, and asymmetries in HD 142666~\citep[millimeter-wavelength continuum observations of][]{cite_rubinsteinetal2018, cite_huangetal2018}.  There has been a lack of detected scattered light from HD 144432~\citep{cite_monnieretal2017}, which may indicate that the disk was truncated by its stellar companions~\citep[see discussion by][]{cite_monnieretal2017}, or that the disk has already lost much of its gas given the old age of its association~\citep[Upper {Scorpius}, $>$10 Myr;][]{cite_pecautetal2012}.

\subsection{SMA Observations and Data Reduction}
\label{sec_sample_obs}

\begin{deluxetable*}{lccccccccc}
\setlength{\tabcolsep}{3.9pt}
\tablecaption{SMA Observations. \label{table_obs}}
\tablehead{
Project & Date  & Star+Disk    & Config.   & \#    & Time on & Baselines & Bandpass & Flux & Gain    \\
Code    &       & System          &           & of Ant.              & Source                          &       & Calibrator    & Calibrator    &  Calibrator(s) \\
    &   &   &   &   & (min)                          & (m)       &     &     &   }
 \colnumbers \startdata
\hline
2020B-S007 & 2021 May 1$^\mathbf{*}$       & HD 142666        & Compact       & 6          & 187                   & 18-77         & 3c279               & Titan           & 1517-243, 1626-298 \\
           & 2021 May 1$^\mathbf{*}$       & HD 144432        & Compact       & 6          & 178                   & 18-77         & 3c279               & Titan           & 1517-243, 1626-298 \\
\hline
2020A-S007 & 2020 Nov. 14 & HD 38120         & Subcompact    & 8          & 232                   & 9-69          & 3c84                & Uranus          & 0532+075, 0607-085 \\
\hline
2019B-S037 & 2020 March 10    & HD 36112         & Compact       & 7          & 225                   & 16-77         & 3c279               & Uranus          & 0555+398           \\
           & 2020 March 9     & HD 34282         & Compact       & 7          & 200                   & 16-77         & 3c279               & Uranus          & 0541-056           \\
           & 2020 Feb. 27$^\mathbf{*}$ & HD 142666        & Compact       & 7          & 178                   & 16-77         & 3c279               & Titan           & 1517-243, 1626-298 \\
           & 2020 Feb. 27$^\mathbf{*}$ & HD 144432        & Compact       & 7          & 178                   & 16-77         & 3c279               & Titan           & 1517-243, 1626-298
\enddata
\tablecomments{Column 1: SMA project code.  Column 2: Date of observations.  Column 3: Star+disk system observed.  Column 4: SMA configuration.  Column 5: Time spent per source.  Column 6: Number of baselines.  Columns 7, 8, and 9: Bandpass, flux, and gain calibrators, respectively. \\ *: HD 142666 and HD 144432 were observed during the same observing sessions.}
\end{deluxetable*}

All disks were observed with the Submillimeter Array\footnote{The Submillimeter Array is a joint project between the Smithsonian Astrophysical Observatory and the Academia Sinica Institute of Astronomy and Astrophysics and is funded by the Smithsonian Institution and the Academia Sinica.}~\citep[SMA;][]{cite_hoetal2004} during projects 2019B-S037, 2020A-S007, and 2020B-S007.
Table~\ref{table_obs} summarizes the observational details.  Four of the disks were observed in the compact configuration with seven antennas, while the fifth (HD 38120) was observed in the subcompact configuration with eight antennas.  HD 142666 and HD 144432 were observed together during two shared tracks, while all other disks were observed during individual tracks.

The observations were conducted with the SMA SWARM correlator, which at the time of the initial observations consisted of 16 $\sim$2 GHz bands, or $\sim$32 GHz in total per source\footnote{The SWARM correlator has since been upgraded and now allows a total spectral bandwidth of $\sim$48 GHz.  HD 38120 and the second shared track targeting HD 142666 and HD 144432 were observed with the upgraded SWARM correlator, but we do not use the additional bandwidth in this study due to the disks' faint molecular line emission.}.
For the 230 GHz receiver, we tuned band S1 in the upper spectral band to the $^{12}$CO J=2--1 line (230.538 GHz).  {For HD 142666 and HD 144432, which were expected to be relatively inclined disks~\citep[based on the broad Keplerian profiles fitted to their $^{12}$CO 3--2 emission by][]{cite_dentetal2005}, we shifted the 230 GHz setup by} $-$0.2 GHz to permit wider coverage of both the C$^{18}$O J=2--1 line (219.560 GHz) and the DCN J=3--2 line (217.239 GHz).  For the 240 GHz receiver, we tuned band S4 in the upper spectral band to the HCO$^+$ J=3--2 line (267.558 GHz).

This setup permitted coverage of molecular lines within four spectral bands: $\sim$213-221 GHz, $\sim$229-237 GHz, $\sim$244-252 GHz, and $\sim$260-268 GHz.\footnote{Each $\sim$8 GHz spectral band was observed in four $\sim$2 GHz bands.  We removed the $\sim$0.1 GHz from the edges of each band {during analysis} to avoid spectral artifacts.}
The native resolution of the observations was $\sim$140 kHz ($\sim$0.16-0.20 km s$^{-1}$ across the bands) for four disks.  For the fifth and faintest disk (HD 38120), we binned the SMA data by 4, reducing the resolution from the native 140 kHz to $\sim$559 kHz ($\sim$0.63-0.79 km s$^{-1}$ across the bands).
%

We calibrated the SMA data using the \textit{MIR} package\footnote{See the \textit{MIR} Cookbook (\url{https://www.cfa.harvard.edu/~cqi/mircook.html}) for standard calibration practices.}.  We performed continuum subtraction for each SMA band with the \textit{Common Astronomy Software Applications} package{~\citep[CASA;][]{cite_casa_1, cite_casa_2}} version 4.7.2.
For \textsc{clean}ing, we constructed a \textsc{clean} mask that encompassed the mm continuum {and $^{12}$CO 2--1} emission, and we used the mask to uniformly \textsc{clean} all channels down to 3$\sigma$.
%

After the continuum subtraction, we imaged with spectral resolution coarser than the native SWARM resolution to enhance the signal-to-noise for any line emission.  We used 0.5 km s$^{-1}$ velocity bins for HD 34282 and HD 36112 and 2 km s$^{-1}$ bins for the fainter, more highly inclined disks HD 142666 and HD 144432.

\subsection{Target Molecular Lines}
\label{sec_sample_mol}

\begin{deluxetable}{lccc}
\tablecaption{Target Molecular Lines. \label{table_mol}}
\tablehead{
Molecular      & Frequency & A$_\mathrm{ul}$      & $E_\mathrm{u}$  \\
Line          &  (GHz)  & (s$^{-1}$)  & (K)  }
\colnumbers 
\startdata
\hline
$^{12}$CO J=2--1             & 230.538   & 6.91$\times$10$^{-7}$   & 16.6 \\
$^{13}$CO J=2--1             & 220.399   & 6.08$\times$10$^{-7}$   & 15.9 \\
C$^{18}$O J=2--1             & 219.560    & 6.01$\times$10$^{-7}$   & 15.8 \\
HCO$^+$ J=3--2               & 267.558   & 1.48$\times$10$^{-3}$   & 25.7 \\
CS J=5--4   & 244.936   & 3.49$\times$10$^{-4}$ & 35.3  \\
HCN J=3--2\tablenotemark{\footnotesize{a}}                   & 265.886   & 8.36$\times$10$^{-4}$   & 25.5 \\
C$_2$H N=3--2, J=7/2--5/2\tablenotemark{\footnotesize{a}}                & 262.004   & 5.31$\times$10$^{-5}$   & 25.1 \\
DCN J=3--2\tablenotemark{\footnotesize{a}}                   & 217.239   & 4.59$\times$10$^{-4}$   & 20.9 \\
DCO$^+$ J=3--2\tablenotemark{\footnotesize{a}}               & 216.113   & 7.66$\times$10$^{-4}$   & 20.7 \\
H$_2$CO 3$_{03}$--2$_{02}$ & 218.222   & 2.82$\times$10$^{-4}$   & 21.0 \enddata 
\tablecomments{Column 1: Molecular emission line.  Column 2: Frequency.  Columns 3 and 4: Einstein coefficients and upper energy levels, respectively, which were taken from the CDMS database~\citep{cite_cdms2001, cite_cdms2005, cite_cdms2016}.}
\tablenotetext{a}{C$_2$H N=3--2, J=7/2--5/2 includes the hyperfine transitions F=4--3 and F=3--2.  HCN J=3--2, DCN J=3--2, and DCO$^+$ J=3--2 include only the hyperfine transitions blended with their brightest transition (at 265.886 GHz, 217.239 GHz, and 216.113 GHz, respectively).}
\end{deluxetable}

Table~\ref{table_mol} describes the spectroscopic characteristics of the target molecular lines covered by the correlator setup.%
\footnote{We also checked lines of CH$_3$CN, CH$_3$OH, {c-C$_3$H$_2$}, H$^{13}$CN, HC$^{15}$N, HC$_3$N, H$_2$CS, HNCO, H$^{13}$CO$^+$, and N$_2$D$^+$ that fell within the total 213-268 GHz range of the correlator.  These lines were either not detected or were at or beyond the edges of the individual SMA bands.}
These lines have been observed in previous surveys that targeted disks around lower mass pre-main-sequence M-type through G-type stars~\citep[collectively known as ``T Tauri'' stars; e.g.,][]{cite_obergetal2010, cite_obergetal2011, cite_huangetal2017, cite_bergneretal2019, cite_bergneretal2020, cite_miotelloetal2019, cite_legaletal2019, cite_peguesetal2020, cite_peguesetal2021b}, permitting direct comparison of our observations to T Tauri disk chemistry.  Observations and models have shown that these molecules reflect a disk's physical and chemical makeup, including the disk's gas mass and surface density~\citep[the CO isotopologues $^{12}$CO J=2--1, $^{13}$CO J=2--1, and C$^{18}$O J=2--1; e.g.,][]{cite_miotelloetal2014, cite_mapsV_2021}, distributions of small carbon-bearing, nitrogen-bearing, and sulfur-bearing organics~\citep[C$_2$H N=3--2, J=7/2--5/2, HCN J=3--2, and CS J=5--4; e.g.,][]{cite_bergneretal2019, cite_legaletal2019, cite_mapsVI_2021}, cold and deuterated chemistry~\citep[DCN J=3--2, DCO$^+$ J=3--2, and H$_2$CO 3$_{03}$--2$_{02}$; e.g.,][]{cite_huangetal2017, cite_peguesetal2020, cite_mapsX_2021}, and ionization~\citep[HCO$^+$ J=3--2; e.g.,][]{cite_cleevesetal2014, cite_mapsXIII_2021}%
\footnote{These lines are denoted as $^{12}$CO 2--1, $^{13}$CO 2--1, C$^{18}$O 2--1, C$_2$H 3--2, HCN 3--2, CS 5--4, DCN 3--2, DCO$^+$ 3--2, H$_2$CO 3--2, and HCO$^+$ 3--2, respectively, throughout the rest of this paper.}.

\subsection{Disk Sample from the Literature}
\label{sec_sample_lit}

\begin{figure*}
\centering
\resizebox{0.5245\hsize}{!}{
    \includegraphics[trim=0pt 0pt 0pt 0pt, clip]{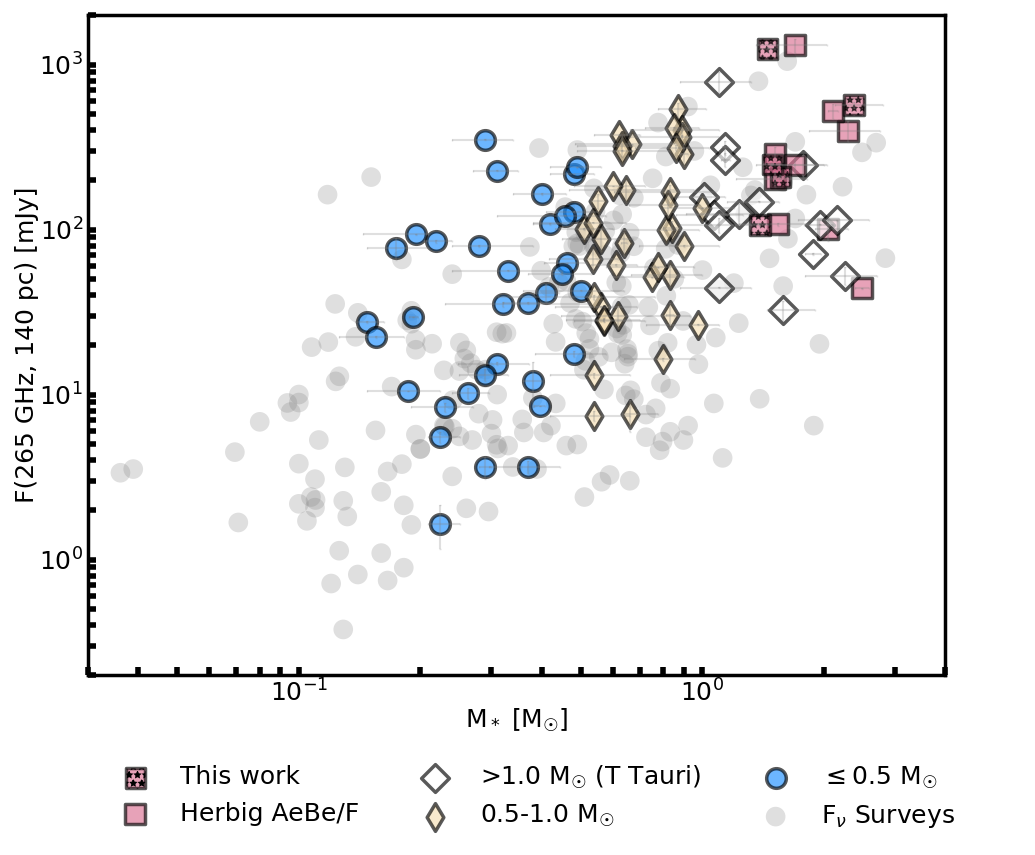}}
\resizebox{0.470\hsize}{!}{
    \includegraphics[trim=60pt 0pt 0pt 0pt, clip]{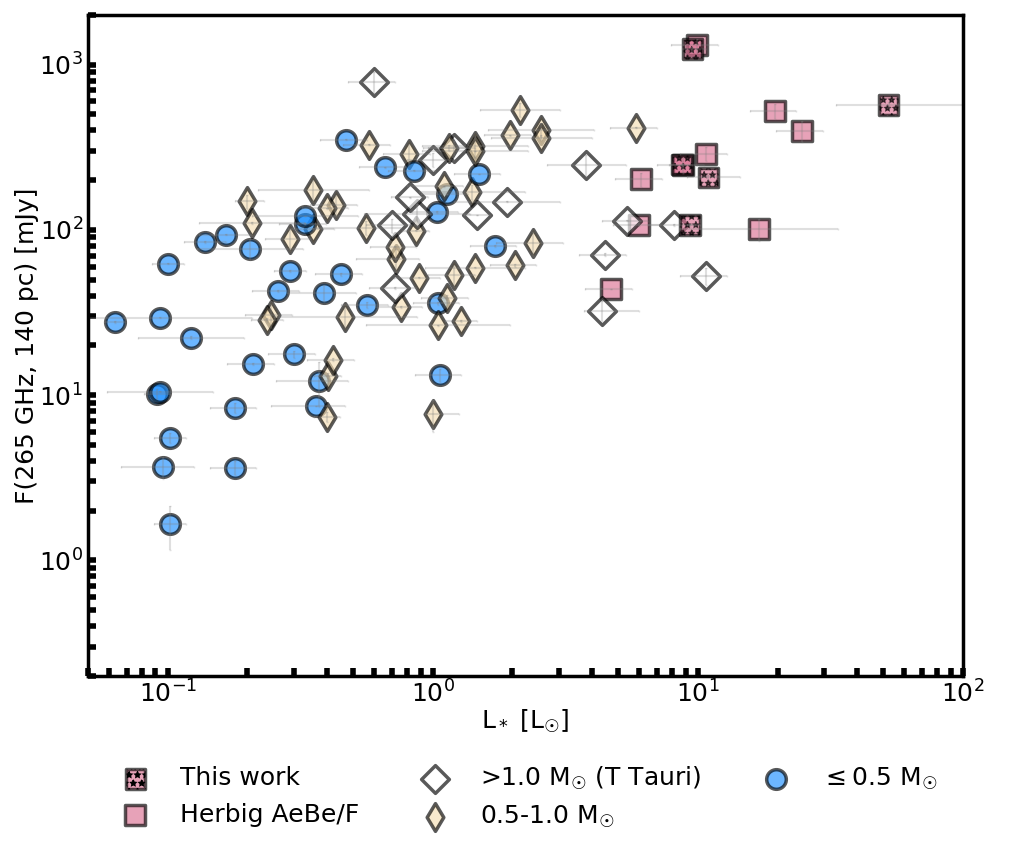}}
\caption{Scaled mm continuum fluxes for Herbig AeBe disks from this work (dotted red squares) as a function of stellar mass (left) and stellar luminosity (right).  The literature sample of {Herbig AeBe/F disks (blank red squares) and T Tauri disks with stellar masses $>$ 1.0 M$_\Sun$ (thick white diamonds), 0.5-1.0 M$_\Sun$ (thin light gold diamonds), and $\leq$ 0.5 M$_\Sun$} (blue circles) are shown for comparison (Section~\ref{sec_sample_lit}, Appendix~\ref{sec_appendix_lit}).  All mm continuum fluxes have been scaled to (1) 265 GHz ($\sim$1.1 mm) using $F_\nu \propto \nu^{2.2}$~\citep[see][]{cite_andrewsetal2020}, where $F_\nu$ is the mm continuum flux at frequency $\nu$, and to (2) 140 pc.  Whenever error in the mm continuum flux, stellar luminosity, or stellar mass is not reported in the literature, we assume the error is 15\%, 20\%, or 20\%, respectively, of the given value.  In the left panel, detected disks from mm continuum surveys of the Chamaeleon I~\citep[][]{cite_pascuccietal2016}, Lupus~\citep[][]{cite_ansdelletal2016}, and Taurus~\citep[][]{cite_andrewsetal2013} star-forming regions are additionally shown in the background as pale gray circles.  These star-forming regions were chosen to span the pre-main-sequence stellar mass distribution.  For visual clarity, we do not show {flux errors} for the disks from the mm continuum surveys.
\label{fig_sample}}
\end{figure*}

To facilitate comparison of chemistry across the pre-main-sequence stellar mass distribution (i.e., from M-type through B-type stars), we have compiled a sample of protoplanetary disks from the literature that have interferometric millimeter-wavelength (mm) observations of at least one of the target lines in Table~\ref{table_mol}%
\footnote{For disks from the literature where H$_2$CO 3--2 was not observed, we use H$_2$CO 4$_{04}$--3$_{03}$ (H$_2$CO 4--3) observations as a proxy {when available}.  Both transitions are expected empirically to give similar fluxes~\citep{cite_peguesetal2020}.}.
Appendix~\ref{sec_appendix_lit} describes the stellar characteristics of the literature sample and lists the total (i.e., disk-integrated and velocity-integrated as applicable) mm continuum and molecular line fluxes and upper limits.

We split the literature sample into four {categories}: (1) $M_* \leq$ 0.5 M$_\Sun$, (2) $M_*$ = 0.5-1.0 M$_\Sun$, (3) {T Tauri disks with} $M_*$ $>$ 1.0 M$_\Sun$, and (4) {disks around F-type, B-type, and A-type pre-main-sequence stars (``Herbig AeBe/F disks''), where $M_*$ is the disk's stellar mass}.
{Figure~\ref{fig_sample} compares mm continuum fluxes for the five disks from this work to (1) the literature disk sample, as well as to (2) disks from mm continuum surveys.
}
{We find that our five target disks and the literature sample are consistent with the generally positive correlation between mm continuum flux and the mass and luminosity of the host star.}

{Figure~\ref{fig_sample} also shows that, though} the {Herbig AeBe/F disks host} the most massive and most luminous stars in the combined sample, they do not all have the largest millimeter-wavelength continuum fluxes.  Assuming mm continuum flux is proportional to disk size by a roughly consistent factor~\citep[see, e.g., empirical relationships in][]{cite_andrewsetal2020},
we note that the disks around less massive stars in our combined sample may have relatively large disks (which, indeed, makes them easier observing targets), and/or the Herbig AeBe/F disks have relatively small or truncated morphologies (possibly due to {disk dust evolution over time, and/or to} their preferentially massive orbital companions, as discussed in Section~\ref{sec_introduction}).

%

\section{Analysis}
\label{sec_analysis}

\subsection{Emission Extraction}
\label{sec_analysis_masks}

To extract the mm continuum emission in each SMA spectral band, we used masks built from two-dimensional Gaussian fits to the mm continuum emission.
We used the model fitted to the combined mm continuum emission (i.e., taken across all four SMA spectral bands) to estimate a characteristic center for each disk, and we used the mask derived from this fit to uniformly extract an unbiased spectrum across all bandwidth{.  The unbiased spectra are plotted in Appendix~\ref{sec_appendix_fullband}, providing a complete view of the SMA bandwidth and the targeted molecular lines for each disk.}

For the {targeted molecular} lines, we extracted emission with Keplerian masks~\citep{cite_kepmask}.  Keplerian masks use the expected Keplerian rotation of a disk to predict spatial distributions of emission within each velocity channel.  These masks can reduce the noise contributed per channel relative to elliptical masks~\citep[e.g.,][]{cite_yenetal2016}.  The edges of these masks were set to 2.5$\times$ the radii in Table~\ref{table_char}~\citep[where $\sim$2.5 is the empirical dust-to-CO disk radius factor measured in][]{cite_andrewsetal2020}.
We used the Keplerian-masked emission across channels to generate image products for each line, including spectra, velocity-integrated emission maps (i.e., zeroth moment maps), and velocity-integrated fluxes.

\subsection{Noise Estimation}
\label{sec_analysis_noise}

We estimated the noise for each image product%
\footnote{Our error estimates do not include 10\% flux calibration uncertainties.}
via bootstrapping over 1000 random samples of emission-free channels/regions.  We used the average root-mean-square (rms; mJy beam$^{-1}$) as an estimate of (1) the noise for the mm continuum emission and (2) the noise per channel for the target line emission.  In both cases, we extracted the random samples from within the corresponding elliptical masks (see Section~\ref{sec_analysis_masks}).

For the mm continuum flux errors and the line flux errors, we took the standard deviation of the noise summed within the elliptical continuum mask (mJy) and within the Keplerian masks (mJy km s$^{-1}$), respectively.
For {the} error in the velocity-integrated emission maps, we used the same methodology as~\cite{cite_bergneretal2018}: we (1) generated velocity-integrated ``noise'' maps from samples of emission-free channels and (2) took the median of the resulting maps as a representation of the map error.

To avoid artifacts at the image edges caused by the primary beam corrections, all random samples were extracted from within a square region -18\farcs0 to +18\farcs0 around the disk center.  We also excluded the inner square region -5\farcs0 to +5\farcs0 when estimating noise for the mm continuum.

%

\section{Results}
\label{sec_results}

\begin{figure*}
\centering
\resizebox{0.95\hsize}{!}{
    \includegraphics{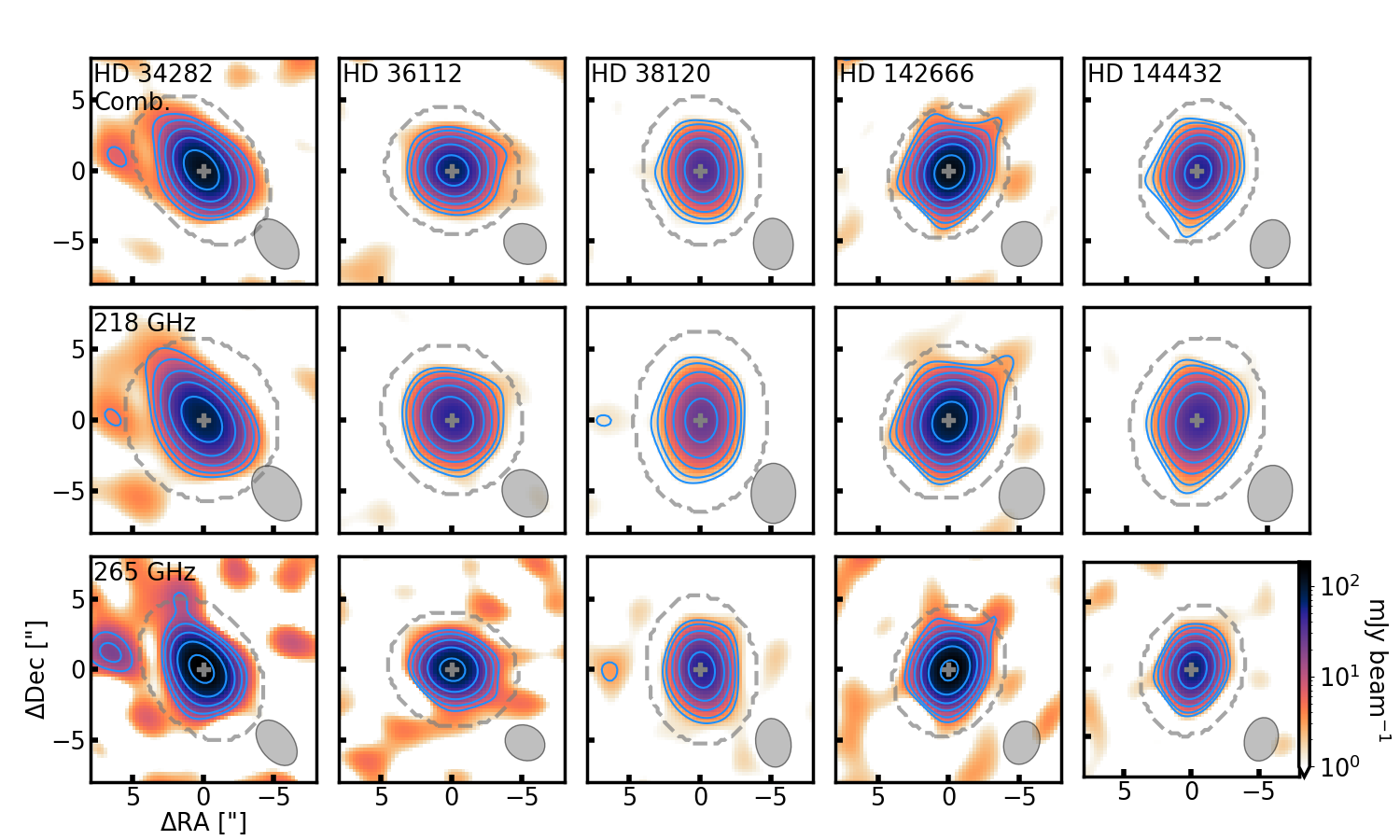}}
\caption{Millimeter-wavelength continuum emission maps for the faintest ($\sim$218 GHz, middle), brightest ($\sim$265 GHz, bottom), and combined (``Comb.''; $\sim$213-268 GHz, top) SMA spectral bands.  Outer edges of the elliptical masks (Section~\ref{sec_analysis_masks}) are outlined in dashed gray.  Contours are [3, 5, 10, 20, 40...]$\times \sigma$, where $\sigma$ is the rms (Section~\ref{sec_analysis_noise}).  `+' signs mark the disk centers estimated from the mm continuum emission (Section~\ref{sec_analysis_masks}).  All panels share the same colorbar (bottom right), which is in log scale to emphasize faint emission.  Synthesized beams are drawn in the lower right corners.
\label{fig_cont}}
\end{figure*}
%

\begin{deluxetable*}{lcccccc}
\tablecaption{Millimeter-Wavelength Continuum Emission. \label{table_contfluxes}}
\tablehead{
Star+Disk      & Repr. Freq.      & Flux        & Peak Em.       & rms    & Synthesized     & Gaus. Param.             \\
System          &    (GHz)       & (mJy)  & (mJy $\times$ beam$^{-1}$) & (mJy $\times$ beam$^{-1}$)   & Beam (P.A.) &   ($\sigma_\mathrm{x}$, $\sigma_\mathrm{y}$, $\theta$)
}
 \colnumbers \startdata
\hline
HD 34282  & Comb. & 130.5 $\pm$ 2.7 & 124.1 $\pm$ 1.3 & 2          & 4$\farcs$0 $\times$ 2$\farcs$6 (36.8$\degree$)  & 1$\farcs$7, 1$\farcs$2, 127.9$\degree$ \\
          & 218   & 100.8 $\pm$ 2.4 & 91.9 $\pm$ 1.0  & 1.5        & \multicolumn{2}{l}{4$\farcs$4 $\times$   2$\farcs$9 (37.0$\degree$)}                     \\
          & 234   & 128.4 $\pm$ 2.6 & 119.3 $\pm$ 0.9 & 1.6        & \multicolumn{2}{l}{4$\farcs$0 $\times$   2$\farcs$5 (33.7$\degree$)}                     \\
          & 249   & 162.8 $\pm$ 3.2 & 152.4 $\pm$ 1.4 & 2.4        & \multicolumn{2}{l}{4$\farcs$0 $\times$   2$\farcs$7 (34.9$\degree$)}                     \\
          & 265   & 205.2 $\pm$ 6.9 & 184.1 $\pm$ 3.0 & 3.6        & \multicolumn{2}{l}{3$\farcs$7 $\times$   2$\farcs$3 (38.0$\degree$)}                     \\
\hline
HD 36112  & Comb. & 83.4 $\pm$ 2.3  & 75.1 $\pm$ 0.9  & 1.3        & 3$\farcs$1 $\times$ 2$\farcs$7 (53.1$\degree$)  & 1$\farcs$3, 1$\farcs$4, 66.5$\degree$  \\
          & 218   & 59.4 $\pm$ 1.0  & 53.7 $\pm$ 0.4  & 0.78       & \multicolumn{2}{l}{3$\farcs$6 $\times$   3$\farcs$0 (38.9$\degree$)}                     \\
          & 234   & 82.3 $\pm$ 1.6  & 74.9 $\pm$ 0.7  & 1.1        & \multicolumn{2}{l}{3$\farcs$1 $\times$   2$\farcs$6 (33.0$\degree$)}                     \\
          & 249   & 102.7 $\pm$ 2.1 & 90.3 $\pm$ 0.9  & 1.3        & \multicolumn{2}{l}{3$\farcs$3 $\times$   2$\farcs$8 (68.2$\degree$)}                     \\
          & 265   & 134.8 $\pm$ 3.9 & 115.0 $\pm$ 1.2 & 2.2        & \multicolumn{2}{l}{2$\farcs$9 $\times$   2$\farcs$5 (63.9$\degree$)}                     \\
\hline
HD 38120  & Comb. & 38.7 $\pm$ 0.9  & 39.3 $\pm$ 0.3  & 0.59       & 3$\farcs$6 $\times$ 2$\farcs$8 (4.8$\degree$)   & 1$\farcs$5, 1$\farcs$2, 84.6$\degree$  \\
          & 218   & 28.6 $\pm$ 0.7  & 28.1 $\pm$ 0.3  & 0.49       & \multicolumn{2}{l}{4$\farcs$3 $\times$   3$\farcs$2 (-0.4$\degree$)}                     \\
          & 234   & 39.1 $\pm$ 1.3  & 39.8 $\pm$ 0.4  & 0.65       & \multicolumn{2}{l}{3$\farcs$8 $\times$   2$\farcs$9 (2.8$\degree$)}                      \\
          & 249   & 43.0 $\pm$ 1.6  & 43.8 $\pm$ 0.3  & 0.69       & \multicolumn{2}{l}{3$\farcs$7 $\times$   2$\farcs$8 (4.4$\degree$)}                      \\
          & 265   & 60.7 $\pm$ 2.5  & 55.8 $\pm$ 0.9  & 0.97       & \multicolumn{2}{l}{3$\farcs$5 $\times$   2$\farcs$5 (9.7$\degree$)}                      \\
\hline
HD 142666 & Comb. & 135.5 $\pm$ 1.4 & 134.6 $\pm$ 0.9 & 1.1        & 3$\farcs$3 $\times$ 2$\farcs$8 (-24.8$\degree$) & 1$\farcs$4, 1$\farcs$2, 114.3$\degree$ \\
          & 218   & 110.0 $\pm$ 1.1 & 106.9 $\pm$ 0.5 & 0.9        & \multicolumn{2}{l}{3$\farcs$7 $\times$   3$\farcs$1 (-21.7$\degree$)}                    \\
          & 234   & 137.3 $\pm$ 1.5 & 134.8 $\pm$ 1.0 & 1.2        & \multicolumn{2}{l}{3$\farcs$4 $\times$   2$\farcs$8 (-21.2$\degree$)}                    \\
          & 249   & 151.6 $\pm$ 1.5 & 143.8 $\pm$ 0.8 & 1.2        & \multicolumn{2}{l}{3$\farcs$3 $\times$   2$\farcs$8 (-23.2$\degree$)}                    \\
          & 265   & 183.1 $\pm$ 2.4 & 174.0 $\pm$ 1.2 & 1.9        & \multicolumn{2}{l}{3$\farcs$1 $\times$   2$\farcs$5 (-21.4$\degree$)}                    \\
\hline
HD 144432 & Comb. & 54.6 $\pm$ 0.7  & 55.4 $\pm$ 0.3  & 0.51       & 3$\farcs$5 $\times$ 2$\farcs$7 (-16.7$\degree$) & 1$\farcs$5, 1$\farcs$2, 106.5$\degree$ \\
          & 218   & 45.9 $\pm$ 0.9  & 44.7 $\pm$ 0.4  & 0.57       & \multicolumn{2}{l}{4$\farcs$0 $\times$   3$\farcs$1 (-14.0$\degree$)}                    \\
          & 234   & 56.4 $\pm$ 1.5  & 56.7 $\pm$ 0.6  & 0.84       & \multicolumn{2}{l}{3$\farcs$6 $\times$   2$\farcs$8 (-13.0$\degree$)}                    \\
          & 249   & 62.3 $\pm$ 1.3  & 59.5 $\pm$ 0.6  & 0.83       & \multicolumn{2}{l}{3$\farcs$5 $\times$   2$\farcs$8 (-14.0$\degree$)}                    \\
          & 265   & 70.2 $\pm$ 1.7  & 68.7 $\pm$ 0.7  & 1.1        & \multicolumn{2}{l}{3$\farcs$3 $\times$   2$\farcs$5 (-16.5$\degree$)}                   
\enddata
\tablecomments{Column 1: Star+disk system.  Column 2: Representative frequency for the SMA spectral band from which the mm continuum emission was measured.  Emission taken across all combined SMA spectral bands is labeled ``Comb.''.  Columns 3, 4, and 5: Total mm continuum flux, peak mm continuum intensity, and rms, respectively, along with the estimated errors (Section~\ref{sec_analysis_noise}).  Note that given the significant change in beam size across all SMA spectral bands, these values and errors were estimated from different elliptical masks (Section~\ref{sec_analysis_masks}).  Column 6: Dimensions of the synthesized beam.  Column 7: Parameters for the Gaussian fits to the combined mm continuum emission (Section~\ref{sec_analysis_masks}).  \\ All errors do not include 10\% flux calibration uncertainties.}
\end{deluxetable*}
%

\begin{deluxetable*}{lccccccc}
\tablecaption{Molecular Line Emission. \label{table_linefluxes}}
\tablehead{
Star+Disk     & Molecular & Int. Flux  & Peak Flux & Velocity  & Channel & Channel rms & Synthesized               \\
 System  & Line & (mJy  & (mJy beam$^{-1}$  & Range & Spacing  & (mJy $\times$ &  Beam (P.A.)  \\
  &  & $\times$ km s$^{-1}$) &   $\times$ km s$^{-1}$)   &  (km s$^{-1}$)  & (km s$^{-1}$)   & beam$^{-1}$)  &   
}
 \colnumbers \startdata
\hline
HD 34282  & $^{12}$CO 2-1             & \textbf{8149 $\pm$ 282} & 7721 $\pm$ 212      & [-6.39, 1.11]  & 0.5        & 145       & 6$\farcs$6 $\times$ 2$\farcs$4 (24.3$\degree$)  \\
HD 36112  &                           & \textbf{7370 $\pm$ 208} & 7142 $\pm$ 161      & [3.02, 8.52]   & 0.5        & 125       & 4$\farcs$8 $\times$ 2$\farcs$6 (16.6$\degree$)  \\
HD 142666 &                           & \textbf{3250 $\pm$ 205} & 3998 $\pm$ 168      & [-1.64, 10.37] & 2.0        & 42        & 3$\farcs$5 $\times$ 2$\farcs$9 (-19.9$\degree$) \\
HD 144432 &                           & \textbf{1034 $\pm$ 226} & 1221 $\pm$ 172      & [-0.88, 11.13] & 2.0        & 44        & 3$\farcs$7 $\times$ 2$\farcs$9 (-13.8$\degree$) \\
\hline
HD 34282  & $^{13}$CO 2-1             & \textbf{2469 $\pm$ 249} & 2661 $\pm$ 184      & [-6.69, 0.82]  & 0.5        & 126       & 7$\farcs$0 $\times$ 2$\farcs$5 (25.5$\degree$)  \\
HD 36112  &                           & \textbf{2017 $\pm$ 176} & 2278 $\pm$ 132      & [3.17, 8.68]   & 0.5        & 109       & 5$\farcs$1 $\times$ 2$\farcs$7 (17.8$\degree$)  \\
HD 142666 &                           & \textbf{1206 $\pm$ 197} & 1293 $\pm$ 161      & [-1.43, 10.59] & 2.0        & 40        & 3$\farcs$6 $\times$ 3$\farcs$0 (-17.2$\degree$) \\
HD 144432 &                           & $\leq$646               & 712 $\pm$ 154       & [-0.68, 11.33] & 2.0        & 41        & 3$\farcs$9 $\times$ 3$\farcs$0 (-14.7$\degree$) \\
\hline
HD 34282  & C$^{18}$O 2-1             & \textbf{912 $\pm$ 249}  & 955 $\pm$ 187       & [-6.22, 0.82]  & 0.5        & 133       & 7$\farcs$0 $\times$ 2$\farcs$5 (25.5$\degree$)  \\
HD 36112  &                           & \textbf{732 $\pm$ 179}  & 750 $\pm$ 135       & [3.18, 8.71]   & 0.5        & 112       & 5$\farcs$1 $\times$ 2$\farcs$7 (17.8$\degree$)  \\
HD 142666 &                           & \textbf{550 $\pm$ 167}  & 761 $\pm$ 140       & [-1.26, 10.70] & 2.0        & 34        & 3$\farcs$6 $\times$ 3$\farcs$1 (-18.3$\degree$) \\
HD 144432 &                           & \textit{$<$552}         & $<$402              & [-0.51, 11.44] & 2.0        & 35        & 3$\farcs$9 $\times$ 3$\farcs$0 (-13.4$\degree$) \\
\hline
HD 34282  & HCO$^+$ 3-2               & \textbf{1325 $\pm$ 428} & 1802 $\pm$ 308      & [-6.29, 1.21]  & 0.5        & 190       & 3$\farcs$3 $\times$ 2$\farcs$3 (45.2$\degree$)  \\
HD 36112  &                           & \textbf{1048 $\pm$ 333} & 981 $\pm$ 205       & [3.11, 8.61]   & 0.5        & 172       & 2$\farcs$8 $\times$ 2$\farcs$3 (80.1$\degree$)  \\
HD 142666 &                           & \textbf{1339 $\pm$ 280} & 1746 $\pm$ 245      & [-2.24, 9.76]  & 2.0        & 55        & 3$\farcs$1 $\times$ 2$\farcs$5 (-18.9$\degree$) \\
HD 144432 &                           & \textit{$<$981}         & $<$725              & [0.51, 12.51]  & 2.0        & 59        & 3$\farcs$3 $\times$ 2$\farcs$5 (-14.1$\degree$) \\
\hline
HD 34282  & CS 5-4                    & $\leq$847               & 750 $\pm$ 208       & [-6.25, 1.27]  & 0.5        & 135       & 3$\farcs$5 $\times$ 2$\farcs$6 (43.4$\degree$)  \\
HD 36112  &                           & $\leq$667               & 691 $\pm$ 182       & [3.07, 8.58]   & 0.5        & 119       & 3$\farcs$1 $\times$ 2$\farcs$4 (84.4$\degree$)  \\
HD 142666 &                           & \textit{$<$629}         & $<$485              & [-2.59, 9.44]  & 2.0        & 40        & 3$\farcs$3 $\times$ 2$\farcs$8 (-17.2$\degree$) \\
HD 144432 &                           & \textit{$<$677}         & $<$491              & [0.17, 12.20]  & 2.0        & 42        & 3$\farcs$5 $\times$ 2$\farcs$7 (-12.3$\degree$) \\
\hline
HD 34282  & HCN 3-2                   & $\leq$1439              & 1719 $\pm$ 260      & [-6.23, 1.26]  & 0.5        & 171       & 3$\farcs$3 $\times$ 2$\farcs$3 (43.5$\degree$)  \\
HD 36112  &                           & \textit{$<$1313}        & $<$568              & [3.15, 8.64]   & 0.5        & 152       & 2$\farcs$9 $\times$ 2$\farcs$3 (80.4$\degree$)  \\
HD 142666 &                           & \textbf{1770 $\pm$ 296} & 2226 $\pm$ 216      & [-1.98, 10.00] & 2.0        & 50        & 3$\farcs$1 $\times$ 2$\farcs$5 (-19.5$\degree$) \\
HD 144432 &                           & \textit{$<$1031}        & 722 $\pm$ 196       & [0.76, 12.75]  & 2.0        & 52        & 3$\farcs$3 $\times$ 2$\farcs$5 (-14.5$\degree$) \\
\hline
HD 34282  & C$_2$H 3-2                & $\leq$1532              & 1228 $\pm$ 350      & [-8.78, 1.20]  & 0.5        & 209       & 6$\farcs$0 $\times$ 2$\farcs$2 (27.3$\degree$)  \\
HD 36112  &                           & \textit{$<$1281}        & $<$802              & [0.58, 8.56]   & 0.5        & 190       & 4$\farcs$5 $\times$ 2$\farcs$4 (25.4$\degree$)  \\
HD 142666 &                           & \textit{$<$921}         & $<$738              & [-4.15, 9.83]  & 2.0        & 53        & 3$\farcs$2 $\times$ 2$\farcs$5 (-19.9$\degree$) \\
HD 144432 &                           & \textit{$<$1079}        & $<$709              & [-3.40, 12.57] & 2.0        & 55        & 3$\farcs$3 $\times$ 2$\farcs$5 (-16.0$\degree$) \\
\hline
HD 34282  & DCN 3-2                   & \textit{$<$581}         & $<$409              & [-6.36, 1.13]  & 0.5        & 96        & 3$\farcs$9 $\times$ 2$\farcs$8 (44.1$\degree$)  \\
HD 36112  &                           & \textit{$<$437}         & $<$288              & [2.95, 8.43]   & 0.5        & 82        & 3$\farcs$3 $\times$ 2$\farcs$8 (61.9$\degree$)  \\
HD 142666 &                           & \textit{$<$504}         & $<$412              & [-2.66, 9.32]  & 2.0        & 35        & 3$\farcs$7 $\times$ 3$\farcs$1 (-17.3$\degree$) \\
HD 144432 &                           & $\leq$566               & 446 $\pm$ 134       & [0.09, 12.06]  & 2.0        & 36        & 3$\farcs$9 $\times$ 3$\farcs$1 (-11.2$\degree$) \\
\hline
HD 34282  & DCO$^+$ 3-2               & $\leq$529               & 567 $\pm$ 128       & [-6.23, 1.29]  & 0.5        & 82        & 3$\farcs$9 $\times$ 2$\farcs$8 (44.1$\degree$)  \\
HD 36112  &                           & $\leq$387               & 332 $\pm$ 89        & [3.12, 8.64]   & 0.5        & 72        & 3$\farcs$3 $\times$ 2$\farcs$8 (61.9$\degree$)  \\
HD 142666 &                           & \textit{$<$493}         & 536 $\pm$ 136       & [-2.01, 10.03] & 2.0        & 33        & 3$\farcs$7 $\times$ 3$\farcs$1 (-17.3$\degree$) \\
HD 144432 &                           & \textit{$<$516}         & $<$382              & [0.75, 12.79]  & 2.0        & 34        & 3$\farcs$9 $\times$ 3$\farcs$1 (-11.2$\degree$) \\
\hline
HD 34282  & H$_2$CO 3$_{03}$-2$_{02}$ & \textit{$<$655}         & $<$496              & [-6.46, 1.06]  & 0.5        & 108       & 7$\farcs$0 $\times$ 2$\farcs$5 (25.4$\degree$)  \\
HD 36112  &                           & \textit{$<$452}         & $<$307              & [3.40, 8.41]   & 0.5        & 92        & 5$\farcs$2 $\times$ 2$\farcs$7 (17.4$\degree$)  \\
HD 142666 &                           & \textit{$<$509}         & $<$413              & [-1.19, 10.84] & 2.0        & 34        & 3$\farcs$6 $\times$ 3$\farcs$1 (-18.3$\degree$) \\
HD 144432 &                           & \textit{$<$571}         & $<$413              & [-0.44, 11.59] & 2.0        & 36        & 3$\farcs$9 $\times$ 3$\farcs$0 (-13.4$\degree$)
\enddata
\tablecomments{Column 1: Star+disk system.  Column 2: Molecular line.  Column 3: Velocity-integrated flux.  ``$\leq$'' and ``$<$'' are 3$\sigma$ upper limits and mark tentative detections and non-detections, respectively.  Column 4: Peak flux.  Column 5: Total velocity range used to estimate columns 3 and 4.  The range is centered on the systemic velocity (Table~\ref{table_char}).  Column 6: Spacings between imaged channels.  Column 7: Channel rms.  Column 8: Dimensions of the synthesized beam. \\ Columns 3, 4, and 7 were estimated within the Keplerian masks.  rms and errors were estimated via bootstrapping (Section~\ref{sec_analysis_noise}).  Errors do not include 10\% absolute flux calibration uncertainties.}
\end{deluxetable*}

\subsection{{Millimeter-Wavelength} Continuum and Molecular Line Detections}
\label{sec_results_detections}

Table~\ref{table_contfluxes} presents the mm continuum fluxes, synthesized beam sizes, and estimated error measured for the individual and combined SMA spectral bands for each disk.  Figure~\ref{fig_cont} shows the mm continuum emission for the lowest frequency, highest frequency, and combined SMA spectral bands.  We detect mm continuum emission at millimeter wavelengths above 3$\sigma$ from all five disks in the sample.

For HD 34282, HD 36112, and HD 142666, $\sim$1.3 mm continuum fluxes have been measured at higher spatial resolution in the literature~\citep{cite_stapperetal2022}.  While our 234 GHz continuum fluxes are consistent with these literature values within 15\% for HD 36112 and HD 142666, our 234 GHz continuum flux for HD 34282 ($\sim$128 mJy) exceeds the literature value by over 15\%~\citep[99 mJy;][]{cite_stapperetal2022}.  This excess could be due {to differences in uv coverage between the observations, or} to inclusion of an extended or external source of emission within the extraction masks (see Section~\ref{sec_results_fluxes}).

We classified detections of the molecular line emission according to the following criteria (where the $\sigma$ and error values are as described in Section~\ref{sec_analysis_noise}):

\begin{enumerate}
    \item Peak emission is $\geq$3$\times$ the noise estimated for the velocity-integrated emission map over a significant spatial area (at least {7\%}\footnote{{7\% is based on a simple geometric argument for the beams, which are too large to fully resolve the observed emission: for a point source image convolved with a Gaussian of standard deviation $\sigma$, the ratio of areas between a circle of diameter $\sigma$ and a circle containing the bulk of the emission (taken to be $\sim$95\%, which is contained within a diameter of 4$\sigma$ by the $\sim$68-95-99 rule) is 1/16, or $\sim$7\%.}} of the beam).
    \item The velocity-integrated line flux is $\geq$3$\sigma$.
\end{enumerate}

Molecular lines that fulfilled both criteria were classified as detections, while lines that fulfilled one criterion were classified as tentative detections.  If neither criterion was met, then the lines were classified as non-detections.

Based on these criteria, we detect $^{12}$CO 2--1 from four disks.  We detect $^{13}$CO 2--1, C$^{18}$O 2--1, and HCO$^+$ 3--2 from three disks, and tentatively detect $^{13}$CO 2--1 from a fourth disk.
We detect HCN 3--2 from one disk and tentatively detect the same line from another disk.  Finally, we tentatively detect CS 5--4 and DCO$^+$ 3--2 from two disks each, and we tentatively detect C$_2$H 3--2 and DCN 3--2 from one disk each.  All other line+disk pairs are not detected at the sensitivity of our observations.

We do not detect any molecular line emission, including $^{12}$CO 2--1, from HD 38120.  This is in contrast to~\cite{cite_dentetal2005}, which observed a bright $^{12}$CO 3--2 spectrum from HD 38120 with the James Clerk Maxwell Telescope (JCMT; see their Figure 2b).
We note that (1)~\cite{cite_dentetal2005} reported a beam size of 1400 au at 100 pc, or an angular resolution of 14\farcs0;
(2)~\cite{cite_dentetal2005} fitted the $^{12}$CO 3--2 spectrum with a single rather than double-peaked Gaussian despite the disk's high inclination angle (Table~\ref{table_char}); and (3) there appears to be absorption affecting the $^{12}$CO 2--1 spectrum (see Appendix~\ref{sec_appendix_fullband}).  It is therefore possible that the $^{12}$CO 3--2 spectrum in~\cite{cite_dentetal2005} was contaminated by CO emission from a surrounding cloud or envelope.

We \textit{do} detect relatively faint mm continuum emission from HD 38120, suggesting that a dust disk does exist in this system.  That being said, observing any molecular gas present in this disk would likely require greater sensitivity than initially expected in this survey.  Since our line flux upper limits for this disk are non-constraining, {in that we are unable to distinguish any of the disk's $^{12}$CO 2--1 emission from possible environmental contamination,} we exclude HD 38120 from subsequent line flux analysis and discussion.

All line fluxes and detection classifications (excluding HD 38120) are presented in Table~\ref{table_linefluxes}.
Appendix~\ref{sec_appendix_fullband} also displays the spectral data across all SMA spectral bands, extracted using elliptical masks (see Section~\ref{sec_analysis_masks}), where our ten target molecular lines are highlighted in each panel.


%

\subsection{Gas and Dust Morphologies}
\label{sec_results_mom0}

\begin{figure*}
\centering
\resizebox{0.79\hsize}{!}{
    \includegraphics[trim=32.5pt 20pt 12.5pt 55pt, clip]{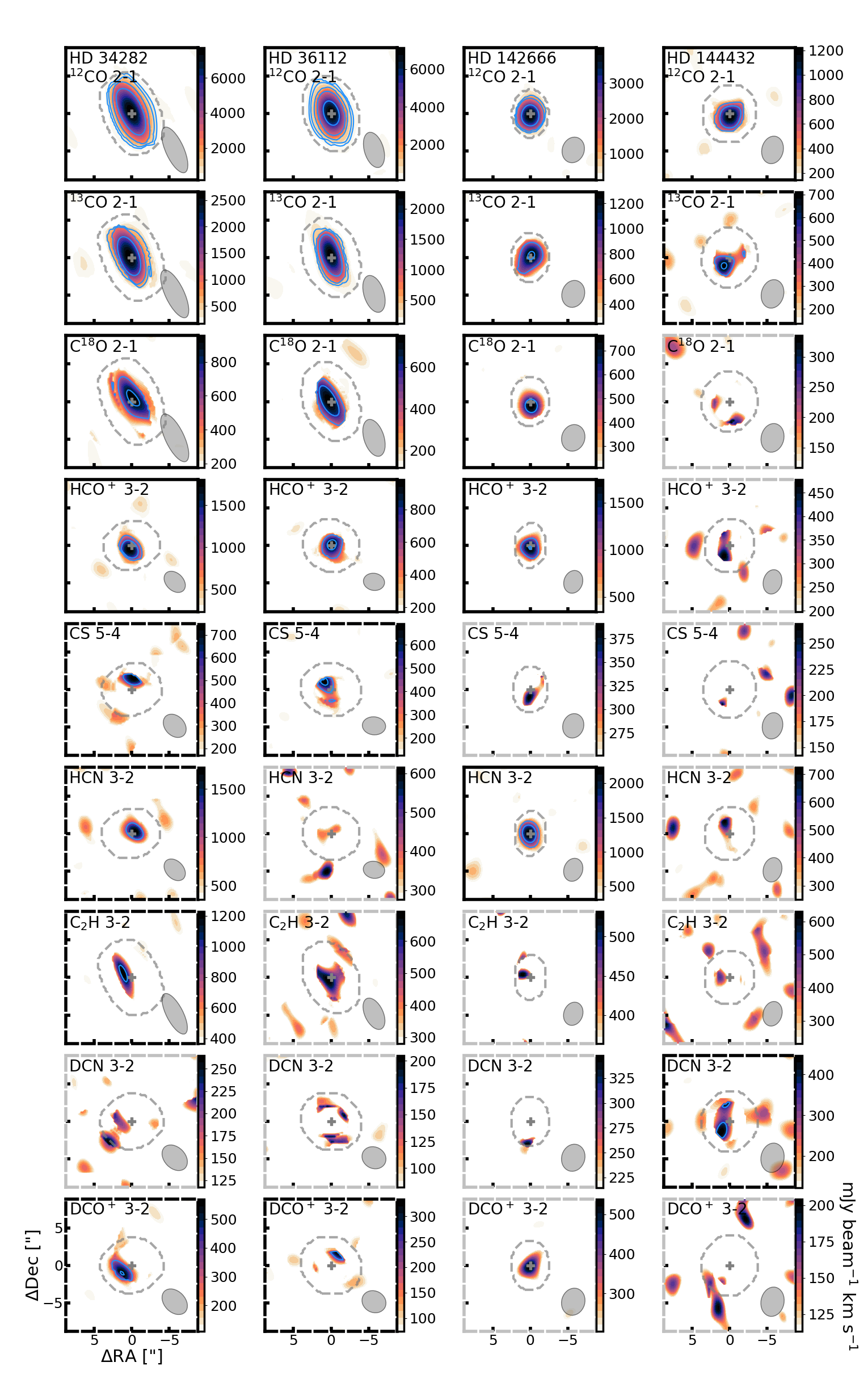}}
{\caption{Velocity-integrated emission maps for the target molecular lines that were detected or tentatively detected from at least one disk in the sample.  H$_2$CO 3--2 is not shown because it was undetected in all disks.  The maps were extracted using Keplerian masks (Section~\ref{sec_analysis_masks}), the edges of which are outlined in dashed gray.  Contours are [3, 5, 10, 20, 40...]$\times \sigma$ (see Section~\ref{sec_analysis_noise} for $\sigma$).  Panels are outlined in solid black, black, and dashed gray for detections, tentative detections, and non-detections respectively.  Colorbars start at 1$\sigma$.  `+' signs mark the estimated disk centers (see Section~\ref{sec_analysis_masks}).  Synthesized beams are drawn in the lower right corners.
\label{fig_mom0}}}
\end{figure*}

\begin{figure*}
\centering
\resizebox{0.725\hsize}{!}{
    \includegraphics[trim=10pt 15pt 77.5pt 57.5pt, clip]{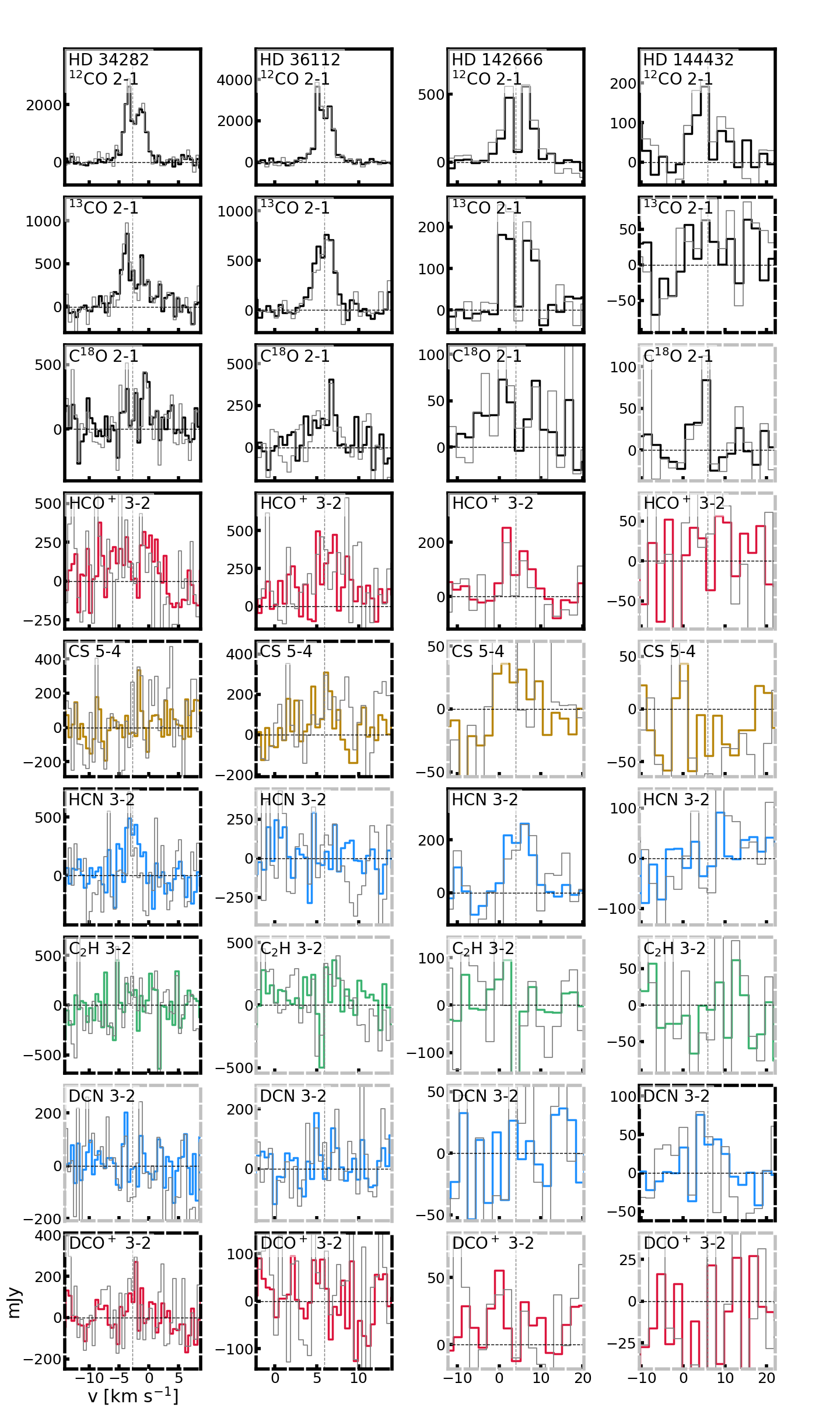}}
{\caption{Keplerian-masked spectra are plotted with thick lines for the CO isotopologues (black), other oxygen carriers (red), sulfur carriers (yellow), hydrocarbons (green), and cyanides (blue) that were detected or tentatively detected from at least one disk in the sample.  H$_2$CO 3--2 is not shown because it was undetected from all disks.  Elliptical-masked spectra are plotted with thin gray lines for comparison.  Panels are outlined in solid black, dashed black, and dashed gray for detections, tentative detections, and non-detections, respectively. Vertical dotted gray lines mark the systemic velocity (Table~\ref{table_char}).
\label{fig_spec}}}
\end{figure*}

Figures~\ref{fig_mom0} and~\ref{fig_spec} display velocity-integrated emission maps and spectra, respectively, for molecular lines that are detected or tentatively detected from at least one disk in our sample.  Channel maps for all detected and tentatively detected lines are given in Appendix~\ref{sec_appendix_chanmaps}.

At the spatial resolution of our observations, all mm continuum emission (Figure~\ref{fig_cont}) and detected molecular line emission (Figure~\ref{fig_mom0}) appear smooth, without any apparent cavities or gaps.  However, we do find persistent asymmetries in the dust and/or bright molecular line emission for certain disks.  Here we qualitatively describe these asymmetries in the context of complex morphologies known from the literature.

\subsubsection{{HD 34282}}
\label{sec_results_mom0_hd34282}

HD 34282 is adjacent to what appears to be an extended or external source of emission.  This neighboring source is centered {eastward} of HD 34282's estimated disk center, roughly $\sim$5 arcseconds away.  The neighbor is seen above 3$\sigma$ in the mm continuum emission {and is brightest in the 265 GHz band} (Figure~\ref{fig_cont}).
There also appears to be a faint ``stream'' of dust connecting HD 34282 and this neighboring source, although the stream is not detected above 3$\sigma$.

\begin{figure*}
\centering
\resizebox{0.99\hsize}{!}{
    \includegraphics[trim=50pt 10pt 30pt 45pt, clip]{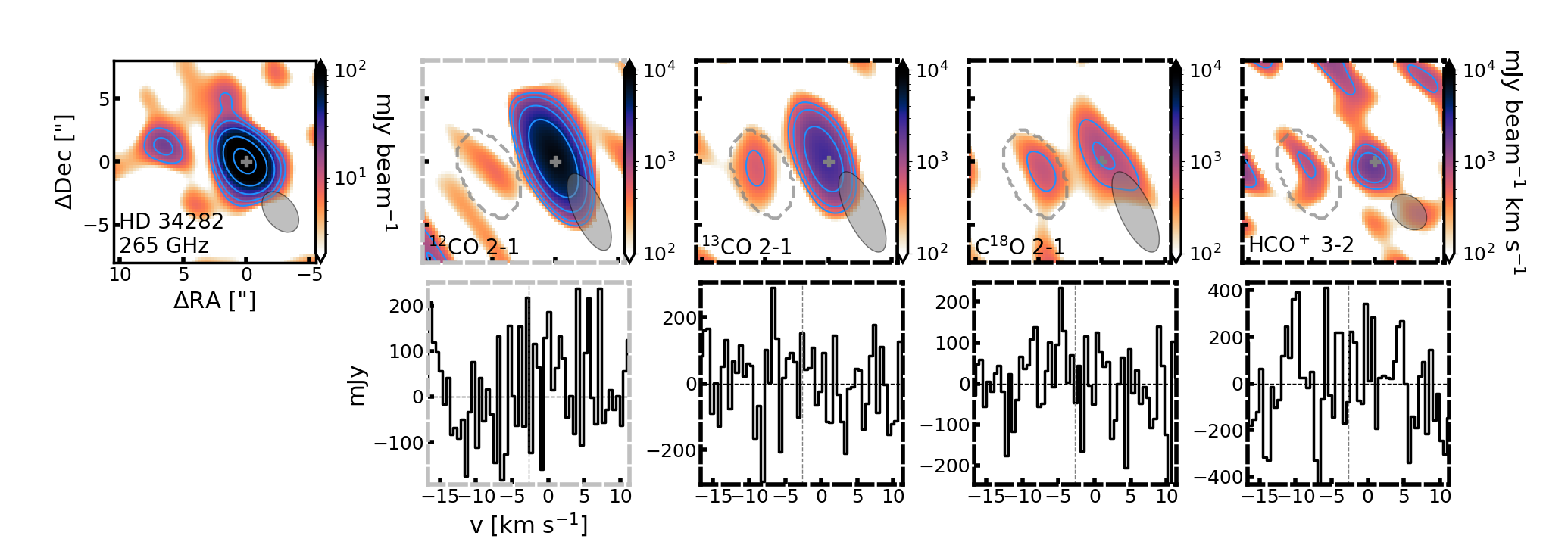}}
\caption{Emission maps (top) and spectra (bottom) for HD 34282's neighboring source of emission.  The brightest mm continuum emission ($\sim$265 GHz) is shown in the leftmost top panel.  Velocity-integrated emission maps of the CO 2--1 isotopologue (top panels 2-4) and HCO$^+$ 3--2 (top panel 5) emission were extracted \textit{without} using masks.  Contours are [3, 5, 10, 20, 40...]$\times \sigma$, where $\sigma$ is the same as in Figure~\ref{fig_mom0}.  `$+$' signs mark the disk centers estimated from the mm continuum emission (Section~\ref{sec_analysis_masks}).  Synthesized beams are drawn in the lower right corners.  Note the difference in beam size across the different panels.  The neighboring source is highlighted in each top panel with a custom mask drawn in dashed gray, which was used to extract the spectra shown in the bottom panels.  The systemic velocity of HD 34282 is marked with a vertical dashed line for each spectrum.
\label{fig_panel_hd34282b}}
\end{figure*}

This neighbor is not visible in the velocity-integrated emission maps of Figure~\ref{fig_mom0}, because those maps were extracted using Keplerian masks catered to HD 34282.  We have thus generated velocity-integrated emission maps for the brightest molecular lines from HD 34282 using \textit{no} masks, which are displayed in Figure~\ref{fig_panel_hd34282b}.  Figure~\ref{fig_panel_hd34282b} shows that this neighbor is detected above 3$\sigma$ in emission maps of $^{13}$CO 2--1, C$^{18}$O 2--1, and HCO$^+$ 3--2, and is seen in $^{12}$CO 2--1 below 3$\sigma$.  Spectra extracted from a custom region containing the neighboring emission do not show Keplerian rotation, at least at the sensitivity of these observations.

HD 34282 has previously been observed in millimeter-wavelength continuum and {CO} emission in the literature.
{\cite{cite_pietuetal2003} mapped the Keplerian rotation of HD 34282 in $^{12}$CO 2--1 emission using the Institut de Radioastronomie Millim\'etrique (IRAM) interferometer (the Plateau de Bure Interferometer, originally denoted PdBI but now upgraded and renamed NOEMA).
\cite{cite_vanderplasetal2017} conducted an ALMA study of $\sim$350 GHz continuum, $^{12}$CO 3--2, and HCO$^+$ 4--3 emission (from ALMA
Project 2013.1.00658.S).
\cite{cite_stapperetal2022} presented an ALMA image of the $\sim$230 GHz continuum emission for HD 34282 (their Figure 2; from the ALMA Project 2015.1.00192.S), while \cite{cite_lawetal2022b} recently conducted an ALMA vertical structure study and presented new images of the $\sim$230 GHz continuum, $^{12}$CO 2--1, $^{13}$CO 2--1, and C$^{18}$O 2--1 emission (from ALMA Projects 2015.1.00192.S and 2017.1.01578.S).  The latter three studies~\citep{cite_vanderplasetal2017, cite_stapperetal2022, cite_lawetal2022b} all presented images of HD 34282 within a field of view of $\lesssim$2\farcs0$\times$2\farcs0.}

{In terms of the dust, faint structure can be seen eastward of HD 34282 in the published $\sim$230 GHz continuum images of~\cite{cite_stapperetal2022} and~\cite{cite_lawetal2022b} within their $\lesssim$2\farcs0$\times$2\farcs0 fields of view (see their Figures 2 and 1, respectively).  However, the same structure is not apparent within the published $\lesssim$2\farcs0$\times$2\farcs0 $\sim$350 GHz continuum image of~\cite{cite_vanderplasetal2017}~\citep[see, e.g., their Figure 4; see also Figure 1 of][]{cite_lawetal2022b}.  We expanded the field of view for the $\sim$350 GHz continuum image from the same observing project as~\cite{cite_vanderplasetal2017} (ALMA {Project} 2013.1.00658.S, extracted from the ALMA archive; Appendix~\ref{sec_appendix_hd34282}). The $\sim$350 GHz continuum image observed in the compact configuration \textit{tentatively} suggests what could be faint structure {eastward} of HD 34282, while the $\sim$350 GHz continuum image observed in the extended configuration shows no such structure.}

{In terms of the gas, model residuals of the $^{12}$CO 2--1 emission in~\cite{cite_pietuetal2003} (see their Figure 5) suggest a neighboring emitting source, with a center that shifts from roughly $\sim$3 arcseconds away to $\sim$5 arcseconds away over the course of their channel maps.  However, published ALMA images of the $^{12}$CO 3--2, HCO$^+$ 4--3, $^{12}$CO 2--1, $^{13}$CO 2--1, and C$^{18}$O 2--1 emission in~\cite{cite_vanderplasetal2017} and~\cite{cite_lawetal2022b} show no emission structure eastward of the disk within $\lesssim$2\farcs0$\times$2\farcs0.  We expanded the field of view of the $^{12}$CO 2--1, $^{13}$CO 2--1, and C$^{18}$O 2--1 image cubes provided by~\cite{cite_lawetal2022b} (derived from ALMA Projects 2015.1.00192.S and 2017.1.01578; not shown).  We also expanded the archival ALMA images of the $^{12}$CO 3--2 and HCO$^+$ 4--3 emission presented in~\cite{cite_vanderplasetal2017} (ALMA Project 2013.1.00658.S, extracted from the ALMA archive; Appendix~\ref{sec_appendix_hd34282}).  The $^{12}$CO 3--2 image observed in the compact configuration tentatively suggests faint structure eastward of HD 34282, cospatial to the faint structure seen for the expanded $\sim$350 GHz continuum image in the same configuration.  However, no such structure is apparent in the other expanded images of line emission.}

{Altogether, we detect what appears to be a neighboring source eastward of HD 34282 in mm continuum emission, which is connected to HD 34282 by a faint `stream' of emission.  Faint $\sim$230 GHz continuum emission structure eastward of HD 34282 can also be seen in $\lesssim$2\farcs0$\times$2\farcs0 images in the literature~\citep{cite_stapperetal2022, cite_lawetal2022b}. 
There is some very \textit{tentative} suggestion of cospatial $^{12}$CO emission from compact observations of this system~\citep[][Figure~\ref{fig_panel_hd34282b}, Appendix~\ref{sec_appendix_hd34282}]{cite_pietuetal2003}.  However, the neighbor is not confidently detected in our brightest CO isotopologue ($^{12}$CO 2--1, Figure~\ref{fig_panel_hd34282b}), and it is not detected in ALMA CO and HCO$^+$ 4--3 images observed in more extended configurations and at higher spatial resolution~\citep{cite_vanderplasetal2017, cite_lawetal2022b}.}

It is possible that this neighboring source is actually a persistent imaging artifact.  That being said, {based on mm continuum emission observations of HD 34282 in this work and in the literature}, we do believe this neighbor to be real.  If so, the exact morphology of the neighbor is likely distorted by the elongated beams of our observations.

We infer that the neighbor is distinct from the spiral arm and point source previously identified within $\sim$1\farcs0 of the disk center~\citep{cite_Deboeretal2020, cite_vanderplasetal2017}.  It is also likely distinct from the orbital companion reported in~\cite{cite_wheelwrightetal2010}, which based on their technique is within $\sim$0.1-2" from HD 34282~\citep{cite_bainesetal2006}.  We speculate that this possible neighbor is either (1) an extended spiral arm or (2) a distant orbital companion.  {It is possible that any gas present for this neighboring source has been dissociated or ionized, and so is not traceable by our target molecules, or that the gas has dissipated or been disrupted through dynamical interactions over time}.

\subsubsection{{HD 142666}}

The mm continuum emission from HD 142666 shows what appear to be asymmetries above 3$\sigma$ to the northwest and northeast of the disk center (Figure~\ref{fig_cont}).  However,~\cite{cite_huangetal2018} observed HD 142666 in 1.25 mm continuum emission and found no indication of dust asymmetries near the outer disk edge.  It is possible that the asymmetries seen in this work are imaging artifacts. 

\subsubsection{{HD 144432}}
\label{sec_results_mom0_hd144432}

HD 144432 has the faintest mm continuum emission in our sample (excluding HD 38120), and only CO isotopologue emission is detected or tentatively detected from the disk.  The dust disk has also previously been undetected in scattered light~\citep{cite_monnieretal2017}.  Notably HD 144432 has a known stellar companion~\citep{cite_maheswaretal2002} at a distance of $\sim$1$\farcs$5~\citep{cite_mulleretal2011}.  \cite{cite_monnieretal2017} postulated that interactions with this companion may have truncated the disk and led to its non-detection in scattered light.  Other possible explanations include that
the disk is old ($\sim$5 Myr; Section~\ref{sec_sample_systems}), and so may have already lost much of its dust and gas over time.
%

%

\begin{figure*}
\centering
\resizebox{0.99\hsize}{!}{
    \includegraphics[trim=0pt 0pt 0pt 0pt, clip]{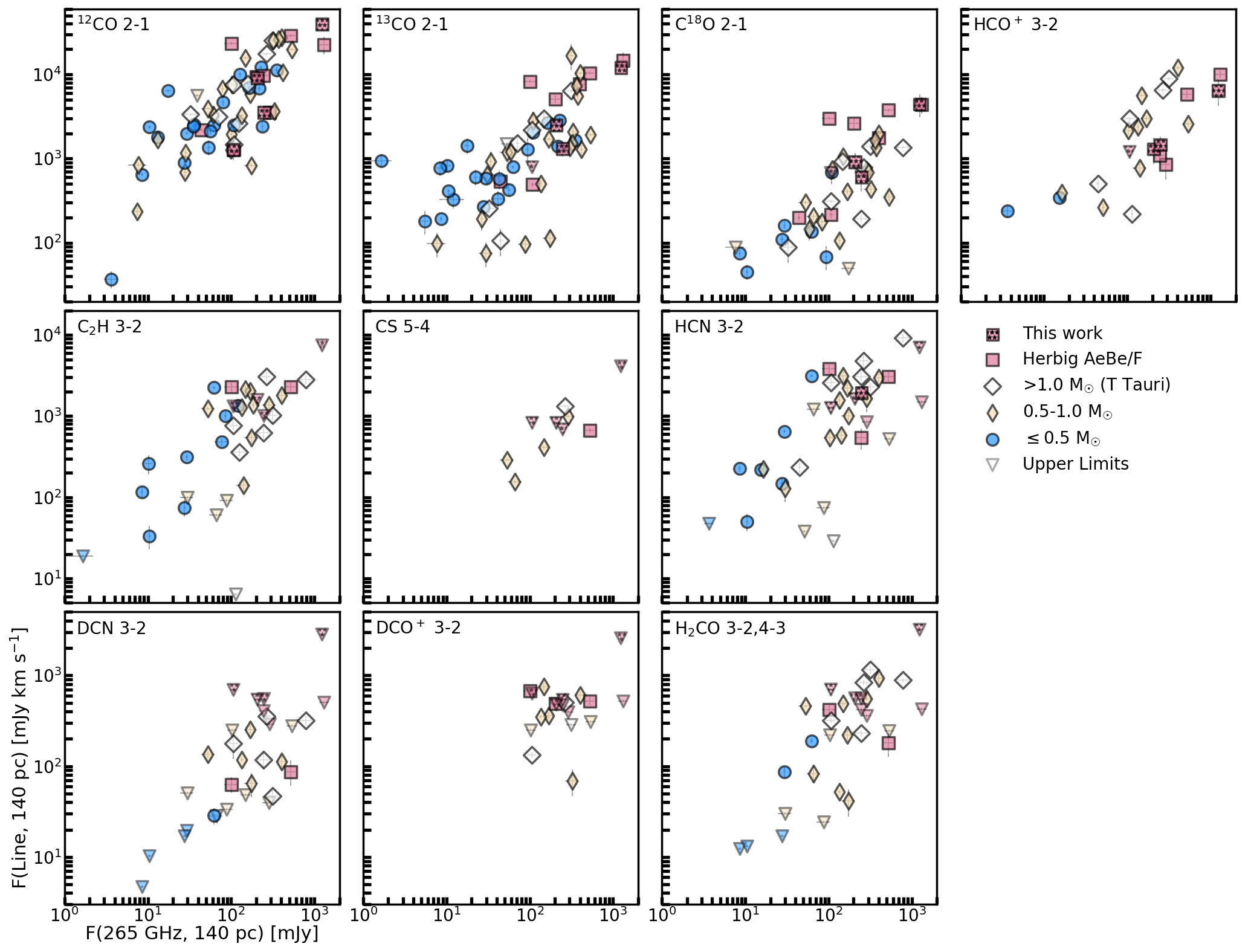}}
\caption{Molecular line fluxes as a function of mm continuum fluxes for Herbig AeBe disks from this work (dotted red squares).  The literature sample of {Herbig AeBe/F disks (blank red squares) and T Tauri disks with stellar masses $>$ 1.0 M$_\Sun$ (thick white diamonds), 0.5-1.0 M$_\Sun$ (thin light gold diamonds), and $\leq$ 0.5 M$_\Sun$} (blue circles) are shown for comparison (Section~\ref{sec_sample_lit}, Appendix~\ref{sec_appendix_lit}).  3$\sigma$ upper limits are shown as faint downward-pointing triangles.  All mm continuum fluxes have been scaled to 265 GHz using $F_\nu \propto \nu^{2.2}$~\citep{cite_andrewsetal2020} for flux $F_\nu$ and frequency $\nu$.  Both mm continuum and molecular line fluxes have been scaled to 140 pc.  Whenever flux errors for a disk are not reported in the literature, we assume the error is 15\% (for mm continuum) or 20\% (for molecular lines) of the given value.
\label{fig_abs_fcont}}
\end{figure*}

\subsection{Molecular Line Fluxes}
\label{sec_results_fluxes}

%

Figure~\ref{fig_abs_fcont} plots scaled molecular line fluxes and upper limits as a function of mm continuum flux for our sample and for the literature sample (Section~\ref{sec_sample_lit}, Appendix~\ref{sec_appendix_lit}).  Most molecular lines (e.g., the CO 2--1 isotopologues, C$_2$H 3--2, and HCN 3--2) generally increase in flux as mm continuum flux increases, with $\sim$one or more orders-of-magnitude of scatter in the flux.  Potential exceptions are the {CS 5--4 and DCO$^+$ 3--2} lines{; for disks with stellar masses below 0.5 M$_\Sun$, few detections or upper limits are known for these lines}.





%

\section{Discussion}
\label{sec_discussion}

\subsection{Disk Inventories across the Pre-Main-Sequence Stellar Mass Distribution}
\label{sec_discussion_inventory}

We now compare molecular line flux ratios across the combined sample of disks from this work and from the literature sample (Section~\ref{sec_sample_lit}, Appendix~\ref{sec_appendix_lit}).  We focus on flux ratios here to help normalize against the different disk sizes across the sample.
{We compare the flux ratios to stellar mass and spectral type (via binning), stellar luminosity, and mm continuum flux as representative properties of the disks and their host stars.}

{We note that stellar age, and therefore timespan of chemical evolution, is another important factor.  So far, however, previous millimeter-wavelength chemistry surveys of mostly T Tauri disks have not found any clear trends with stellar age~\citep[e.g.,][]{cite_bergneretal2019, cite_legaletal2019, cite_peguesetal2020}.  Noting this finding, and noting the large intrinsic and explicit uncertainties in existing stellar age estimates for pre-main-sequence stars~\citep[e.g., discussion by][]{cite_hillenbrandetal2004}, we do not consider stellar age in this work.}

\begin{figure*}
\centering
\resizebox{0.95\hsize}{!}{
    \includegraphics[trim=25pt 15pt 180pt 50pt, clip]{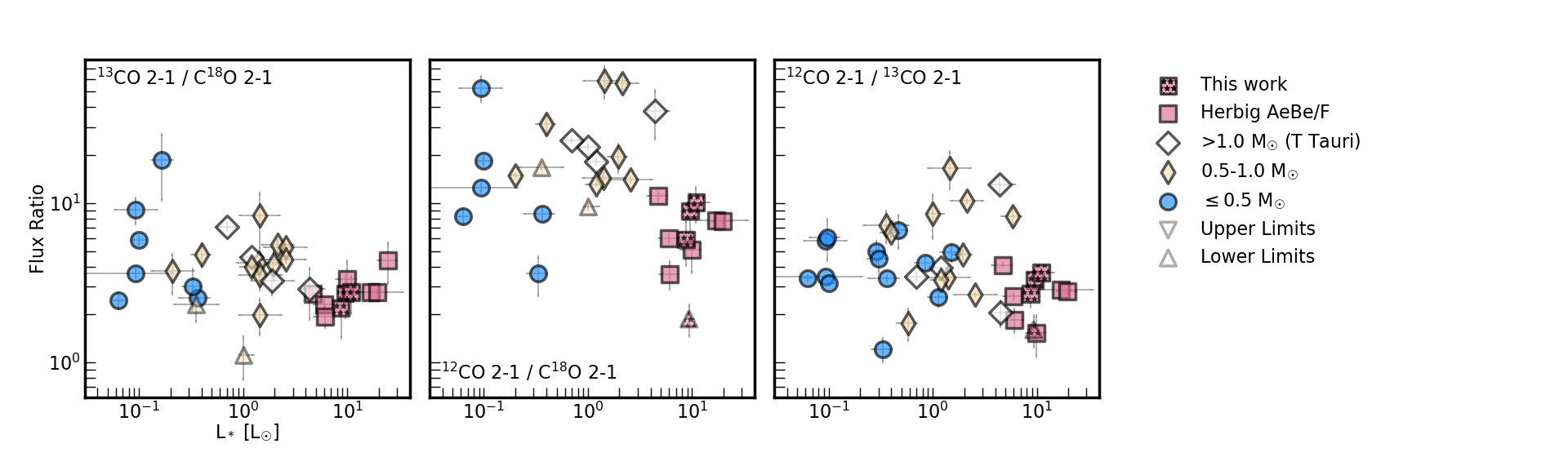}}
\resizebox{0.95\hsize}{!}{
    \includegraphics[trim=25pt 15pt 180pt 40pt, clip]{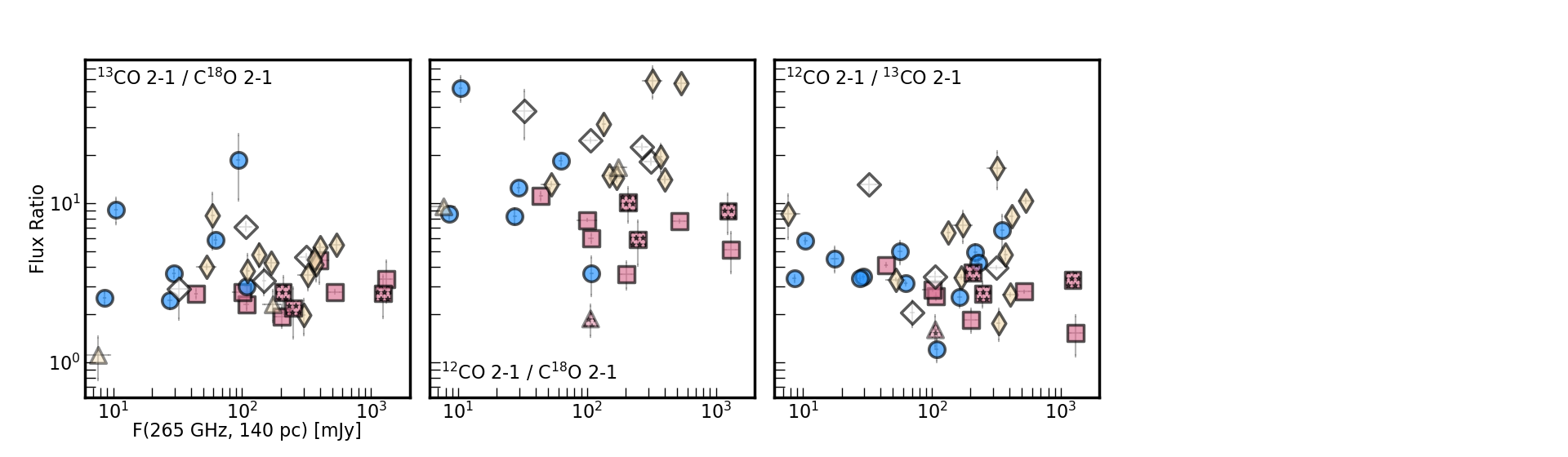}}
\resizebox{0.95\hsize}{!}{
    \includegraphics[trim=15pt 25pt 180pt 40pt, clip]{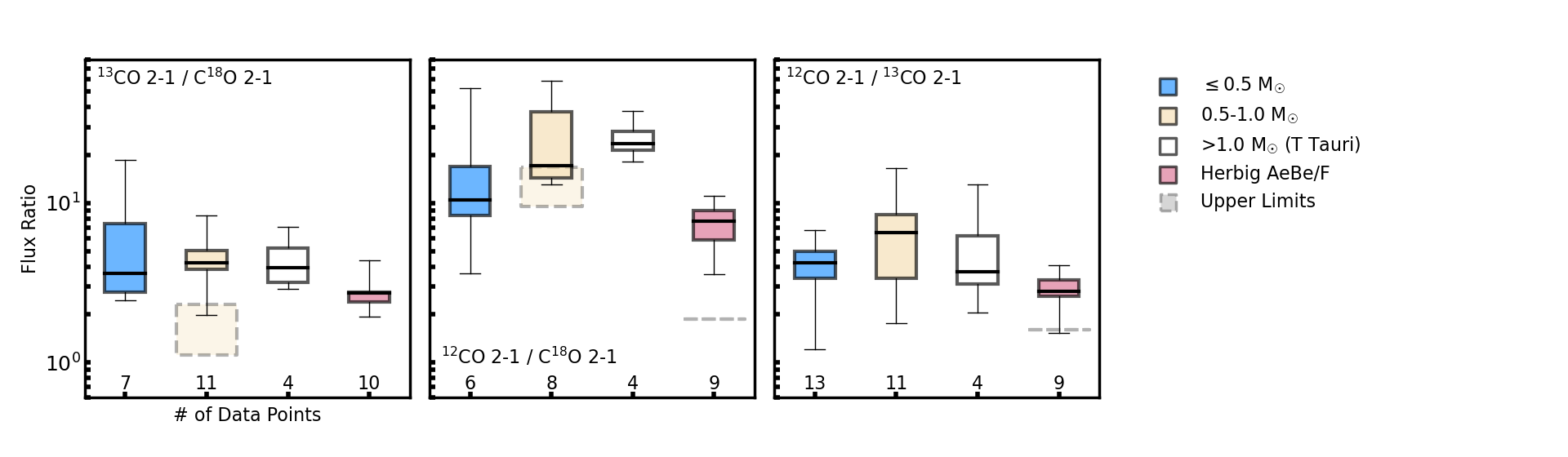}}
\caption{\textit{Top and middle rows:} CO 2--1 isotopologue line flux ratios as a function of stellar luminosity (top row) and mm continuum flux (middle row) for Herbig Ae disks from this work (dotted red squares).  The literature sample of {Herbig AeBe/F disks (blank red squares) and T Tauri disks with stellar masses $>$ 1.0 M$_\Sun$ (thick white diamonds), 0.5-1.0 M$_\Sun$ (thin light gold diamonds), and $\leq$ 0.5 M$_\Sun$} (blue circles) are shown for comparison (Section~\ref{sec_sample_lit}, Appendix~\ref{sec_appendix_lit}).  3$\sigma$ upper limits are shown as triangles and point in the direction of the limit.  Points are not shown when both line fluxes are upper limits. \textit{Bottom row:} Boxplot summaries of the CO 2--1 isotopologue line flux ratios for the combined samples, using the same color scheme as above.  The boxplots contain detected fluxes only.  Dashed boxes illustrate the span of any upper limits.  The numbers of data points included in the boxplots are written along the x-axis.  When $<$4, individual data points are drawn instead.  The ``box'' portions of the boxplots illustrate the 25$^\mathrm{th}$-50$^\mathrm{th}$-75$^\mathrm{th}$ percentiles, while the ``whiskers'' span all data.
\label{fig_flux_gas}}
\end{figure*}

\begin{figure*}
\centering
\resizebox{0.975\hsize}{!}{
    \includegraphics[trim=15pt 15pt 50pt 50pt, clip]{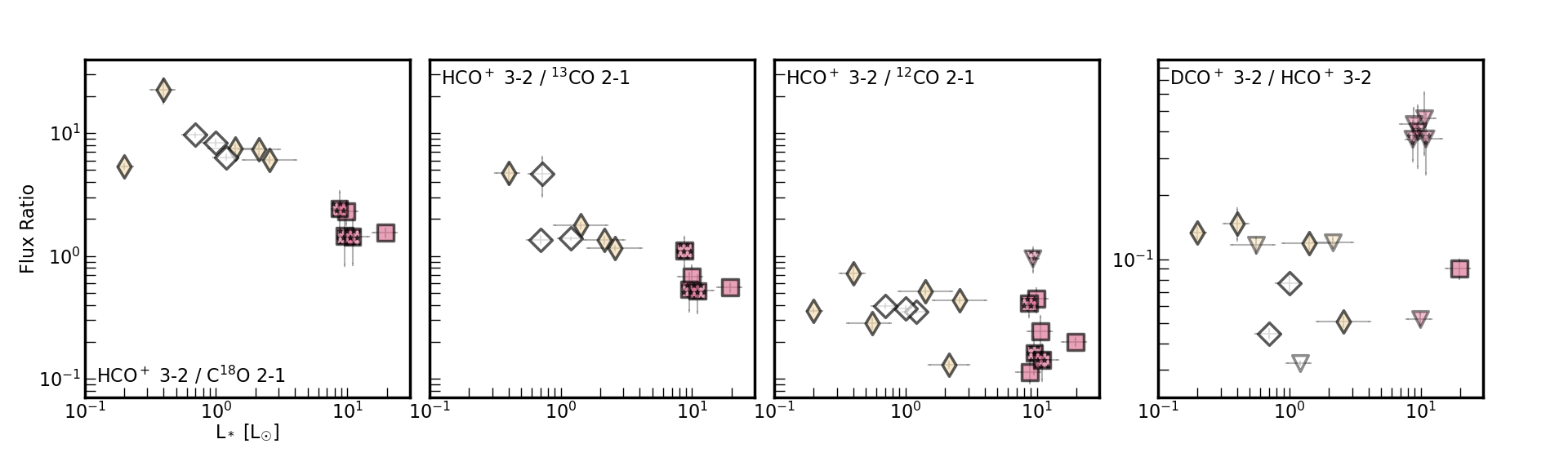}}
\resizebox{0.975\hsize}{!}{
    \includegraphics[trim=15pt 15pt 50pt 40pt, clip]{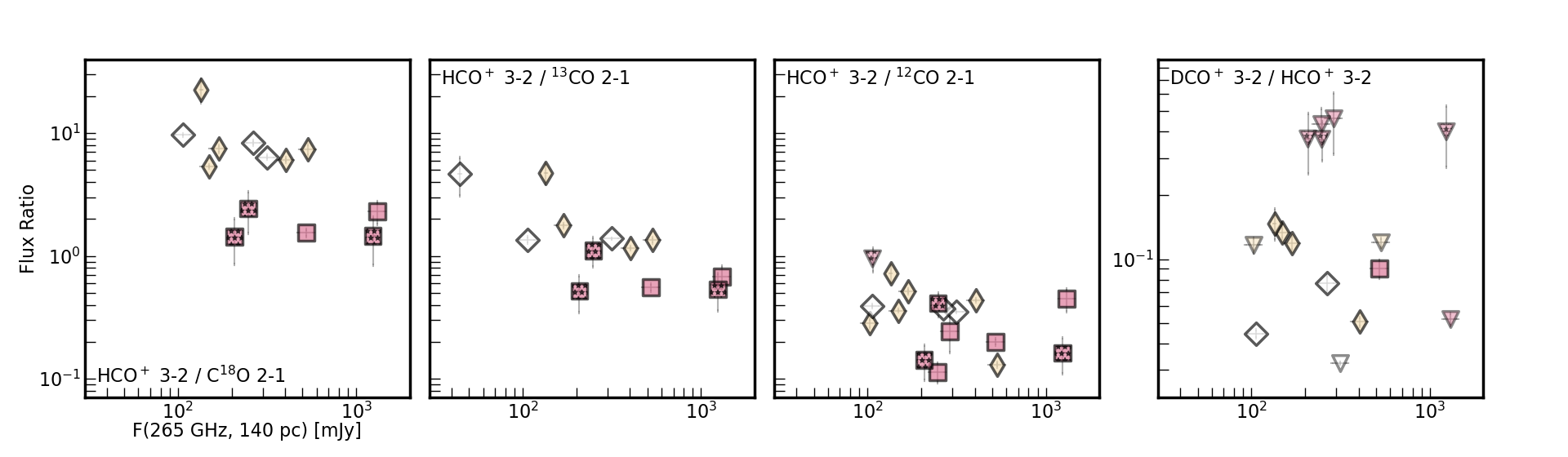}}
\resizebox{0.975\hsize}{!}{
    \includegraphics[trim=125pt 260pt 260pt 365pt, clip]{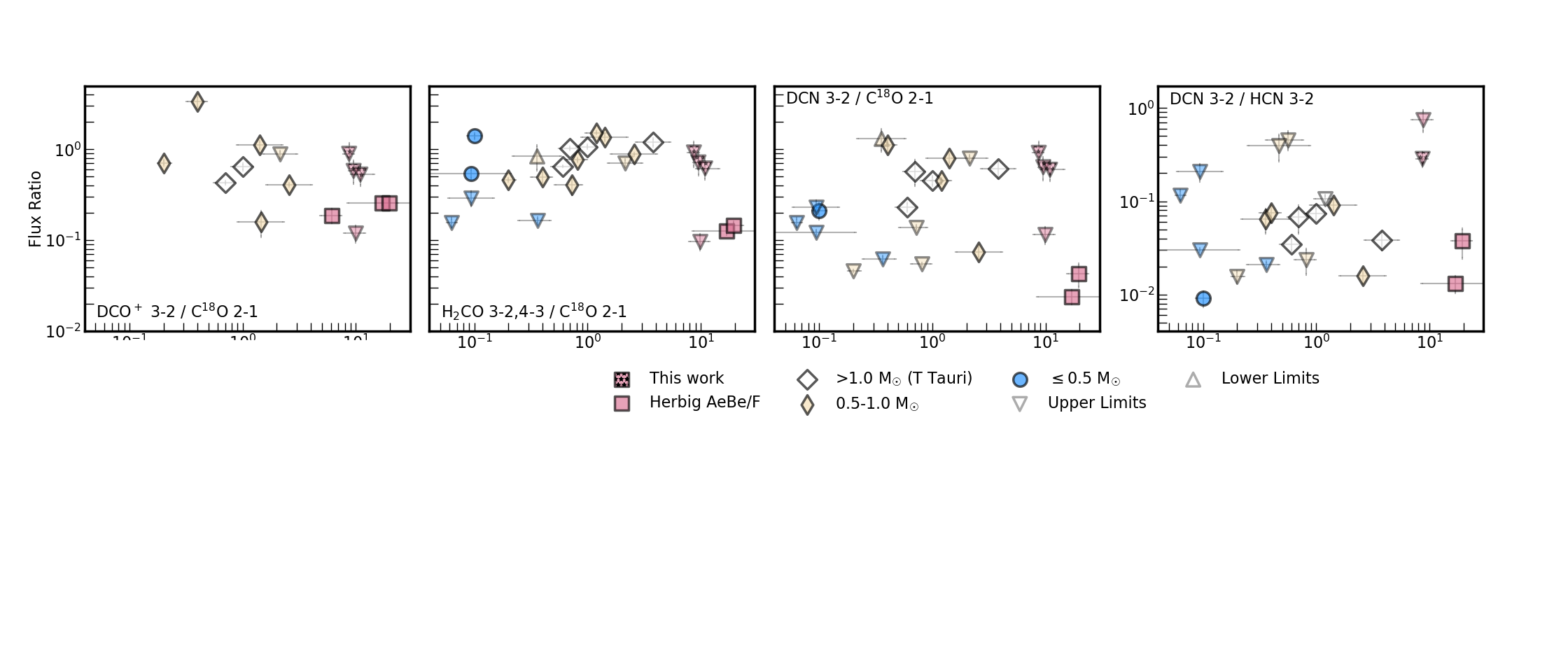}}
\resizebox{0.975\hsize}{!}{
    \includegraphics[trim=5pt 15pt 50pt 40pt, clip]{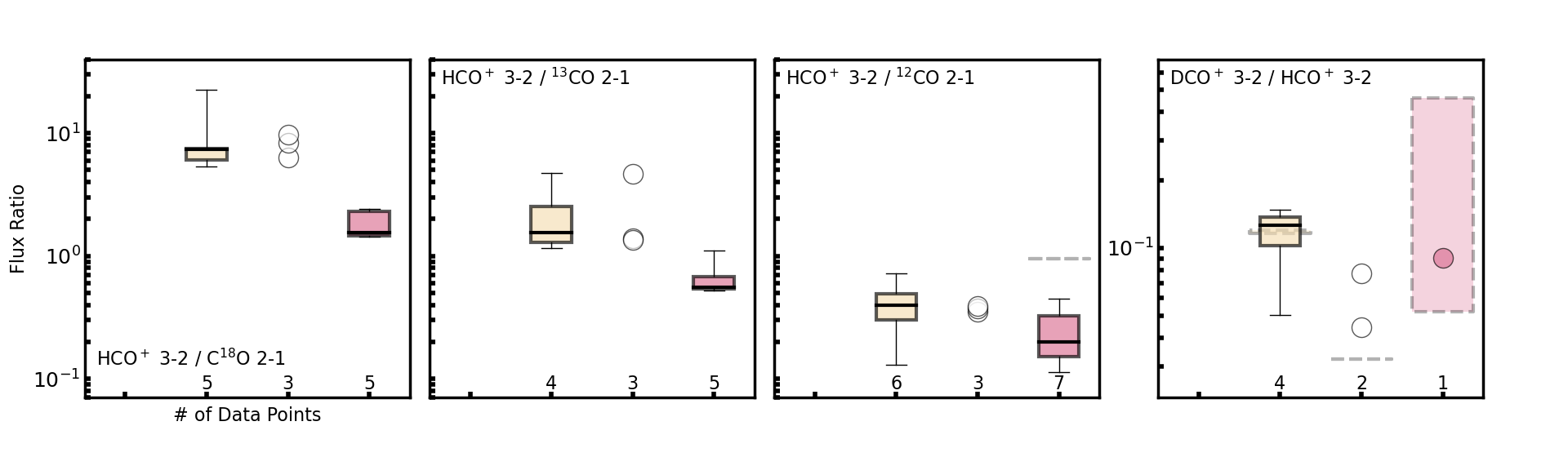}}
\resizebox{0.975\hsize}{!}{
    \includegraphics[trim=245pt 265pt 100pt 375pt, clip]{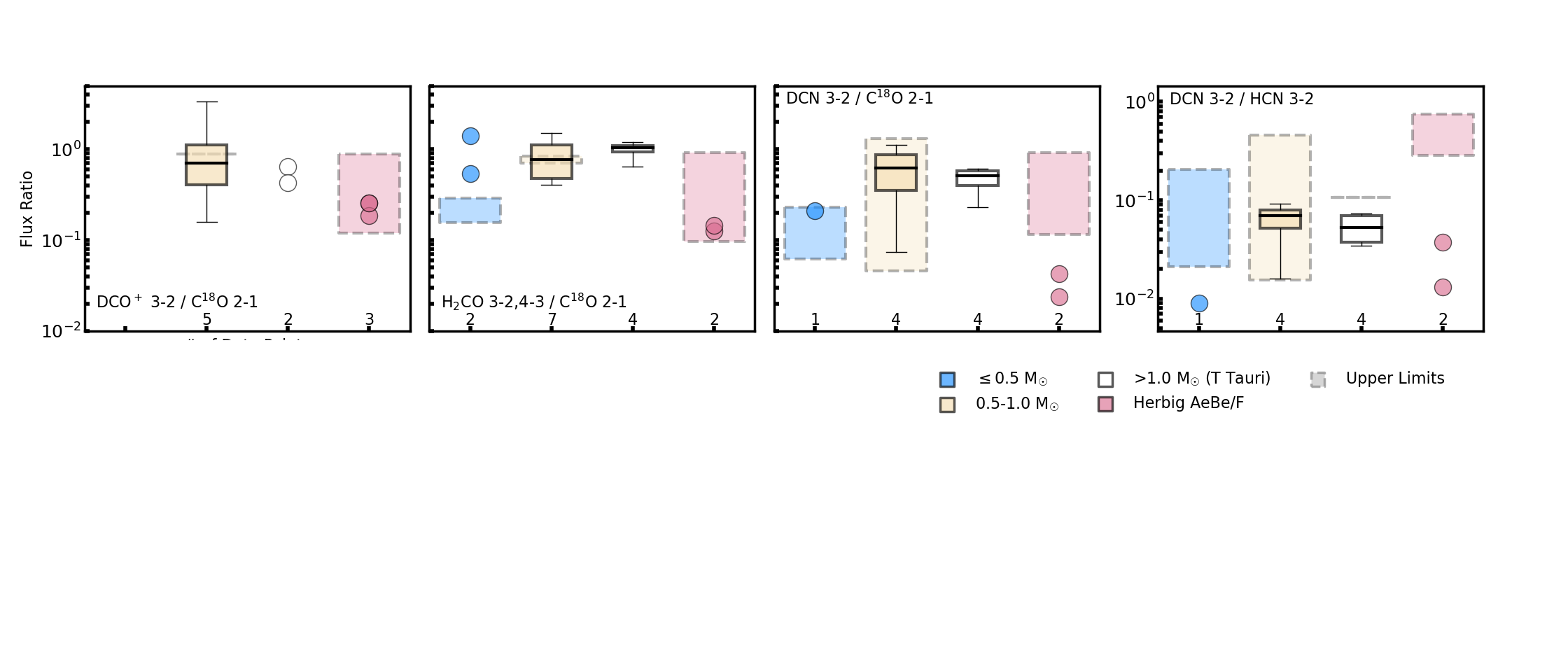}}
\caption{\textit{Top and middle rows:} HCO$^+$ 3--2, {DCO$^+$ 3--2,} and CO 2--1 isotopologue line flux ratios as a function of stellar luminosity (top row) and mm continuum flux (middle row) for Herbig Ae disks from this work (dotted red squares).  The literature sample of {Herbig AeBe/F disks (blank red squares) and T Tauri disks with stellar masses $>$ 1.0 M$_\Sun$ (thick white diamonds), 0.5-1.0 M$_\Sun$ (thin light gold diamonds), and $\leq$ 0.5 M$_\Sun$} (blue circles) are shown for comparison (Section~\ref{sec_sample_lit}, Appendix~\ref{sec_appendix_lit}).  3$\sigma$ upper limits are shown as triangles and point in the direction of the limit.  Points are not shown when both line fluxes are upper limits. \textit{Bottom row:} Boxplot summaries of the HCO$^+$ 3--2, {DCO$^+$ 3--2,} and CO 2--1 isotopologue line flux ratios for the combined samples, using the same color scheme as above.  The boxplots contain detected fluxes only.  Dashed boxes illustrate the span of any upper limits.  The numbers of data points included in the boxplots are written along the x-axis.  When $<$4, individual data points are drawn instead.  The ``box'' portions of the boxplots illustrate the 25$^\mathrm{th}$-50$^\mathrm{th}$-75$^\mathrm{th}$ percentiles, while the ``whiskers'' span all data.
\label{fig_flux_ion}}
\end{figure*}

\begin{figure*}
\centering
\resizebox{0.975\hsize}{!}{
    \includegraphics[trim=15pt 15pt 50pt 50pt, clip]{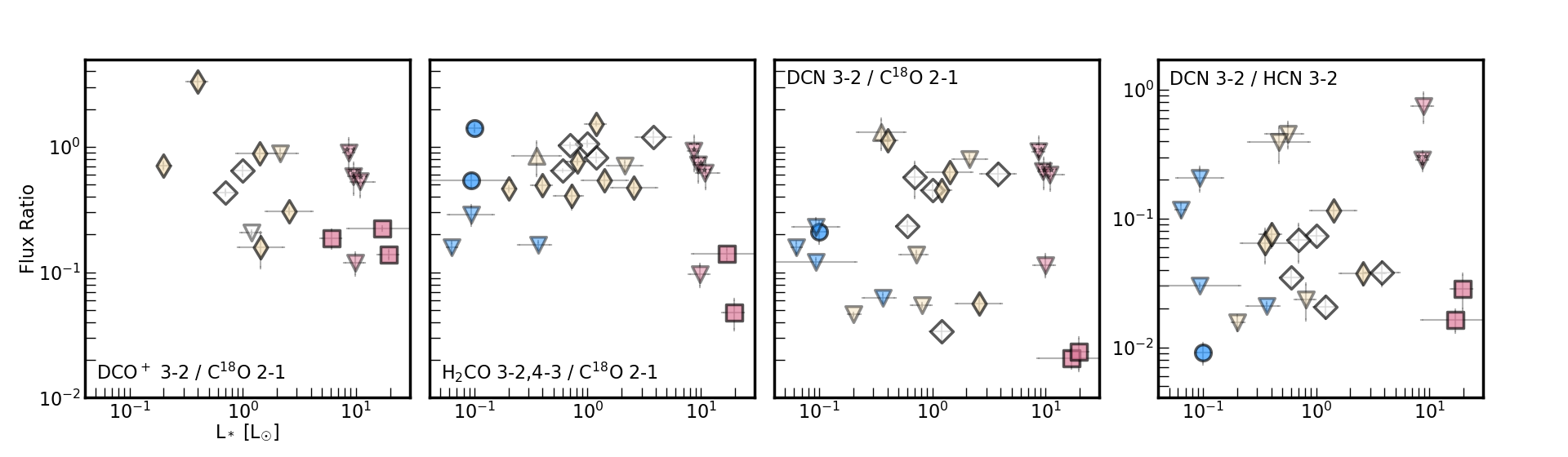}}
\resizebox{0.975\hsize}{!}{
    \includegraphics[trim=15pt 15pt 50pt 40pt, clip]{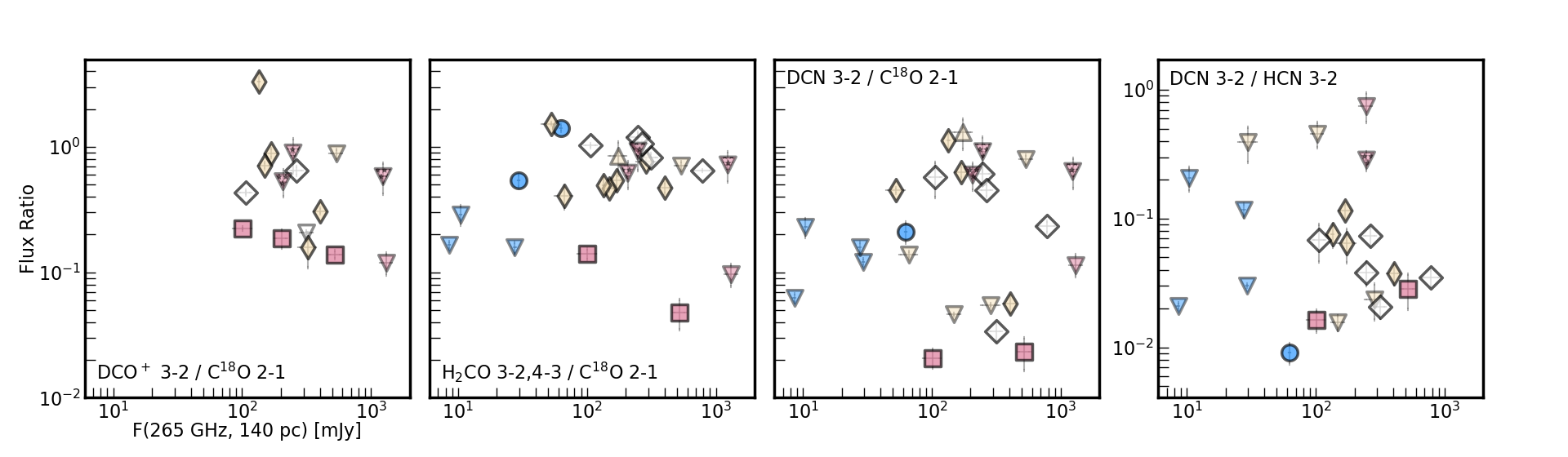}}
\resizebox{0.975\hsize}{!}{
    \includegraphics[trim=125pt 260pt 260pt 365pt, clip]{plot_HBsurv_ratio_kep_legend.png}}
\resizebox{0.975\hsize}{!}{
    \includegraphics[trim=5pt 15pt 50pt 40pt, clip]{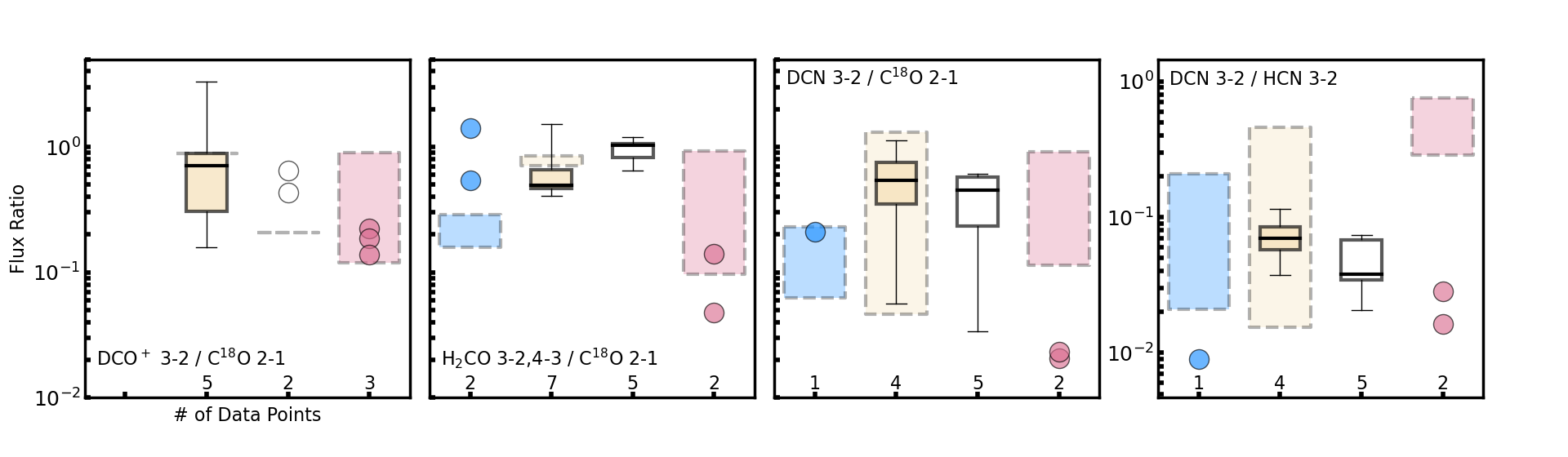}}
\resizebox{0.975\hsize}{!}{
    \includegraphics[trim=245pt 265pt 100pt 375pt, clip]{plot_HBsurv_box_ratio_legend.png}}
\caption{\textit{Top and middle rows:} {Line flux ratios involving DCO$^+$ 3--2, H$_2$CO 3--2,4--3, and DCN 3--2} as a function of stellar luminosity (top row) and mm continuum flux (middle row) for Herbig Ae disks from this work (dotted red squares).  The literature sample of {Herbig AeBe/F disks (blank red squares) and T Tauri disks with stellar masses $>$ 1.0 M$_\Sun$ (thick white diamonds), 0.5-1.0 M$_\Sun$ (thin light gold diamonds), and $\leq$ 0.5 M$_\Sun$} (blue circles) are shown for comparison (Section~\ref{sec_sample_lit}, Appendix~\ref{sec_appendix_lit}).  3$\sigma$ upper limits are shown as triangles and point in the direction of the limit.  Points are not shown when both line fluxes are upper limits. \textit{Bottom row:} Boxplot summaries of {line flux ratios involving DCO$^+$ 3--2, H$_2$CO 3--2,4--3, and DCN 3--2} for the combined samples, using the same color scheme as above.  The boxplots contain detected fluxes only.  Dashed boxes illustrate the span of any upper limits.  The numbers of data points included in the boxplots are written along the x-axis.  When $<$4, individual data points are drawn instead.  The ``box'' portions of the boxplots illustrate the 25$^\mathrm{th}$-50$^\mathrm{th}$-75$^\mathrm{th}$ percentiles, while the ``whiskers'' span all data.
\label{fig_flux_cold}}
\end{figure*}

\subsubsection{CO Isotopologues}
\label{sec_discussion_gas}

CO is the second most abundant {gas-phase} molecule in protoplanetary disks after H$_2$.  Its isotopologue emission has a complex dependence on disk structure:
$^{12}$CO emission is optically thick and traces the temperature and extent of the disk's surface layer; $^{13}$CO emission is less optically thick and traces the temperature and extent of a vertically lower emission layer; and finally C$^{18}$O emission is generally optically thin, emits near the disk midplane, and can serve as a tracer of disk gas mass{~\citep[e.g.,][]{cite_miotelloetal2014, cite_pinteetal2018, cite_mapsIV_2021, cite_mapsV_2021, cite_lawetal2022b}}.

Chemically, astrochemical disk models exploring a range of physical conditions have predicted that chemical processing, and therefore depletion, of CO should be most efficient in colder disks~\citep{cite_bosmanetal2018}.  As a result, the warmer Herbig AeBe/F disks should be \textit{most} abundant in CO.  {The study by~\cite{cite_mapsV_2021} agreed with this prediction; they} estimated larger CO gas masses {and higher CO-derived gas-to-dust ratios} for the Herbig Ae disks HD 163296 and MWC 480, relative to T Tauri disks, via high-resolution CO observations.

Figure~\ref{fig_flux_gas} plots CO isotopologue line flux ratios as a function of {mm continuum flux and stellar luminosity} for the combined disk sample.
There is a potential decrease in the line flux ratios with respect to stellar {luminosity}, with $\sim$one order of magnitude in scatter.
When considered with respect to mm continuum flux, the line flux ratios {appear scattered, but are consistent with a globally flat trend across both Herbig AeBe/F and T Tauri disks.}

Notably, most of the Herbig AeBe/F disks appear in the lowest regimes of the overall scatter for all three ratios.
%
{Let} us assume the optically thick $^{12}$CO 2--1 and $^{13}$CO 2--1 emission trace the temperatures of their emitting layers.  Then the {CO isotopologue flux ratios for the} Herbig AeBe/F disks, which are closest to {unity} in value compared to the rest of the combined sample, may {indicate vertical emitting layers with overall warmer (and therefore more numerically similar) temperatures.}
{Warmer layers would be consistent with the findings of~\cite{cite_mapsIV_2021} and~\cite{cite_lawetal2022a}, which altogether derived generally warmer CO temperature profiles from observations for Herbig AeBe disks when compared to T Tauri disks.}

{Let us also assume that the C$^{18}$O 2--1 emission is generally optically thin over most of the outer disk~\citep[e.g., see models, observations, and discussion by][]{cite_miotelloetal2014, cite_pinteetal2018, cite_mapsIV_2021, cite_lawetal2022b}, and can trace the disk gas mass.}
{From Figure~\ref{fig_abs_fcont}, we see that the Herbig AeBe/F disks have the largest C$^{18}$O 2--1 fluxes as a function of mm continuum flux.  Noting again from~\cite{cite_bosmanetal2018}} that CO depletion is expected to \textit{decrease} in warmer layers, we {can infer that the Herbig AeBe/F disks are generally most abundant in CO relative to disk mass, consistent with the results of~\cite{cite_mapsV_2021}}.

\subsubsection{Signposts of Ionization}
\label{sec_discussion_ion}

Based on both theory and observations, HCO$^+$ 3--2 is one of the simplest molecular ions found in protoplanetary disks~\citep[e.g.,][]{cite_dutreyetal1997, cite_aikawaetal2001}.  Its formation depends on the degree of ionization within the disk, which in turn depends on the presence/absence of stellar X-ray radiation~\citep[e.g.,][]{cite_aikawaetal2001}.  Models have predicted that HCO$^+$ is the dominant molecular ion {relative to H$^+_3$} in the warm CO-rich regions of the disk~\citep[specifically, where CO is at least a few orders of magnitude more abundant than free electrons;][]{cite_aikawaetal2015}.  We thus expect HCO$^+$ to be most prominent in the warm upper layers of the disk, where (1) CO is most abundant and (2) the disk material is most exposed to X-ray radiation from the central star~\citep{cite_mapsXIII_2021}.  Furthermore, since HCO$^+$ 3--2 emission is optically thick, we expect its emission to trace its emitting layer within the disk.

Figure~\ref{fig_flux_ion} plots ratios of HCO$^+$ 3--2{, DCO$^+$ 3--2,} and CO isotopologue fluxes as a function of {stellar luminosity and mm continuum flux} for the combined disk sample.  {Here $^{12}$CO 2--1 and $^{13}$CO 2--1 serve as probes of the emitting surface layers, the optically thin C$^{18}$O 2--1{~\citep[e.g., see models, observations, and discussion by][]{cite_miotelloetal2014, cite_pinteetal2018, cite_mapsIV_2021, cite_lawetal2022b}} as a normalizing metric for the HCO$^+$ 3--2 flux, and DCO$^+$ 3--2 as the deuterated counterpart of HCO$^+$ 3--2.}
 
The existing data above $>$0.5 M$_\Sun$ suggest a general decrease in the HCO$^+$ 3--2 flux relative to CO isotopologue fluxes as a function {of both stellar luminosity and mm continuum flux}.  Furthermore, Figure~\ref{fig_abs_fcont} {and the HCO$^+$ 3--2 / C$^{18}$O 2--1 flux ratios in Figure~\ref{fig_flux_ion} show} that the absolute HCO$^+$ 3--2 fluxes for the Herbig AeBe/F disks are distinctly offset from {the T Tauri disks with existing data}, and have lower HCO$^+$ 3--2 line fluxes when compared to T Tauri disks with the same mm continuum flux.  Together, these trends suggest a general decrease in HCO$^+$ 3--2 for Herbig AeBe/F disks relative to {T Tauri disks with M$_*$ $>$ 0.5 M$_\Sun$}.



One possible explanation for this trend is that the optically thick HCO$^+$ 3--2 emission is tracing a different emitting layer {relative to CO} for the Herbig AeBe/F disks {than for} the T Tauri disks.  {For T Tauri disks, models have predicted that HCO$^+$ is vertically cospatial with CO~\citep{cite_mapsXIII_2021}.  Observationally, \cite{cite_panequecarrenoetal2023} estimated and compared vertical emission surfaces for CO 2--1 isotopologues and HCO$^+$ 1--0 in two Herbig Ae disks (HD 163296 and MWC 480).  They found the HCO$^+$ 1--0 emission surfaces were not cospatial with $^{12}$CO 2--1 but were instead close to the midplane.  For HD 163296, they also noted that the HCO$^+$ 1--0 emission was likely optically thin, based on a low brightness temperature profile.  It is unclear if these results can be extrapolated to the HCO$^+$ 3--2 in this work, given the differences in excitation and optical depth between the two line transitions.}

Another possible explanation is that Herbig AeBe/F disks are relatively less ionized than T Tauri disks, as predicted from some disk models in the literature~\citep[e.g.,][]{cite_walshetal2015}.  Unlike convective T Tauri stars, Herbig AeBe stars are fully radiative and do not have a coronal layer.  They are thus not strong X-ray emitters like their T Tauri counterparts~\citep[e.g.,][]{cite_cohenetal1984}, and so their disks should receive less stellar X-ray radiation.  Less ionization could lead to less abundant HCO$^+$ in the Herbig AeBe/F disks.  This is consistent with the conclusions of~\cite{cite_mapsXIII_2021}, who found relatively low HCO$^+$ abundances for the Herbig Ae disk MWC 480, and attributed this result to relatively fewer high-energy X-rays incident on this disk.

Figure~\ref{fig_flux_ion} also shows the DCO$^+$ 3--2 / HCO$^+$ 3--2 line flux ratios.  We note, however, that unlike DCO$^+$ 3--2 emission, HCO$^+$ 3--2 emission is expected to be optically thick.  Therefore the HCO$^+$ 3--2 emission is likely not tracing its molecular abundance.  Keeping that in mind, we note that the sole existing Herbig AeBe/F disk DCO$^+$ 3--2 / HCO$^+$ 3--2 detection is consistent with the median flux ratio for the Tauri disks.  Intriguingly, this single flux ratio is from MWC 480, which was measured by~\cite{cite_mapsXIII_2021} to have a low HCO$^+$ abundance relative to the T Tauri disks in their sample.  This could suggest that DCO$^+$ is similarly less ionized in Herbig AeBe/F disks, as discussed for HCO$^+$ in Section~\ref{sec_discussion_ion}, although more observations are needed to further investigate this possibility.

%

\subsubsection{Cold Chemistry}
\label{sec_discussion_cold}

Three of our target molecules are expected to trace cold molecular gas: DCN, DCO$^+$, and H$_2$CO.  Based on theoretical models, deuterated chemical pathways are expected to be most efficient at cold temperatures, although warm pathways exist as well.  Production of the deuterated molecule DCO$^+$ is expected to peak within the cold disk midplane at temperatures $\lesssim$30K, while DCN production is expected to peak in the warm inner regions~{\cite[e.g., results and discussion by][]{cite_millaretal1989, cite_aikawaetal1999, cite_aikawaetal2001, cite_huangetal2017, cite_obergetal2021a, cite_munozromeroetal2023}}. The simple oxygen carrier H$_2$CO is predicted to form through both gas-phase pathways in the warm inner disk and through CO ice hydrogenation on grain surfaces beyond the CO snowline~\citep[e.g.,][]{cite_hiraokaetal1994, cite_fockenbergetal2002, cite_hiraokaetal2002, cite_watanabeetal2002}.
%
Notably \cite{cite_peguesetal2020} and~\cite{cite_mapsVI_2021} estimated lower H$_2$CO column densities for the two Herbig Ae disks in their samples (HD 163296 and MWC 480) relative to colder T Tauri disks, possibly due (at least in part) to relatively small reservoirs of CO ice around the warmer stars.

Figure~\ref{fig_flux_cold} {plots DCN 3--2, DCO$^+$ 3--2, and H$_2$CO 3--2,4--3 line fluxes relative to C$^{18}$O 2--1 line fluxes as a function of stellar luminosity and mm continuum flux for the combined disk sample.  We also show DCN 3--2 / HCN 3--2, although we note that HCN 3--2 emission is expected to be optically thick.}
For the Herbig AeBe/F disks, only a few disks have been detected in {some or all of} these three {lines (see also Figure~\ref{fig_abs_fcont})}.  The rest of the existing Herbig AeBe/F disk data are upper limits from tentative and non-detections.

{The few existing detected DCO$^+$ 3--2, DCN 3--2, and H$_2$CO 3--2,4--3 line fluxes relative to C$^{18}$O 2--1 for the Herbig AeBe/F disks are below the median values for the T Tauri disks.  When considered as a function of mm continuum flux, these few detections occupy the lowest regimes of the overall scatter across all observations.}
{These results are} consistent with the hypothesis that these molecules with dominant cold pathways are less abundant in the warmer Herbig AeBe/F disks.  {The few detected, relatively low line flux ratios may also be a byproduct of less CO depletion (and therefore more C$^{18}$O 2--1 emission) in the Herbig AeBe/F disks (see Section~\ref{sec_discussion_gas}).}

{We emphasize, however, that only a few Herbig AeBe/F disk detections exist for these lines.  Upper limits for the Herbig AeBe/F disks are either consistent with T Tauri disk fluxes and upper limits, or are lower, and are not all constraining.}  {Surveys of these lines toward Herbig AeBe/F disks} at higher sensitivity are needed to {further investigate} these tentative trends.

%

%

\section{Summary}
\label{sec_summary}

Using the SMA, we have surveyed millimeter-wavelength ($\sim$213-268 GHz) molecular line emission from protoplanetary disks around four Herbig Ae stars (HD 34282, HD 36112, HD 142666, and HD 144432) and one Herbig Be star (HD 38120), and we have compared our results to disk chemistry around Herbig AeBe/F stars and T Tauri stars from the literature.  We summarize our main findings below:

\begin{enumerate}
    \item All five disks are detected in mm continuum emission, while four are detected in molecular line emission (Section~\ref{sec_results_detections}).  The fifth disk, HD 38120, appears to be cloud contaminated near its systemic velocity and likely has fainter line emission than was previously expected from single-dish observations in the literature.
    \item Focusing on the other four disks in the sample, $^{12}$CO 2--1 is detected from all four disks (Section~\ref{sec_results_detections}).  $^{13}$CO 2--1, C$^{18}$O 2--1, HCO$^+$ 3--2 are detected from three disks, and $^{13}$CO 2--1 is tentatively detected from a fourth disk.  HCN 3--2 is detected from one disk and tentatively detected from another.  Finally, CS 5--4 and DCO$^+$ 3--2 are tentatively detected from two disks each, and C$_2$H 3--2 and DCN 3--2 are tentatively detected from one disk each.  H$_2$CO 3--2 is not detected at the sensitivity of our observations.
    \item Based on gas and dust morphologies from this work and from{~\cite{cite_pietuetal2003},~\cite{cite_vanderplasetal2017},~\cite{cite_stapperetal2022},~\cite{cite_lawetal2022b}}, and the ALMA archive, HD 34282 appears to be adjacent to a faint source of emission. {This source is detected in mm continuum emission and is either tentatively or not detected in molecular line emission}  (Section~\ref{sec_results_mom0_hd34282}).  This neighbor is located roughly $\sim$5 arcseconds {eastward} of HD 34282. 
    {We} speculate that the neighbor could be an extended, clumpy spiral arm or a distant orbital companion, {for which the gas is either not traceable in our target molecular lines or has dispersed}.
    \item HD 144432 has previously been undetected in scattered light~\citep[i.e., emission from $\sim$micron-sized dust grains;][]{cite_monnieretal2017}.  Here we detect HD 144432 in emission from $\sim$millimeter-sized dust grains (Section~\ref{sec_results_mom0_hd144432}).  It is possible this disk is compact, truncated~\citep[see discussion by][]{cite_monnieretal2017}, or has lost material {with} age.
    \item We have compiled a database of known fluxes and upper limits for protoplanetary disks from the literature that have been detected in at least one of our ten target molecular lines (Section~\ref{sec_sample_lit}, Appendix~\ref{sec_appendix_lit}).
    \begin{enumerate}
        \item Across the combined sample, {CO 2--1 isotopologue} flux ratios appear {closest} to unity {for Herbig AeBe/F disks} relative to T Tauri disks ({Section~\ref{sec_discussion_gas}}).  The CO line flux ratios {may} indicate their vertical emitting layers have relatively similar, {warmer} temperatures in Herbig AeBe/F disks, {and more abundant CO relative to disk dust mass overall}.  Relatively {diminished} HCO$^+$ 3--2 {flux ratios may} result from less ionization in Herbig AeBe/F disks, as Herbig AeBe stars are fainter X-ray emitters than their T Tauri counterparts {(Section~\ref{sec_discussion_ion})}.
        \item DCN 3--2, DCO$^+$ 3--2, and H$_2$CO 3--2 have been detected in only a few Herbig AeBe/F disks so far.  Observations from this work for these lines yielded only upper limits (Section~\ref{sec_discussion_cold}).  These low detection rates, and flux ratios for the few existing detections, are consistent with studies from the literature, which have predicted smaller regimes of cold chemistry in disks around the hotter Herbig AeBe/F stars.
    \end{enumerate}
\end{enumerate}

We stress that these conclusions are based on the small sample of disk chemistry surveyed so far (Appendix~\ref{sec_appendix_lit}).  We look forward to any future follow-up or new observations of protoplanetary disks, particularly Herbig AeBe/F disks, that could test these hypotheses over a larger sample of the pre-main-sequence stellar mass distribution.
%

\acknowledgments

{We thank the anonymous referee for their helpful comments that greatly improved the structure and clarity of this manuscript.}

C.J.L. acknowledges funding from the National Science Foundation Graduate Research Fellowship under Grant DGE1745303.

Support for J.H. was provided by NASA through the NASA Hubble Fellowship grant \#HST-HF2-51460.001- A awarded by the Space Telescope Science Institute, which is operated by the Association of Universities for Research in Astronomy, Inc., for NASA, under contract NAS5-26555.

The Submillimeter Array is a joint project between the Smithsonian Astrophysical Observatory and the Academia Sinica Institute of Astronomy and Astrophysics and is funded by the Smithsonian Institution and the Academia Sinica.

This paper makes use of the following ALMA data:

{ADS/JAO.ALMA\#2013.1.00658.S,}

{ADS/JAO.ALMA\#2015.1.00192.S,}

{ADS/JAO.ALMA\#2017.1.01578.S.}

ALMA is a partnership of ESO (representing its member states), NSF (USA) and NINS (Japan), together with NRC (Canada), MOST and ASIAA (Taiwan), and KASI (Republic of Korea), in cooperation with the Republic of Chile. The Joint ALMA Observatory is operated by ESO, AUI/NRAO and NAOJ.  The National Radio Astronomy Observatory is a facility of the National Science Foundation operated under cooperative agreement by Associated Universities, Inc.

This work has made use of data from the European Space Agency (ESA) mission {\it Gaia} (\url{https://www.cosmos.esa.int/gaia}), processed by the {\it Gaia} Data Processing and Analysis Consortium (DPAC, \url{https://www.cosmos.esa.int/web/gaia/dpac/consortium}). Funding for the DPAC has been provided by national institutions, in particular the institutions participating in the {\it Gaia} Multilateral Agreement.

\software{
All data reduction scripts and computer code used for this research were written in {Python} (version 3.8).  {Data calibration and imaging were performed using the \textit{MIR} package (\url{https://www.cfa.harvard.edu/~cqi/mircook.html}) and CASA~\citep{cite_casa_1, cite_casa_2}}. 
 All plots were generated using {Python}'s {Matplotlib} package~\citep{cite_matplotlib}.  Colormaps were generated using the {cmasher} package~\citep{cite_cmasher}.  This research also made use of {Astropy} (\url{http://www.astropy.org}), a community-developed core {Python} package for Astronomy~\citep{cite_astropy2013, cite_astropy2018}, as well as the {NumPy}~\citep{cite_numpy} Python package.}

\bibliography{projectbib}

\clearpage

\appendix

\section{The Literature Sample of Protoplanetary Disks}
\label{sec_appendix_lit}

\subsection{Stellar Characteristics for the Literature Sample.}

Table~\ref{table_litdisk} lists the stellar characteristics of the literature sample (Section~\ref{sec_sample_lit}).

{Please refer to the original references for retrieval and discussion of all star+disk systems and their characteristics.}

\startlongtable
\begin{deluxetable*}{lccccccc}
\tablecaption{Stellar Characteristics of the Literature Sample. \label{table_litdisk}}
\tablehead{
Star+Disk              & Distance & Spectral     & $L_*$     & $M_*$     & $F_{\nu}$ & $\nu$   & References \\
System              & (pc) &   Type   & (L$_\Sun$)     & (M$_\Sun$) &     (mJy)    & (GHz)   &  (Distance; Spectral Type; $L_*$; $M_*$; $F_\nu$)
}
 \colnumbers \startdata
\hline
AA Tau             & 135   & K5        & 0.56$_{-0.20}^{+0.20}$   & 0.84$_{-0.04}^{+0.04}$ & 113$\pm$11      & 267               & 1; 2; 3; 4; 5                                   \\
AK Sco             & 140   & F5        & $\sim$4.70               & 2.49$_{-0.10}^{+0.10}$ & 33$\pm$0.07     & 231               & 1; 6; 7; 8; 8                                   \\
AS 205             & 142   & K5        & 2.14$_{-0.62}^{+0.88}$   & 0.87$_{-0.09}^{+0.15}$ & 528$\pm$53      & 267               & 1; 9; 10; 10; 11                                \\
AS 209             & 121   & K5        & 1.41$_{-0.52}^{+0.83}$   & 0.83$_{-0.23}^{+0.24}$ & 229$\pm$23      & 267               & 1; 12; 10; 10; 11                               \\
CI Tau             & 160   & K5.5      & $\sim$0.81               & 0.90$_{-0.02}^{+0.02}$ & $\sim$161       & 231               & 1; 13; 14; 4; 15                                \\
CM Cha             & 192   & K7        & 0.86$_{-0.09}^{+0.11}$   & 0.81$_{-0.18}^{+0.17}$ & 39$\pm$0.20     & 231               & 1; 16; 17; 18; 18                               \\
CQ Tau             & 149   & F5        & $\sim$10.70              & 1.51$_{-0.07}^{+0.07}$ & 258$\pm$26      & 267               & 1; 19; 7; 4; 5                                  \\
CW Tau             & 132   & K3.0      & 2.40$_{-0.66}^{+0.69}$   & 0.64$_{-0.01}^{+0.01}$ & 160$\pm$27      & 337               & 1; 13; 20; 4; 21                                \\
\enddata
\end{deluxetable*}
%

\subsection{Molecular Line Fluxes for the Literature Sample.}

Tables~\ref{table_litmol_12CO} through~\ref{table_litmol_HCOplus} list molecular line fluxes and upper limits for the ten target molecular lines (Table~\ref{table_mol}) observed from the literature disk sample (Section~\ref{sec_sample_lit}), in alphabetical order by molecular line.  The fluxes here are written exactly as they were reported in the literature.  They are scaled separately by distance and/or frequency in our figures for consistent comparison.

{Please refer to the original references for retrieval of all line fluxes, how the fluxes were measured, and discussion of the star+disk systems.}

\startlongtable
\begin{deluxetable}{lcc}
\tablecaption{$^{12}$CO 2--1 Fluxes for the Literature Sample. \label{table_litmol_12CO}}
\tablehead{
Star+Disk              & $^{12}$CO 2--1 Flux & Reference \\
 System    &  (mJy km s$^{-1}$)    &
}
 \colnumbers \startdata
\hline
AA Tau             & 8200$\pm$160 (S)    & \cite{cite_obergetal2010}                      \\ 
AK Sco             & 2210$\pm$10 (A)     & \cite{cite_czekalaetal2015}                    \\ 
AS 205             & 19230$\pm$270 (A)   & \cite{cite_salyketal2014}                      \\ 
AS 209             & 7790$\pm$45 (A)     & \cite{cite_mapsI_2021}                  \\ 
CM Cha             & 670$\pm$40 (A)      & \cite{cite_villenaveetal2021}                  \\ 
CQ Tau             & 3100$\pm$180 (S)    & \cite{cite_obergetal2010}                      \\ 
DM Tau             & 14870$\pm$120 (S)   & \cite{cite_obergetal2010}                      \\ 
FP Tau             & 781$\pm$11 (A)      & \cite{cite_peguesetal2021b}                    \\
\enddata
\tablecomments{Columns 1 and 2: $^{12}$CO 2--1 line fluxes for sources from the literature sample (Section~\ref{sec_sample_lit}).  3$\sigma$ upper limits are marked with $<$.  Fluxes from ALMA, the SMA, and the IRAM PdBI are marked with (A), (S), and (I), respectively.  When fluxes from multiple instruments are available, we select the flux from the instrument with the highest spatial resolution to decrease the chance of contamination from external sources of emission.  Column 3: Reference.}
\end{deluxetable}
%


\startlongtable
\begin{deluxetable}{lcc}
\tablecaption{$^{13}$CO 2--1 Fluxes from the Literature Sample. \label{table_litmol_13CO}}
\tablehead{
Star+Disk              & $^{13}$CO 2--1 Flux & Reference \\
 System    &  (mJy km s$^{-1}$)    &
}
 \colnumbers \startdata
\hline
AK Sco             & 540$\pm$10 (A)      & \cite{cite_czekalaetal2015}                    \\ 
AS 205             & 1860$\pm$40 (A)     & \cite{cite_salyketal2014}                      \\ 
AS 209             & 2269$\pm$38 (A)     & \cite{cite_mapsI_2021}                  \\ 
CX Tau             & $\sim$400 (I)       & \cite{cite_pietuetal2014}                      \\ 
DE Tau             & $\sim$400 (I)       & \cite{cite_pietuetal2014}                      \\ 
DS Tau             & $\sim$500 (I)       & \cite{cite_pietuetal2014}                      \\ 
EX Lup             & 760$\pm$200 (A)     & \cite{cite_ansdelletal2018}                    \\ 
FP Tau             & 232$\pm$8 (A)       & \cite{cite_peguesetal2021b}                    \\
\enddata
\tablecomments{Column 1: Star+disk system from the literature sample (Section~\ref{sec_sample_lit}).  Column 2: $^{13}$CO 2--1 line fluxes for protoplanetary disks from the literature sample (Section~\ref{sec_sample_lit}).  3$\sigma$ upper limits are marked with $<$.  Fluxes from ALMA, the SMA, and the IRAM PdBI are marked with (A), (S), and (I), respectively.  When fluxes from multiple instruments are available, we select the flux from the instrument with the highest spatial resolution to decrease the chance of contamination from external sources of emission.  Only detected $^{13}$CO 2--1 fluxes are shown from~\cite{cite_ansdelletal2018}.  Column 3: Reference.}
\end{deluxetable}
%


\startlongtable
\begin{deluxetable}{lcc}
\tablecaption{C$^{18}$O 2--1 Fluxes from the Literature Sample. \label{table_litmol_C18O}}
\tablehead{
Star+Disk              & C$^{18}$O 2--1 Flux & Reference \\
 System    &  (mJy km s$^{-1}$)    &
}
 \colnumbers \startdata
\hline
AK Sco             & 200$\pm$10 (A)      & \cite{cite_czekalaetal2015}                    \\ 
AS 205             & 340$\pm$30 (A)      & \cite{cite_salyketal2014}                      \\ 
AS 209             & 538$\pm$27 (A)      & \cite{cite_mapsI_2021}                  \\ 
CI Tau             & 549$\pm$17 (A)      & \cite{cite_peguesetal2020}                     \\ 
CW Tau             & $\sim$200 (I)       & \cite{cite_pietuetal2014}                      \\ 
DM Tau             & 998$\pm$17 (A)      & \cite{cite_peguesetal2020}                     \\ 
DO Tau             & 210$\pm$12 (A)      & \cite{cite_peguesetal2020}                     \\ 
FP Tau             & 91$\pm$7 (A)        & \cite{cite_peguesetal2021b}                    \\
\enddata
\tablecomments{Column 1: Star+disk system from the literature sample (Section~\ref{sec_sample_lit}).  Column 2: C$^{18}$O 2--1 line fluxes for protoplanetary disks from the literature sample (Section~\ref{sec_sample_lit}).  3$\sigma$ upper limits are marked with $<$.  Fluxes from ALMA, the SMA, and the IRAM PdBI are marked with (A), (S), and (I), respectively.  When fluxes from multiple instruments are available, we select the flux from the instrument with the highest spatial resolution to decrease the chance of contamination from external sources of emission.  Only detected C$^{18}$O 2--1 fluxes are shown from~\cite{cite_ansdelletal2018}.  Column 3: Reference.}
\end{deluxetable}
%


\begin{deluxetable}{lcc}
\tablecaption{C$_2$H 3--2 Fluxes from the Literature Sample. \label{table_litmol_CCH32a}}
\tablehead{
Star+Disk              & C$_2$H 3--2 Flux & Reference \\
 System    &  (mJy km s$^{-1}$)    &
}
 \colnumbers \startdata
\hline
AS 209             & 2709$\pm$290 (A)    & \cite{cite_bergneretal2019}                    \\ 
CI Tau             & 1041$\pm$110 (A)    & \cite{cite_bergneretal2019}                    \\ 
DM Tau             & 2013$\pm$200 (A)    & \cite{cite_bergneretal2019}                    \\ 
DO Tau             & $<$62 (A)           & \cite{cite_bergneretal2019}                    \\ 
FP Tau             & 140$\pm$18 (A)      & \cite{cite_peguesetal2021b}                    \\ 
GM Aur             & 801$\pm$141 (A)     & \cite{cite_mapsVI_2021}                 \\ 
HD 143006          & 435$\pm$56 (A)      & \cite{cite_bergneretal2019}                    \\ 
HD 163296          & 4396$\pm$440 (A)    & \cite{cite_bergneretal2019}                    \\
\enddata
\tablecomments{Column 1: Star+disk system from the literature sample (Section~\ref{sec_sample_lit}).  Column 2: C$_2$H 3--2 line fluxes for protoplanetary disks from the literature sample (Section~\ref{sec_sample_lit}).  3$\sigma$ upper limits are marked with $<$.  Fluxes from ALMA, the SMA, and the IRAM PdBI are marked with (A), (S), and (I), respectively.  When fluxes from multiple instruments are available, we select the flux from the instrument with the highest spatial resolution to decrease the chance of contamination from external sources of emission.  Column 3: Reference.}
\end{deluxetable}
%


\begin{deluxetable}{lcc}
\tablecaption{CS 5--4 Fluxes from the Literature Sample. \label{table_litmol_CS}}
\tablehead{
Star+Disk              & CS 5--4 Flux & Reference \\
 System    &  (mJy km s$^{-1}$)    &
}
 \colnumbers \startdata
\hline
CI Tau             & 756$\pm$17 (A)      & \cite{cite_legaletal2019}                      \\ 
DM Tau             & 388$\pm$9 (A)       & \cite{cite_legaletal2019}                      \\ 
DO Tau             & 159$\pm$6 (A)       & \cite{cite_legaletal2019}                      \\ 
LkCa 15            & 1050$\pm$14 (A)     & \cite{cite_legaletal2019}                      \\ 
MWC 480            & 536$\pm$19 (A)      & \cite{cite_legaletal2019}                      \\ 
PDS 70             & 450$\pm$6 (A)       & \cite{cite_facchinietal2021}                    
\enddata
\tablecomments{Column 1: Star+disk system from the literature sample (Section~\ref{sec_sample_lit}).  Column 2: C$_2$H 3--2 line fluxes for protoplanetary disks from the literature sample (Section~\ref{sec_sample_lit}).  3$\sigma$ upper limits are marked with $<$.  Fluxes from ALMA, the SMA, and the IRAM PdBI are marked with (A), (S), and (I), respectively.  When fluxes from multiple instruments are available, we select the flux from the instrument with the highest spatial resolution to decrease the chance of contamination from external sources of emission.  Column 3: Reference.}
\end{deluxetable}
%


\begin{deluxetable}{lcc}
\tablecaption{DCN 3--2 Fluxes from the Literature Sample. \label{table_litmol_DCN32}}
\tablehead{
Star+Disk              & DCN 3--2 Flux & Reference \\
 System    &  (mJy km s$^{-1}$)    &
}
 \colnumbers \startdata
\hline
AA Tau             & $<$270 (S)          & \cite{cite_obergetal2010}                      \\ 
AS 205             & $<$270 (S)          & \cite{cite_obergetal2011}                      \\ 
AS 209             & 340$\pm$30 (A)      & \cite{cite_huangetal2017}                      \\ 
CI Tau             & 27$\pm$10 (A)       & \cite{cite_bergneretal2020}                    \\ 
CQ Tau             & $<$255 (S)          & \cite{cite_obergetal2010}                      \\ 
DM Tau             & $<$46 (A)           & \cite{cite_bergneretal2020}                    \\ 
DO Tau             & $<$29 (A)           & \cite{cite_bergneretal2020}                    \\ 
FP Tau             & $<$6 (A)            & \cite{cite_peguesetal2021b}                    \\
\enddata
\tablecomments{Column 1: Star+disk system from the literature sample (Section~\ref{sec_sample_lit}).  Column 2: DCN 3--2 line fluxes for protoplanetary disks from the literature sample (Section~\ref{sec_sample_lit}).  3$\sigma$ upper limits are marked with $<$.  Fluxes from ALMA, the SMA, and the IRAM PdBI are marked with (A), (S), and (I), respectively.  When fluxes from multiple instruments are available, we select the flux from the instrument with the highest spatial resolution to decrease the chance of contamination from external sources of emission.  Column 3: Reference.}
\end{deluxetable}
%


\begin{deluxetable}{lcc}
\tablecaption{DCO$^+$ 3--2 Fluxes from the Literature Sample. \label{table_litmol_DCOplus}}
\tablehead{
Star+Disk              & DCO$^+$ 3--2 Flux & Reference \\
 System    &  (mJy km s$^{-1}$)    &
}
 \colnumbers \startdata
\hline
AA Tau             & $<$270 (S)          & \cite{cite_obergetal2010}                      \\ 
AS 205             & $<$300 (S)          & \cite{cite_obergetal2011}                      \\ 
AS 209             & 480$\pm$30 (A)      & \cite{cite_huangetal2017}                      \\ 
CQ Tau             & $<$345 (S)          & \cite{cite_obergetal2010}                      \\ 
DM Tau             & 710$\pm$40 (S)      & \cite{cite_obergetal2010}                      \\ 
GM Aur             & $<$225 (S)          & \cite{cite_obergetal2010}                      \\ 
HD 142527          & $<$405 (S)          & \cite{cite_obergetal2011}                      \\ 
HD 163296          & 1290$\pm$40 (A)     & \cite{cite_huangetal2017}                      \\
\enddata
\tablecomments{Column 1: Star+disk system from the literature sample (Section~\ref{sec_sample_lit}).  Column 2: DCO$^+$ 3--2 line fluxes for protoplanetary disks from the literature sample (Section~\ref{sec_sample_lit}).  3$\sigma$ upper limits are marked with $<$.  Fluxes from ALMA, the SMA, and the IRAM PdBI are marked with (A), (S), and (I), respectively.  When fluxes from multiple instruments are available, we select the flux from the instrument with the highest spatial resolution to decrease the chance of contamination from external sources of emission.  Column 3: Reference.}
\end{deluxetable}
%


\begin{deluxetable}{lcc}
\tablecaption{H$_2$CO 3--2 Fluxes from the Literature Sample. \label{table_litmol_H2CO}}
\tablehead{
Star+Disk              & H$_2$CO 3--2 Flux & Reference \\
System    &  (mJy km s$^{-1}$)    &
}
 \colnumbers \startdata
\hline
AA Tau             & 160$\pm$80 (S)      & \cite{cite_obergetal2010}                      \\ 
AS 205             & $<$240 (S)          & \cite{cite_obergetal2011}                      \\ 
AS 209             & 292$\pm$28 (A)      & \cite{cite_mapsVI_2021}                       \\ 
CI Tau             & 420$\pm$12 (A)      & \cite{cite_peguesetal2020}                     \\ 
CQ Tau             & $<$315 (S)          & \cite{cite_obergetal2010}                      \\ 
DM Tau             & 462$\pm$17 (A)      & \cite{cite_peguesetal2020}                     \\ 
DO Tau             & 85$\pm$17 (A)       & \cite{cite_peguesetal2020}                     \\ 
FP Tau             & $<$15 (A)           & \cite{cite_peguesetal2021b}                    \\
\enddata
\tablecomments{Column 1: Star+disk system from the literature sample (Section~\ref{sec_sample_lit}).  Column 2: H$_2$CO 3--2 line fluxes for protoplanetary disks from the literature sample (Section~\ref{sec_sample_lit}).  3$\sigma$ upper limits are marked with $<$.  Fluxes from ALMA, the SMA, and the IRAM PdBI are marked with (A), (S), and (I), respectively.  When fluxes from multiple instruments are available, we select the flux from the instrument with the highest spatial resolution to decrease the chance of contamination from external sources of emission.  H$_2$CO 3--2 fluxes that are actually H$_2$CO 4--3 fluxes used as proxies are marked with a `$^*$'.  Column 3: Reference.}
\end{deluxetable}
%


\begin{deluxetable}{lcc}
\tablecaption{HCN 3--2 Fluxes from the Literature Sample. \label{table_litmol_HCN}}
\tablehead{
Star+Disk              & HCN 3--2 Flux & Reference \\
 System    &  (mJy km s$^{-1}$)    &
}
 \colnumbers \startdata
\hline
AA Tau             & 590$\pm$130 (S)     & \cite{cite_obergetal2010}                      \\ 
AS 205             & $<$510 (S)          & \cite{cite_obergetal2011}                      \\ 
AS 209             & 2965$\pm$297 (A)    & \cite{cite_mapsVI_2021}                       \\ 
CI Tau             & 1271$\pm$390 (S)    & \cite{cite_bergneretal2019}                    \\ 
CQ Tau             & $<$750 (S)          & \cite{cite_obergetal2010}                      \\ 
DM Tau             & 2966$\pm$400 (S)    & \cite{cite_bergneretal2019}                    \\ 
DO Tau             & $<$1252 (S)         & \cite{cite_bergneretal2019}                    \\ 
FP Tau             & 272$\pm$17 (A)      & \cite{cite_peguesetal2021b}                    \\
\enddata
\tablecomments{Column 1: Star+disk system from the literature sample (Section~\ref{sec_sample_lit}).  Column 2: HCN 3--2 line fluxes for protoplanetary disks from the literature sample (Section~\ref{sec_sample_lit}).  3$\sigma$ upper limits are marked with $<$.  Fluxes from ALMA, the SMA, and the IRAM PdBI are marked with (A), (S), and (I), respectively.  When fluxes from multiple instruments are available, we select the flux from the instrument with the highest spatial resolution to decrease the chance of contamination from external sources of emission.  Column 3: Reference.
\\ $\dagger$: From~\cite{cite_qietal2008} as reported in~\cite{cite_cleevesetal2015}.
}
\end{deluxetable}
%


\begin{deluxetable}{lcc}
\tablecaption{HCO$^+$ 3--2 Fluxes from the Literature Sample. \label{table_litmol_HCOplus}}
\tablehead{
Star+Disk              & HCO$^+$ 3--2 Flux & Reference \\
 System    &  (mJy km s$^{-1}$)    &
}
 \colnumbers \startdata
\hline
AA Tau             & 2320$\pm$190 (S)    & \cite{cite_obergetal2010}                      \\ 
AS 205             & 2500$\pm$120 (S)    & \cite{cite_obergetal2011}                      \\ 
AS 209             & 4020$\pm$160 (S)    & \cite{cite_obergetal2011}                      \\ 
CQ Tau             & 750$\pm$240 (S)     & \cite{cite_obergetal2010}                      \\ 
DM Tau             & 5340$\pm$120 (S)    & \cite{cite_obergetal2010}                      \\ 
GM Aur             & 6960$\pm$60 (A)     & \cite{cite_huangetal2020a}                     \\ 
HD 142527          & 7790$\pm$530 (S)    & \cite{cite_obergetal2011}                      \\ 
IM Lup             & 9670$\pm$290 (S)    & \cite{cite_obergetal2011}                      \\
\enddata
\tablecomments{Column 1: Star+disk system from the literature sample (Section~\ref{sec_sample_lit}).  Column 2: HCO$^+$ 3--2 line fluxes for protoplanetary disks from the literature sample (Section~\ref{sec_sample_lit}).  3$\sigma$ upper limits are marked with $<$.  Fluxes from ALMA, the SMA, and the IRAM PdBI are marked with (A), (S), and (I), respectively.  When fluxes from multiple instruments are available, we select the flux from the instrument with the highest spatial resolution to decrease the chance of contamination from external sources of emission.  Column 3: Reference.}
\end{deluxetable}

\clearpage

\section{Unbiased Spectra for all SMA Bands}
\label{sec_appendix_fullband}

\begin{figure*}
\centering
\resizebox{0.99\hsize}{!}{
    \includegraphics[trim=0pt 0pt 0pt 0pt, clip]{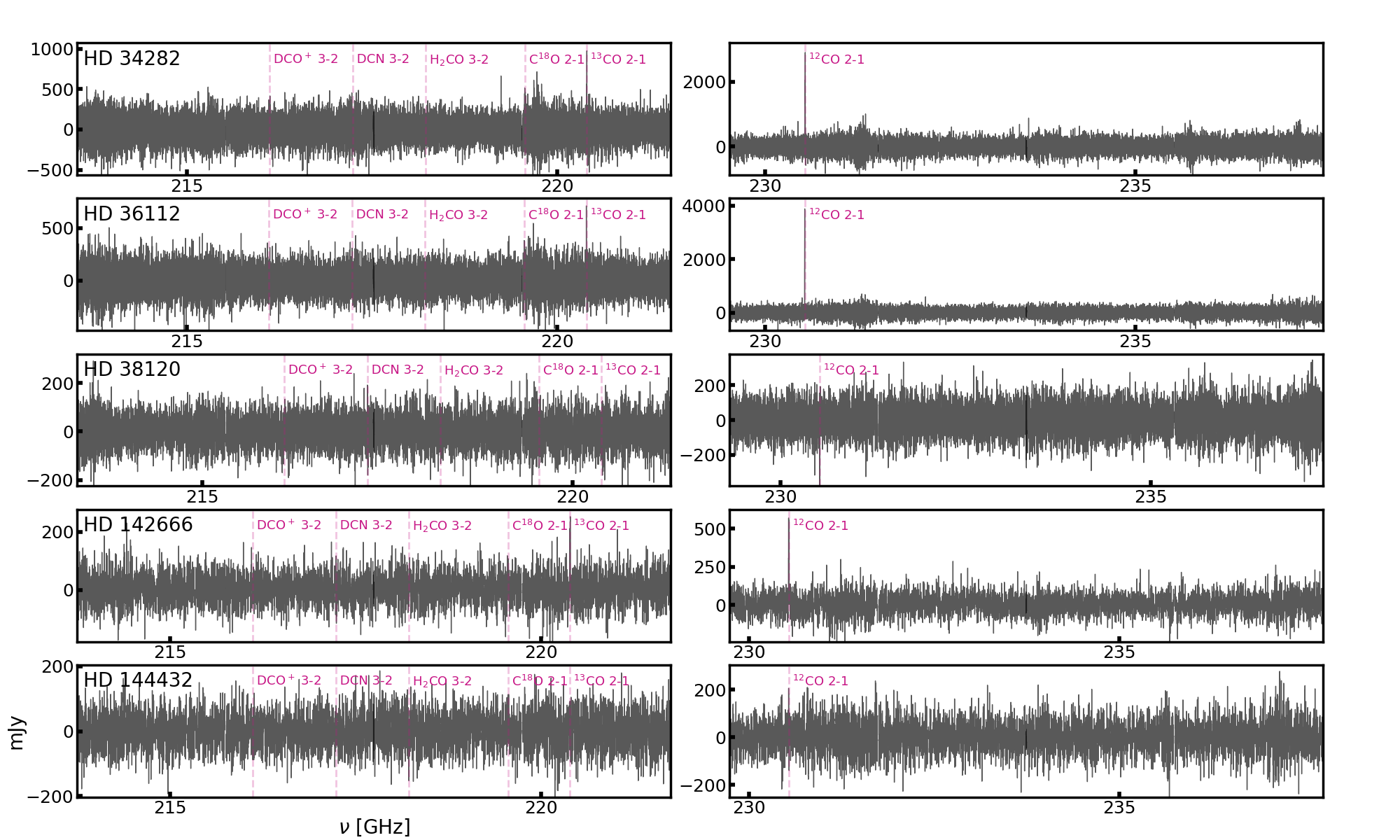}}
%
\resizebox{0.99\hsize}{!}{
    \includegraphics[trim=0pt 0pt 0pt 10pt, clip]{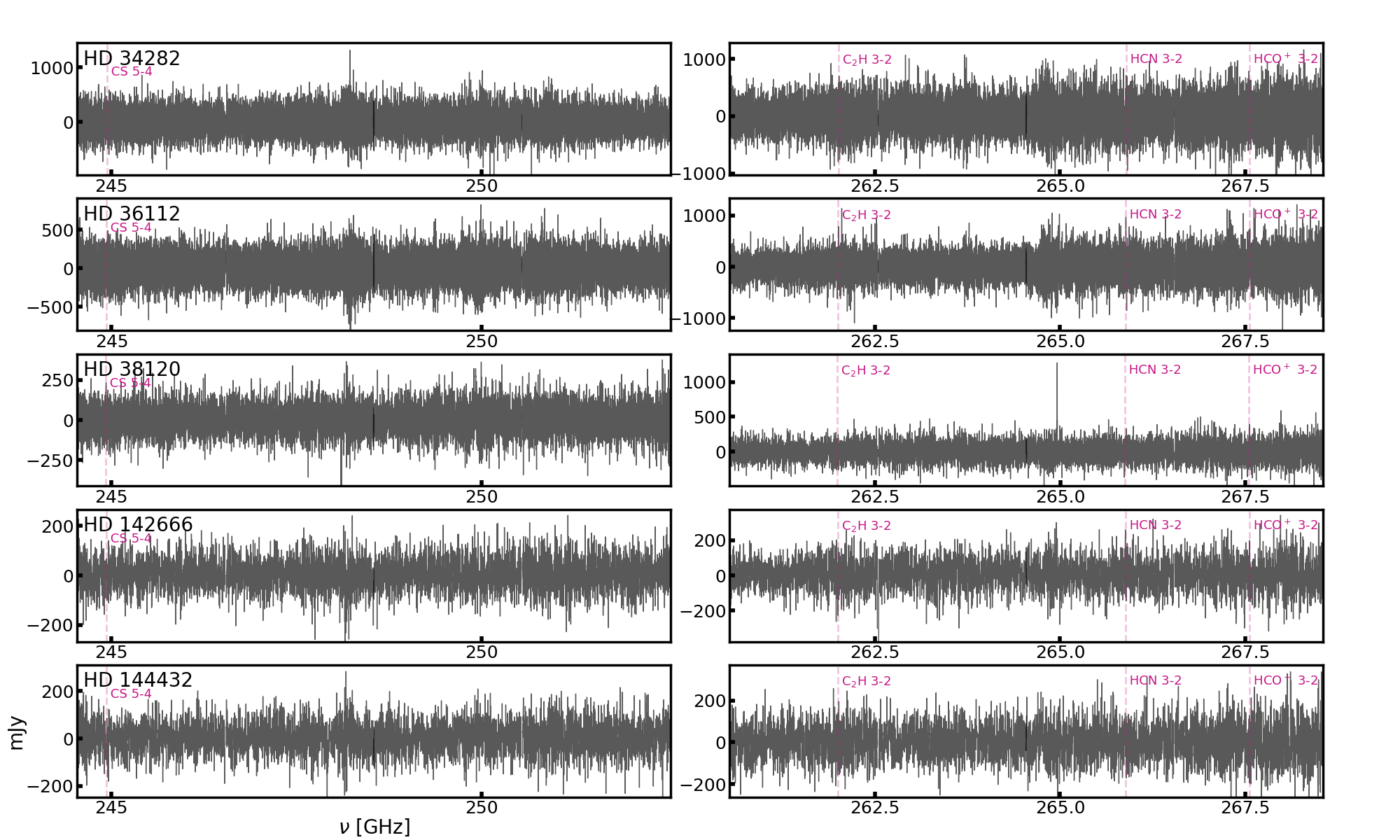}}
\caption{Unbiased spectra (i.e., spectra extracted with uniform elliptical masks; Section~\ref{sec_analysis_masks}) for all disks (split by row) and for all bands observed with the 230 GHz (top panels) and 240 GHz (bottom panels) SMA receiver.  The spectra per band are allowed to overlap in the panels at the bands' edges; in our analysis, however, we excluded the $\sim$0.1 GHz of each band edge to avoid artifacts.  The ten target molecular lines (Table~\ref{table_mol}) are marked with dashed pink lines.
\label{fig_rxs}}
\end{figure*}

Figure~\ref{fig_rxs} displays the spectral data across all SMA spectral bands for the 230 GHz and 240 GHz SMA receivers.

\clearpage
\section{Channel Maps}
\label{sec_appendix_chanmaps}

\subsection{Channel Maps for HD 34282}

\begin{figure*}
\centering
\resizebox{0.99\hsize}{!}{
    \includegraphics[trim=70pt 25pt 20pt 45pt, clip]{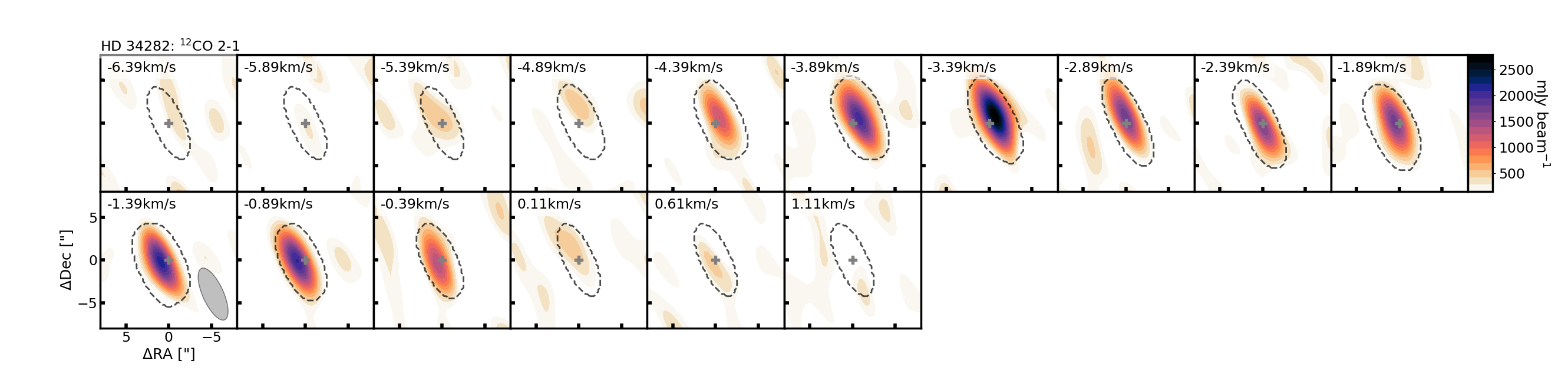}}
\resizebox{0.99\hsize}{!}{
    \includegraphics[trim=70pt 25pt 20pt 45pt, clip]{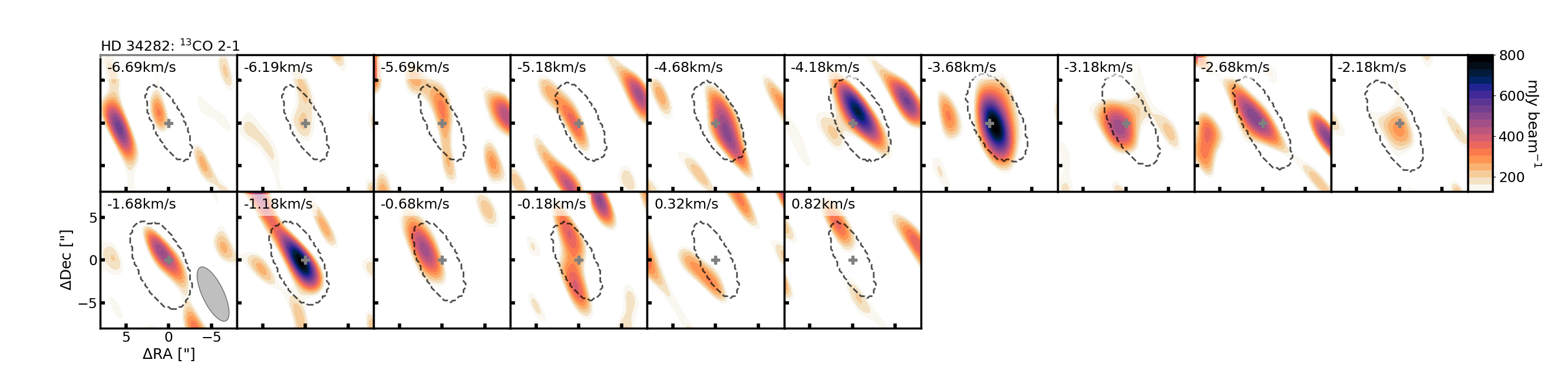}}
\resizebox{0.99\hsize}{!}{
    \includegraphics[trim=70pt 25pt 20pt 45pt, clip]{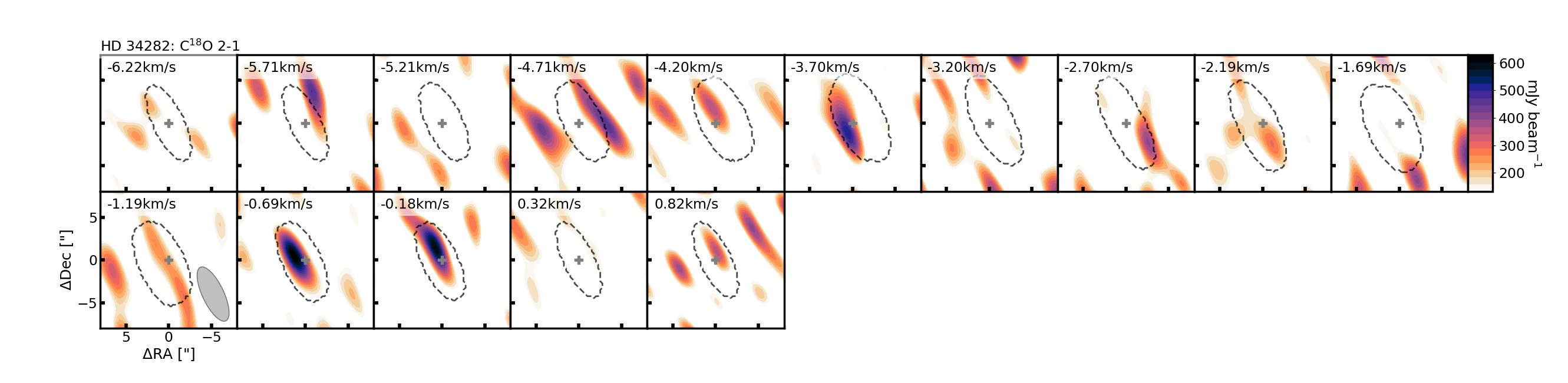}}
\resizebox{0.99\hsize}{!}{
    \includegraphics[trim=70pt 25pt 20pt 45pt, clip]{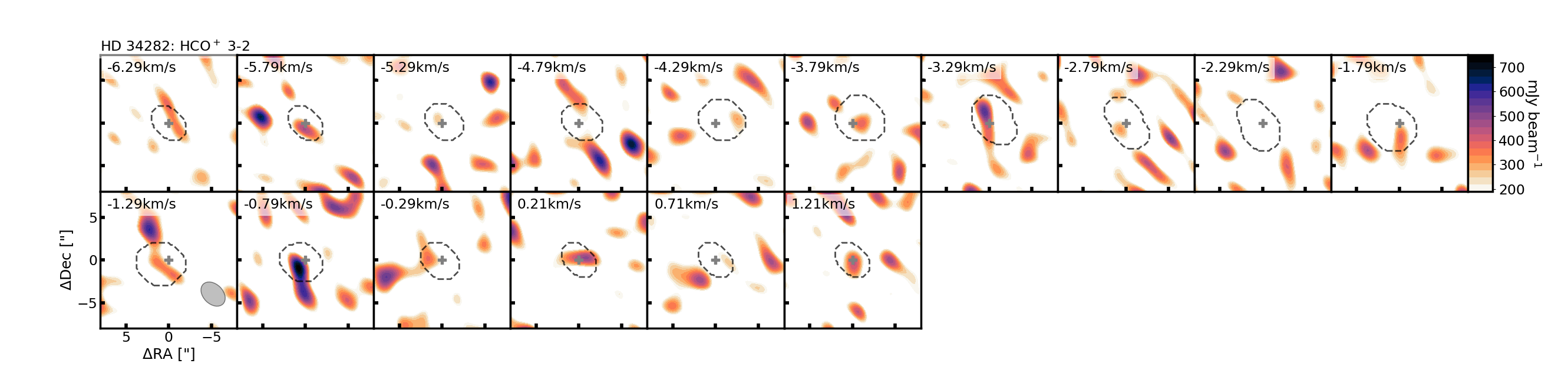}}
\resizebox{0.99\hsize}{!}{
    \includegraphics[trim=70pt 25pt 20pt 45pt, clip]{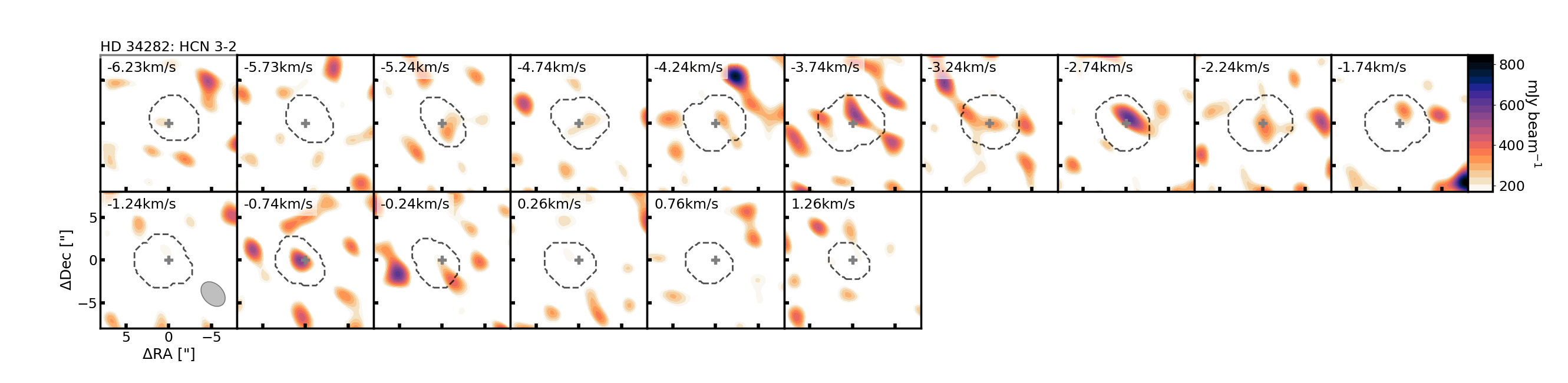}}
\caption{$^{12}$CO 2--1 (first panel), $^{13}$CO 2--1 (second panel), C$^{18}$O 2--1 (third panel), HCO$^+$ 3--2 (fourth panel), and HCN 3--2 (fifth panel) emission from HD 34282.  Emission is shown above 1$\sigma$, where $\sigma$ is the channel rms (Table~\ref{table_linefluxes}).  The estimated disk center is marked with a gray `$+$'.  Keplerian masks are overlaid with dashed lines, and synthesized beams are drawn in the lower left corner of the entire figure.
\label{fig_chanmaps_hd34282_p1}}
\end{figure*}

\begin{figure*}
\centering
\resizebox{0.99\hsize}{!}{
    \includegraphics[trim=70pt 25pt 20pt 45pt, clip]{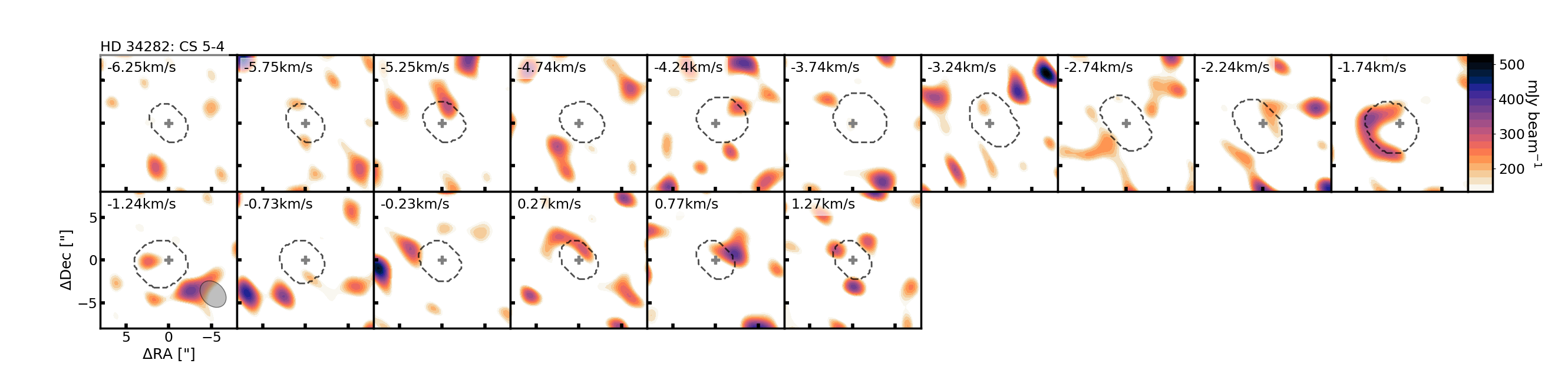}}
\resizebox{0.99\hsize}{!}{
    \includegraphics[trim=70pt 25pt 20pt 45pt, clip]{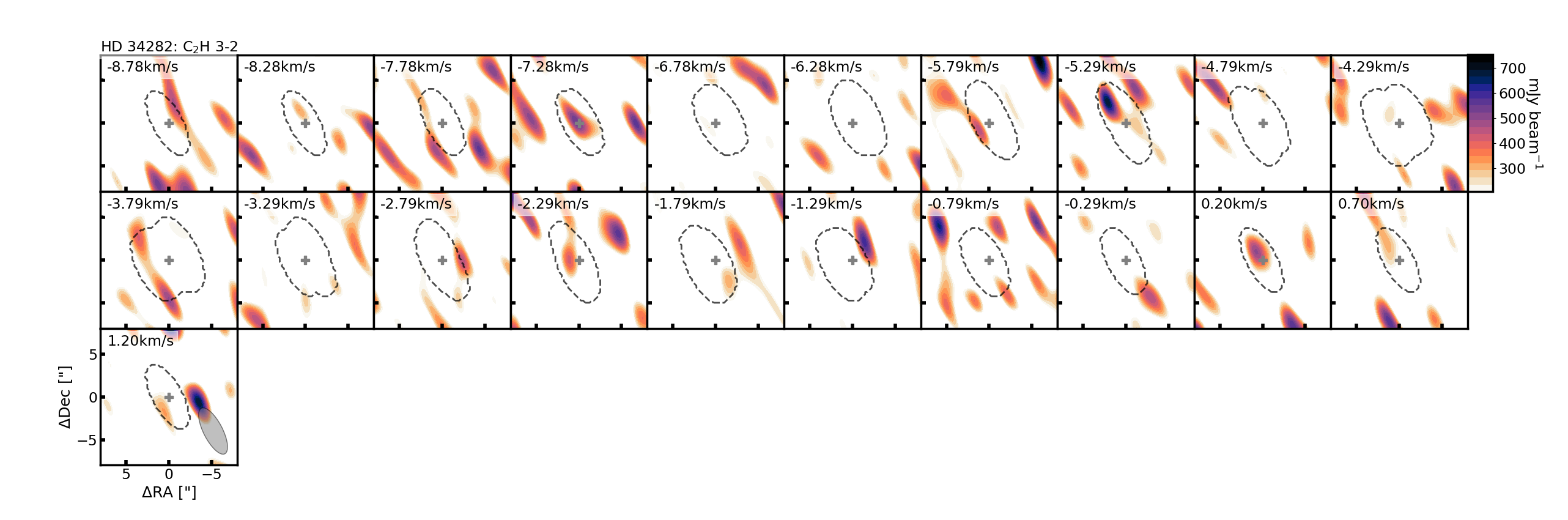}}
\resizebox{0.99\hsize}{!}{
    \includegraphics[trim=70pt 25pt 20pt 45pt, clip]{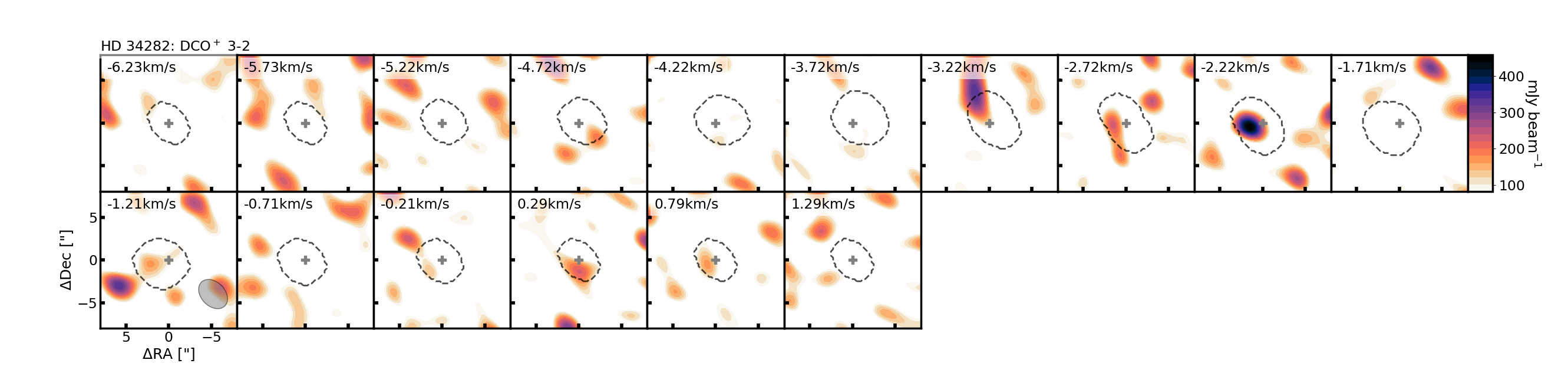}}
\caption{CS 5--4 (first panel), C$_2$H 3--2 (second panel), and DCO$^+$ 3--2 (third panel) emission from HD 34282.  Emission is shown above 1$\sigma$, where $\sigma$ is the channel rms (Table~\ref{table_linefluxes}).  The estimated disk center is marked with a gray `$+$'.  Keplerian masks are overlaid with dashed lines, and synthesized beams are drawn as gray ellipses in the lower left corner of the entire figure.
\label{fig_chanmaps_hd34282_p2}}
\end{figure*}

Figures~\ref{fig_chanmaps_hd34282_p1} and~\ref{fig_chanmaps_hd34282_p2} display channel maps for molecular line emission detected or tentatively detected from HD 34282.

%

\subsection{Channel Maps for HD 36112}

\begin{figure*}
\centering
\resizebox{0.99\hsize}{!}{
    \includegraphics[trim=70pt 25pt 20pt 45pt, clip]{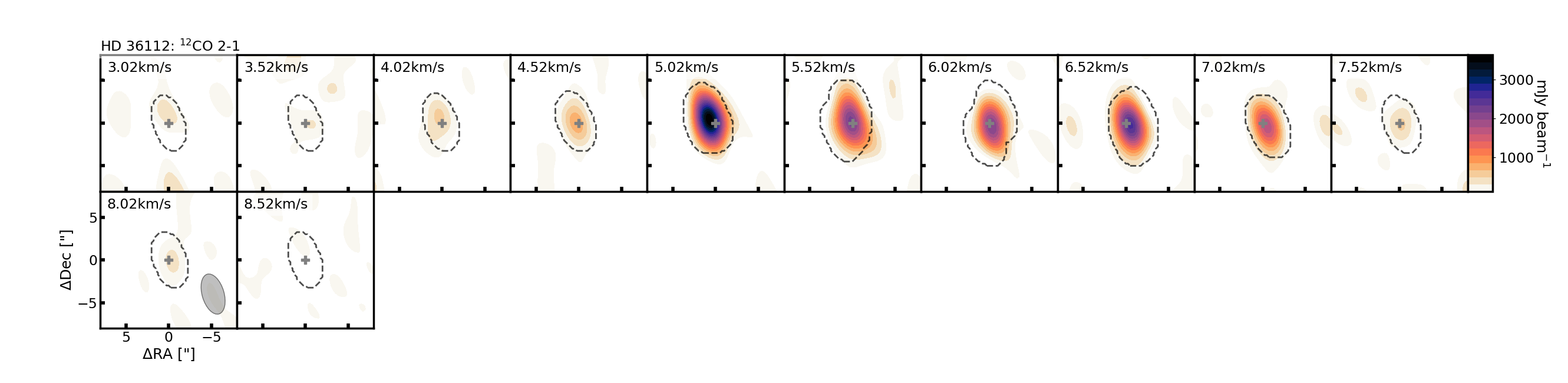}}
\resizebox{0.99\hsize}{!}{
    \includegraphics[trim=70pt 25pt 20pt 45pt, clip]{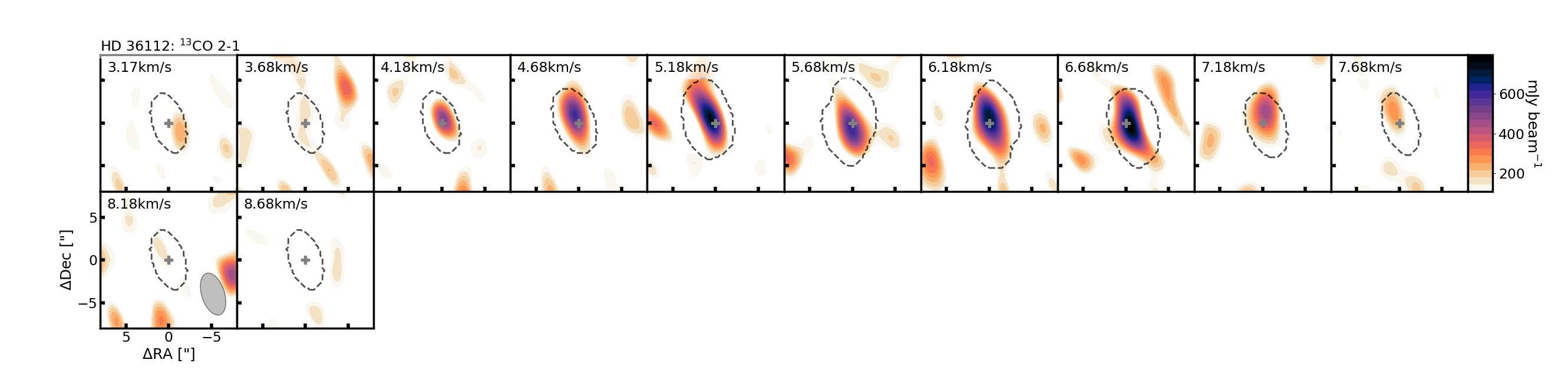}}
\resizebox{0.99\hsize}{!}{
    \includegraphics[trim=70pt 25pt 20pt 45pt, clip]{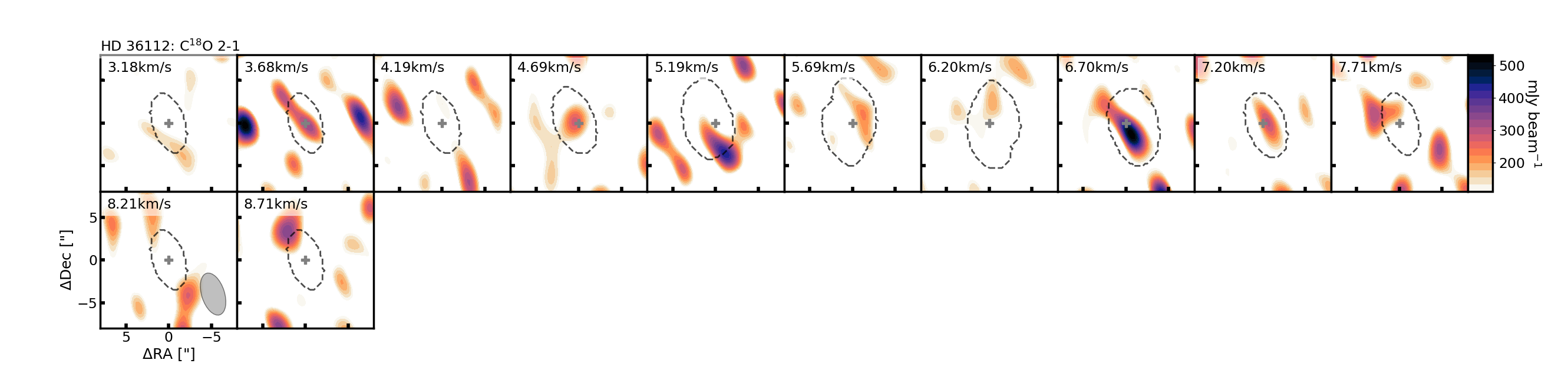}}
\resizebox{0.99\hsize}{!}{
    \includegraphics[trim=70pt 25pt 20pt 45pt, clip]{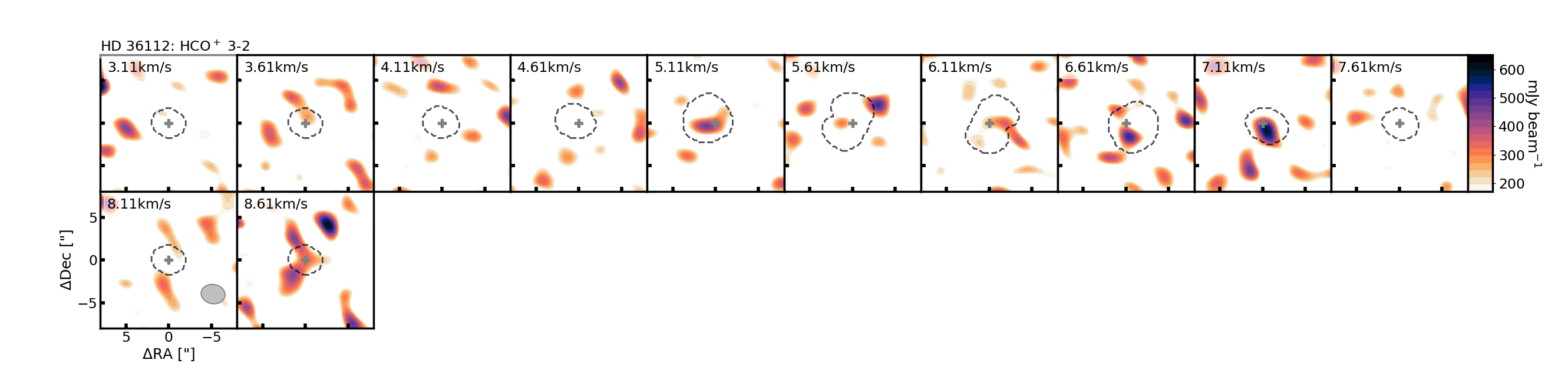}}
\resizebox{0.99\hsize}{!}{
    \includegraphics[trim=70pt 25pt 20pt 45pt, clip]{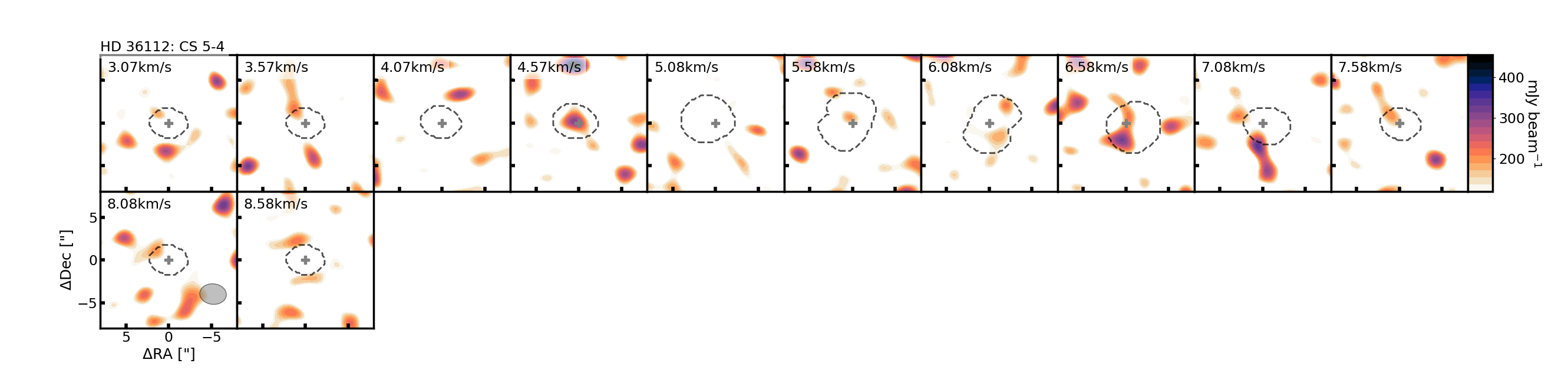}}
\resizebox{0.99\hsize}{!}{
    \includegraphics[trim=70pt 25pt 20pt 45pt, clip]{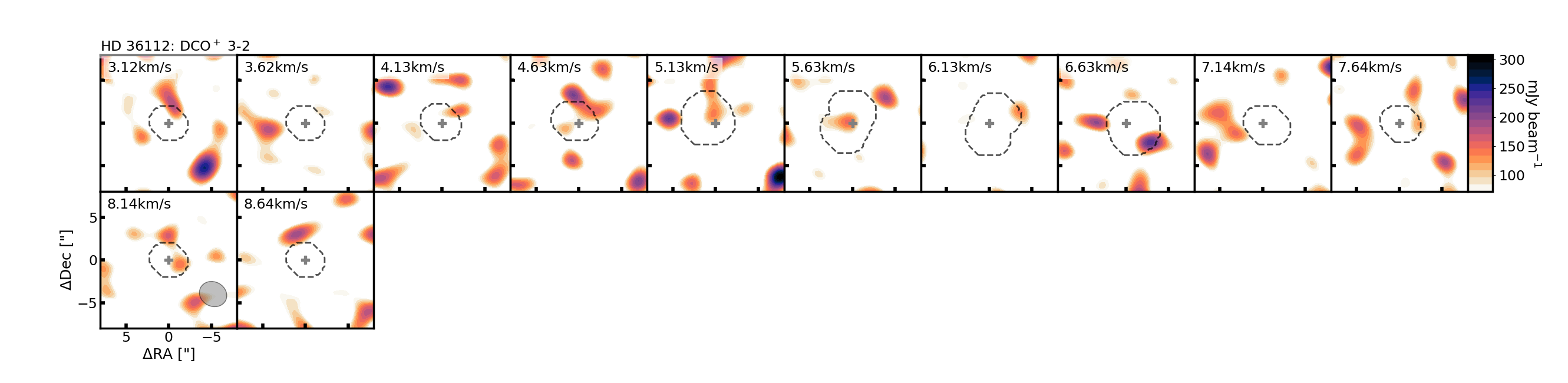}}
\caption{$^{12}$CO 2--1 (first panel), $^{13}$CO 2--1 (second panel), C$^{18}$O 2--1 (third panel), HCO$^+$ 3--2 (fourth panel), CS 5--4 (fifth panel), and DCO$^+$ 3--2 (sixth panel) emission from HD 36112.  Emission is shown above 1$\sigma$, where $\sigma$ is the channel rms (Table~\ref{table_linefluxes}).  The estimated disk center is marked with a gray `$+$'.  Keplerian masks are overlaid with dashed lines, and synthesized beams are drawn as gray ellipses in the lower left corner of the entire figure.
\label{fig_chanmaps_hd36112_p1}}
\end{figure*}

Figure~\ref{fig_chanmaps_hd36112_p1} displays channel maps for molecular line emission detected or tentatively detected from HD 36112.

%

\subsection{Channel Maps for HD 142666}

\begin{figure*}
\centering
\resizebox{0.99\hsize}{!}{
    \includegraphics[trim=70pt 25pt 240pt 45pt, clip]{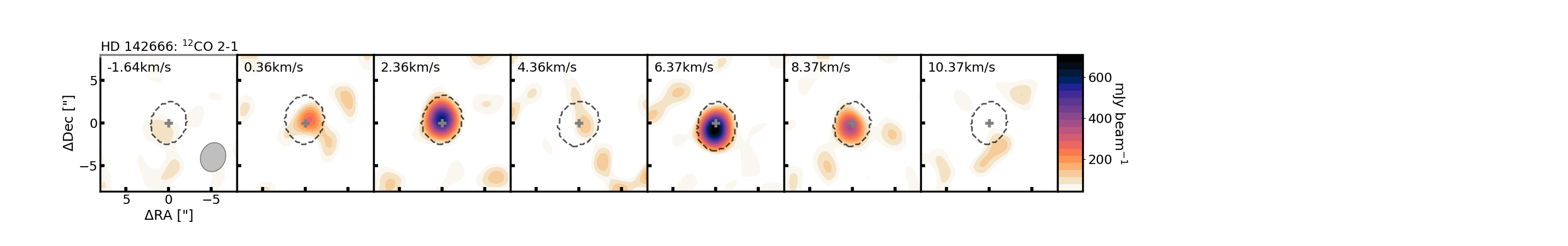}}
\resizebox{0.99\hsize}{!}{
    \includegraphics[trim=70pt 25pt 240pt 45pt, clip]{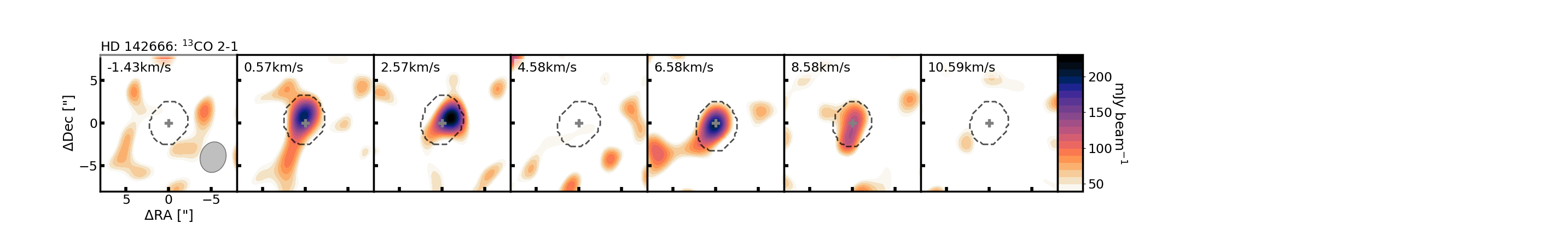}}
\resizebox{0.99\hsize}{!}{
    \includegraphics[trim=70pt 25pt 240pt 45pt, clip]{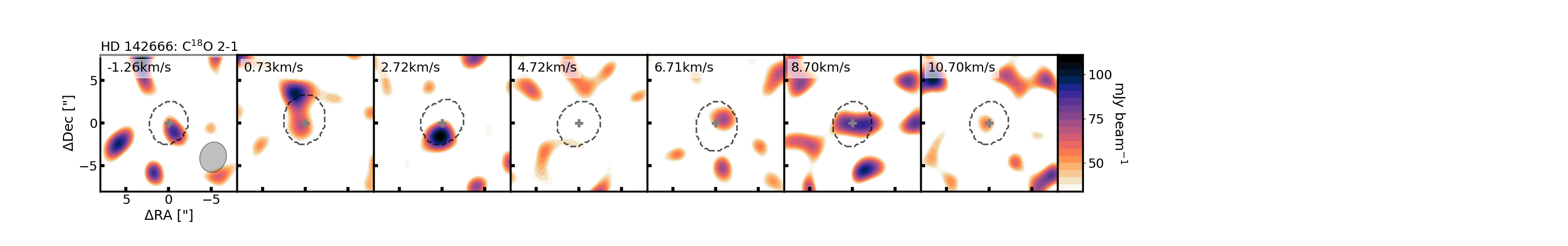}}
\resizebox{0.99\hsize}{!}{
    \includegraphics[trim=70pt 25pt 240pt 45pt, clip]{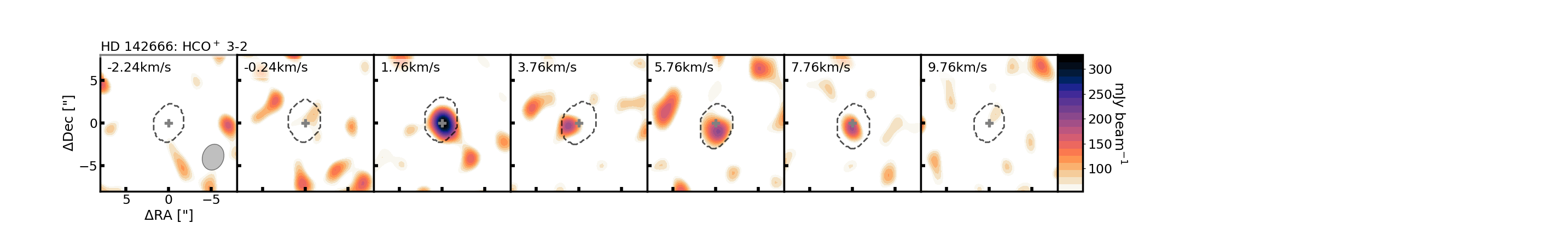}}
\resizebox{0.99\hsize}{!}{
    \includegraphics[trim=70pt 25pt 240pt 45pt, clip]{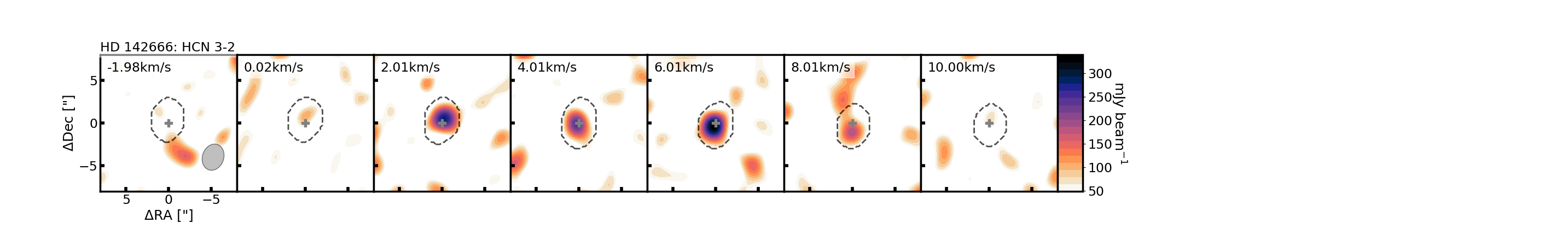}}
\caption{$^{12}$CO 2--1 (first panel), $^{13}$CO 2--1 (second panel), C$^{18}$O 2--1 (third panel), HCO$^+$ 3--2 (fourth panel), and HCN 3--2 (fifth panel) emission from HD 142666.  Emission is shown above 1$\sigma$, where $\sigma$ is the channel rms (Table~\ref{table_linefluxes}).  The estimated disk center is marked with a gray `$+$'.  Keplerian masks are overlaid with dashed lines, and synthesized beams are drawn as gray ellipses in the lower left corner of the entire figure.
\label{fig_chanmaps_hd142666_p1}}
\end{figure*}

Figure~\ref{fig_chanmaps_hd142666_p1} displays channel maps for molecular line emission detected or tentatively detected from HD 142666.

%

\subsection{Channel Maps for HD 144432}

\begin{figure*}
\centering
\resizebox{0.99\hsize}{!}{
    \includegraphics[trim=70pt 25pt 240pt 45pt, clip]{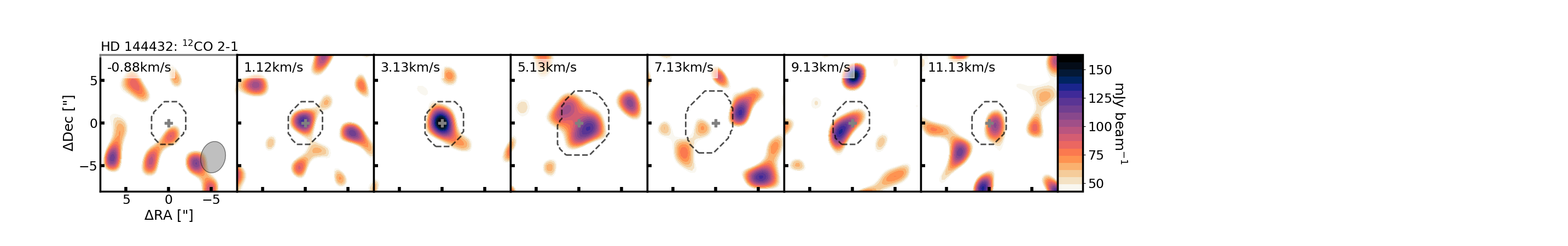}}
\resizebox{0.99\hsize}{!}{
    \includegraphics[trim=70pt 25pt 240pt 45pt, clip]{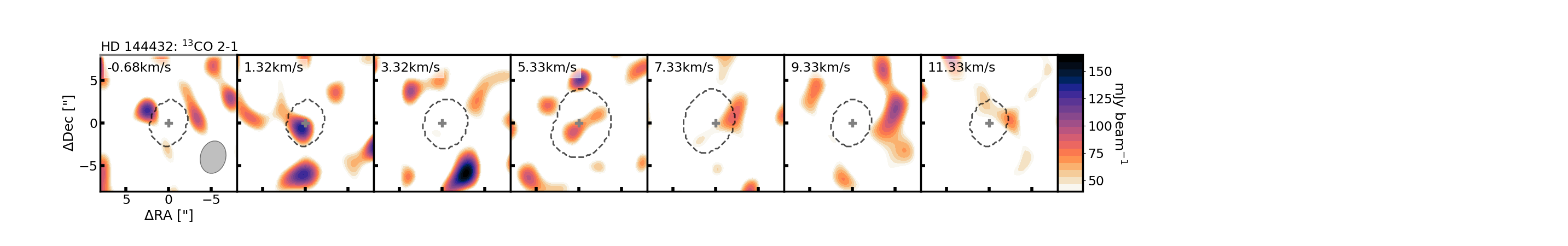}}
\resizebox{0.99\hsize}{!}{
    \includegraphics[trim=70pt 25pt 240pt 45pt, clip]{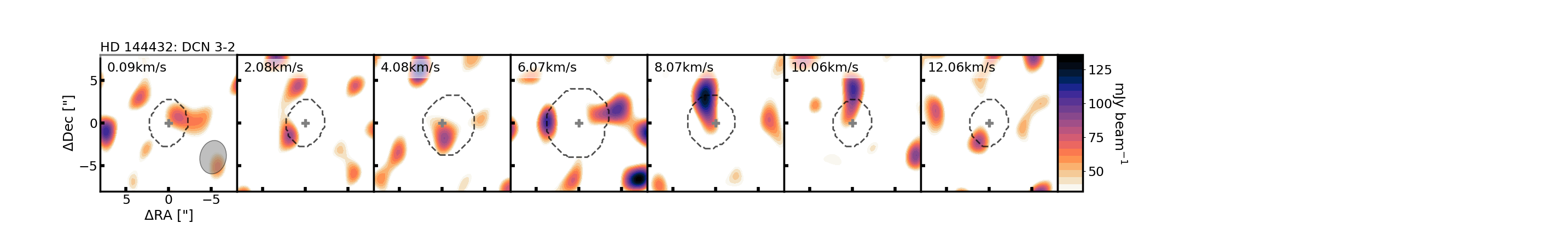}}
\caption{$^{12}$CO 2--1 (first panel), $^{13}$CO 2--1 (second panel), and DCN 3--2 (third panel) emission from HD 144432.  Emission is shown above 1$\sigma$, where $\sigma$ is the channel rms (Table~\ref{table_linefluxes}).  The estimated disk center is marked with a gray `$+$'.  Keplerian masks are overlaid with dashed lines, and synthesized beams are drawn as gray ellipses in the lower left corner of the entire figure.
\label{fig_chanmaps_hd144432_p1}}
\end{figure*}

Figure~\ref{fig_chanmaps_hd144432_p1} displays channel maps for molecular line emission detected or tentatively detected from HD 144432.

%

\clearpage


\section{ALMA Archival Observations of HD 34282}
\label{sec_appendix_hd34282}

\begin{figure*}
\centering
\resizebox{0.99\hsize}{!}{
    \includegraphics[trim=40pt 35pt 20pt 30pt, clip]{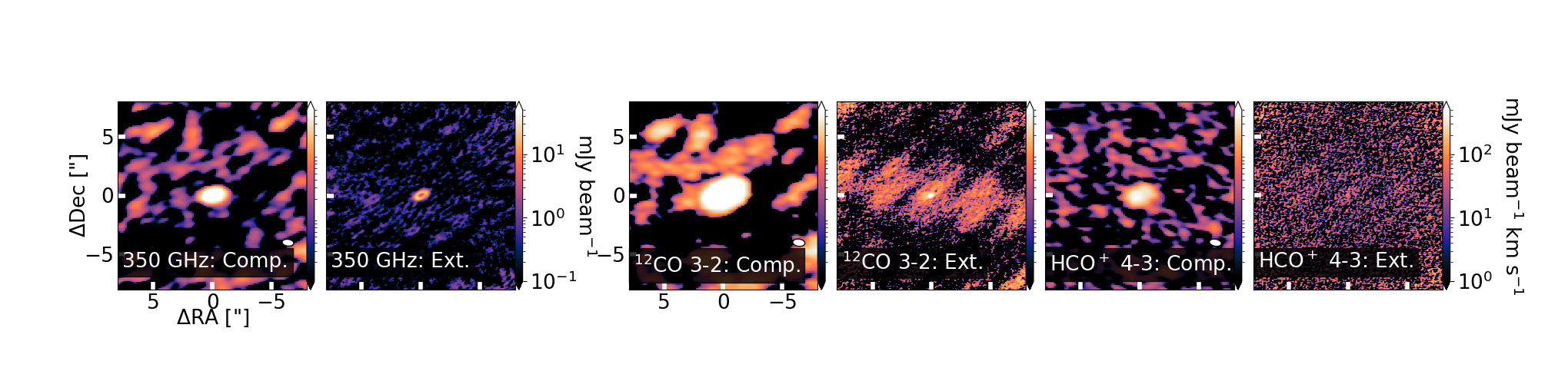}}
\caption{Images of HD 34282 emission extracted directly from the ALMA archive for the ALMA observing project 2013.1.00658.S.  The emission shown is for the 350 GHz continuum (leftmost two panels), $^{12}$CO 3--2 emission (middle two panels), and HCO$^+$ 4--3 emission (rightmost two panels).  The emission was observed in either the compact (``Comp.'') or extended (``Ext.'') configurations.  Synthesized beams are drawn in white on the bottom right of each panel.  The continuum images share the same colorbar, as do the line emission images.  Both colorbars are in log scale to emphasize faint emission.
\label{fig_hd34282_archive}}
\end{figure*}

Figure~\ref{fig_hd34282_archive} displays images of HD 34282 emission extracted directly from the ALMA archive for the observing project 2013.1.00658.S.

\end{document}